\documentclass[aps,pra,superscriptaddress,longbibliography]{revtex4-2}

\usepackage{amsmath}
\usepackage{physics}
\usepackage{amssymb}
\usepackage{mathrsfs}  

\usepackage{graphicx}
\usepackage{float}

\usepackage[normalem]{ulem}

\newcommand{\kk}{\mathbf{k}}
\newcommand{\rr}{\mathbf{r}}
\newcommand{\Psih}{\hat{\Psi}}
\newcommand{\Psihd}{\hat{\Psi}^\dagger}

\newcommand{\rhoh}{\hat{\rho}}

\newcommand{\ah}{\hat{a}}
\newcommand{\ahd}{\hat{a}^\dagger}
\newcommand{\bh}{\hat{b}}
\newcommand{\bhd}{\hat{b}^\dagger}

\begin{document}
\title{How to exploit driving and dissipation to stabilize and manipulate quantum many-body states}
\author{Iacopo Carusotto}


\affiliation{Pitaevskii BEC Center, INO-CNR and Dipartimento di Fisica, University of Trento (Italy)}
\email{iacopo.carusotto@ino.cnr.it}


\begin{abstract}
We review the basic concepts of quantum fluids of light and the different techniques that have been developed to exploit driving and dissipation to stabilize and manipulate interesting many-body states. In the weakly interacting regime, this approach has allowed to study, among other, superfluid light, non-equilibrium Bose-Einstein condensation, photonic analogs of Hall effects, and is opening the way towards the realization of a new family of analog models of gravity. In the strongly interacting regime, the recent observations of Mott insulators and baby Laughlin fluids of light open promising avenues towards the study of novel strongly correlated many-body states.
\end{abstract}
%

\maketitle

\section{Introduction}


In the last decades, a most promising platform for many-body physics has emerged thanks to the dramatic advances in the laser cooling techniques for dilute atomic gases. This has allowed to reach unprecedented temperatures in the sub-nanokelvin range and has allowed the clean observation of textbook many-body phenomena such as Bose-Einstein condensation (BEC), the crossover from BEC to a Bardeen-Cooper-Schrieffer (BCS) superconductive state of Fermi gases, and first realizations of strongly interacting states of matter such Mott insulators and fractional quantum Hall fluids~\cite{pitaevskii2016bose,gross2017quantum,bloch2008many,cooper2019topological}. As compared to traditional condensed matter systems which are subject to sizable fabrication disorder and display additional effects, atomic gases enjoy an extraordinary level of cleanness and control at the microscopic level, which permits a quantitative comparison with ab initio theoretical calculations. Still, the available values of the temperature are typically still too high to observe the most exciting strongly correlated states of matter.

More or less in the same years, the merging of ideas from quantum optics and many-body physics have led to the development of the concept of fluid of light, namely an assembly of photons  confined in suitable cavity devices where they display a finite effective mass and sizable interactions mediated by the optical nonlinearity of the medium~\cite{carusotto2013quantum}. As a result of these features, the photon gas start displaying the collective properties of a standard quantum fluid such as superfluidity and condensation and, by now, an active research is being devoted to the quest for strongly interacting states of the fluid of light~\cite{chang2014quantum,carusotto2020photonic}. As a key difference compared to atomic gases, fluids of light in cavity configurations are typically characterized by a driven-dissipative nature, where the state of the fluid is not determined by a thermal equilibrium condition, but rather by a dynamical balance of pumping and losses~\cite{bloch2022non,sieberer2016keldysh}.

This was initially considered as a serious drawback, as it made it difficult to realize the most celebrated textbook models of many-body physics. But over the years it has become clear that the non-equilibrium nature of the fluid opens the way to new physical phenomena and offers a wide variety of novel experimental tools for the stabilization, the manipulation and the diagnostics of the fluid. In particular, very different states can be realized by simply tuning the properties of the laser beams used to generate the fluid of light. As the open nature of the fluid of light naturally provides a way to exchange energy and particles with the external world, it can be exploited to reduce the level of thermal-like excitation of the fluid. Rather than trying to cool the system towards lower and lower temperatures, the idea is to design driving and loss mechanisms that autonomously stabilize the system into the desired quantum many-body state.

Quite interestingly, these efforts in the non-equilibrium stabilization of fluids of light go on par with analogous developments in the context of cold atoms aiming at exploiting driving and dissipation to stabilize interesting many-body states of atomic matter~\cite{Diehl:NatPhys2008,Bardyn:PRL2012,Budich:PRA2015} or atomic gases coupled to optical cavities~\cite{Mivehvar:AdP2021,Ferri:PRX2021,Marsh:PRX2024}. As key difference, however, the typical dissipation schemes considered for atoms do not involve a net flux of atoms through the system from/to external reservoirs. Stimulated by recent pioneering experiments~\cite{Chen:Nature2022,Labouvie:PRL2016,Benary:NJP2022}, it will be a exciting new frontier to explore the new non-equilibrium physics of driven-dissipative cold atomic gases that can exchange particles with one or more external reservoirs.

The goal of this article is to summarize the main ideas in the physics of fluids of light and introduce the reader to the different strategies that have been developed  in the context for the stabilization of different quantum many-body states. As several reviews are already available on this topic~\cite{carusotto2013quantum,chang2014quantum,carusotto2020photonic,bloch2022non,sieberer2016keldysh}, our focus will be on a pedagogical presentation of the main theoretical tools that can be used to describe the system under different pumping conditions and to design schemes to exploit the driven-dissipative nature for novel investigations of many-body physics. Reference to some most relevant experiments will also be given, yet without the ambition of providing a complete bibliography on this wide and active field of research. In particular, we hope that our presentation will be useful to those researchers that wish to export ideas of quantum fluids of light into the field of cold atoms, so to realize driven-dissipative schemes that autonomously stabilize strongly correlated states of matter in ultracold atomic gases.

The structure of the article is the following. In Sec.\ref{sec:QFL_general} we give the basic concepts in the field of fluids of light and we summarize the basic blocks of their theoretical description. Sec.\ref{sec:weakly} is focussed to the regime of weakly interacting fluids where a mean-field theory provides an accurate description of the system: after a brief summary of the theoretical description, we will outline a few specific configurations where the driven-dissipative nature of the fluid of light opens new exciting possibilities of stabilizing and manipulating many-body states. In Sec.\ref{sec:strongly}, we turn to strongly interacting systems, with a special attention being paid to the stabilization and the manipulation of driven-dissipative fractional quantum Hall states. In Sec.\ref{sec:conclusions} we give our conclusions and we sketch our vision of the perspectives of this field, with a special emphasis on trasferring ideas of quantum fluids of light back to atomic gases in novel driven-dissipative regimes.

\section{Basic concepts of quantum fluids of light}
\label{sec:QFL_general}

In our intuitive picture, we are used to associate light to propagating electromagnetic waves or, in a quantum description, to a stream of corpuscular photons that travel across space at a very fast speed. The concept of {\em Quantum Fluid of Light (QFL)} defies this picture. Merging ideas from condensed matter physics and optics, it deals with the collective behaviours that assemblies of photons display when they are endowed of an effective mass and sizable interactions~\cite{carusotto2013quantum} and are then manipulated as a standard fluid of many interacting particles.

This Section starts with a brief review of these basic concepts and, then, elaborates on the intrinsically driven-dissipative nature of the fluid of light in cavity configurations and on the tools that are needed for its theoretical description.


\subsection{Conservative dynamics}

\subsubsection{Photon mass in cavity configurations}

\begin{figure}
    \centering
    \includegraphics[width=0.58\textwidth]{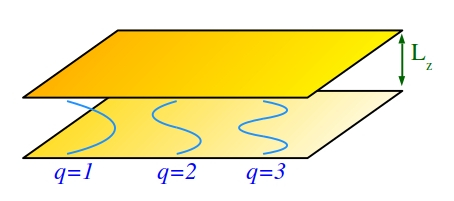}\vspace{0.25cm}\\
    \includegraphics[width=0.45\textwidth]{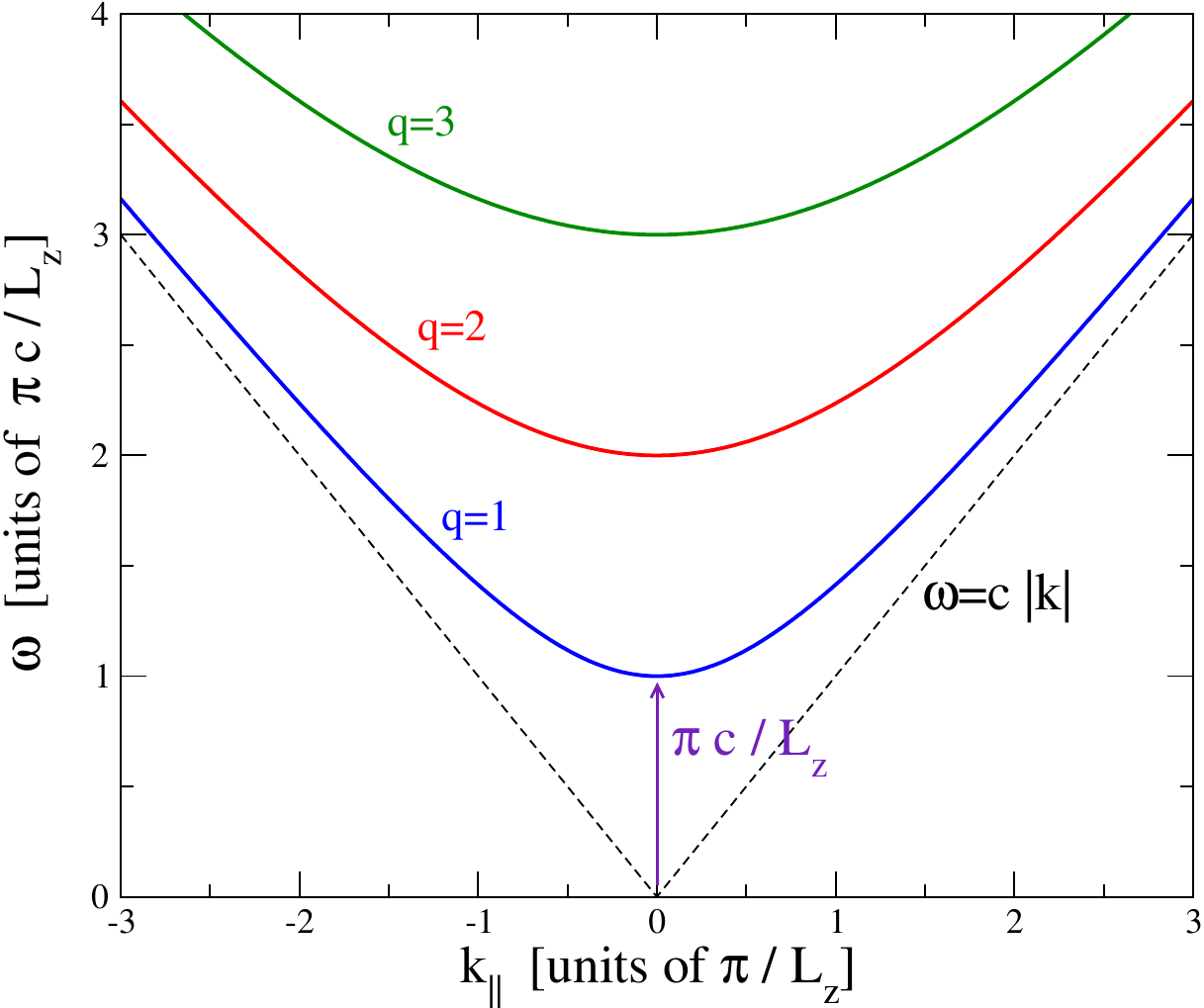}\hspace{1cm}
    \parbox[b]{0.4\textwidth}{ \includegraphics[width=0.35\columnwidth]{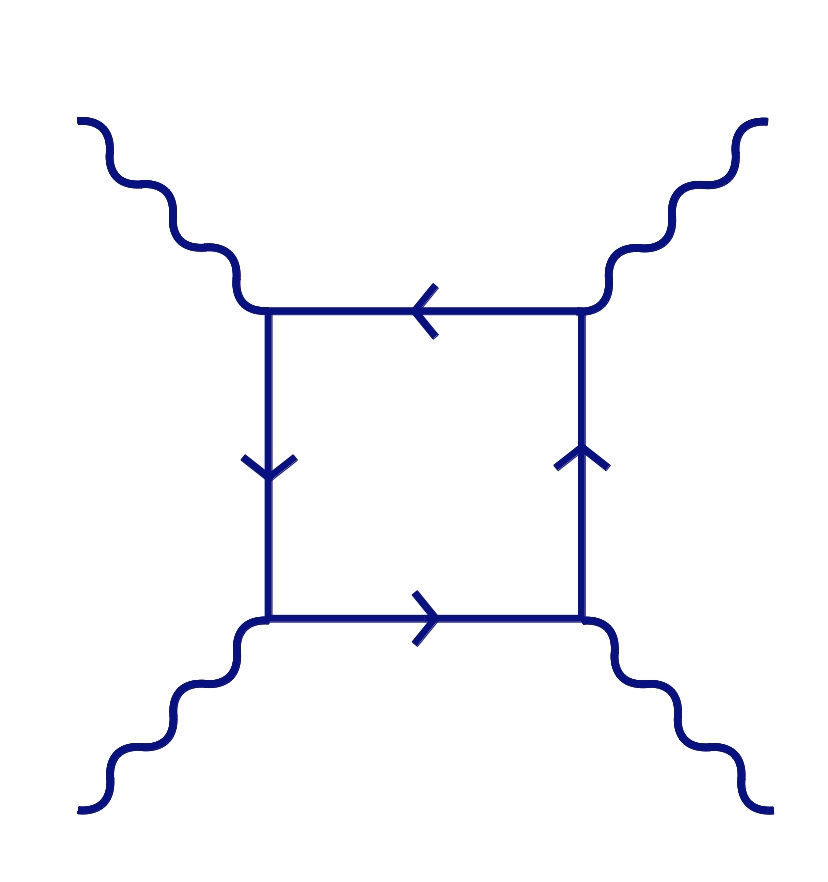}\vspace{0.4cm}}
    \caption{Light confinement in a planar microcavity (top) results in a relativistic dispersion for the in-plane motion of photons (bottom left).
    Feynman diagram describing the Heisenberg-Euler photon-photon interaction processes mediated by virtual electron-positron pairs in Quantum Electrodynamics (bottom right).
    \label{fig:basics}}
\end{figure}

Within special relativity, massless particles travel at a constant speed $c$ and a finite mass is required to put particles at rest. In vacuo, the photons that constitute light beams are massless particles. The situation can dramatically change in spatially confined geometries, where photons can acquire a finite effective mass for their motion along some dimensions thanks to the spatial confinement along the other dimensions. 

The simplest case is the one sketched in the top panel of Fig.\ref{fig:basics}, namely a cavity enclosed by a pair of plane-parallel metallic mirrors located along the $xy$ plane and separated by a distance $L_z$ along $z$. As the tangential component of the electric field has to vanish at the metallic mirrors, the $z$ component of the wavevector is quantized as $k_z^{(q)}=\pi q / L_z$ according to the positive integer number $q>0$ characterizing the number of field nodes along the $z$ direction. On the other hand, given the translational symmetry along the $xy$ plane, the motion along these directions is free and the corresponding component $\kk_\parallel$ of the wavevector can have arbitrary values.

As a result, forgetting for simplicity the polarization degrees of freedom, electromagnetic waves in a planar cavity filled of a material of refractive index $n$ can be described in terms of a massive relativistic-like dispersion,
\begin{equation}
\left( \hbar \omega^{(q)}(\kk_\parallel)\right)^2= \left({m^{(q)} c^2}\right)^2 + \frac{\hbar^2  c^2}{n^2} \,\kk_\parallel^2
 \label{eq:dispersion}
\end{equation}
where the finite effective mass of each $q$ branch
\begin{equation}
m^{(q)} c^2 = \frac{\hbar c k_z^{(q)}}{n}=\hbar \omega^{(q)}_o
 \label{eq:mass}
\end{equation}
arises from the zero-point confinement along $z$ and grows with the quantum number $q$. As usual in special relativity, the mass parameter $m^{(q)}$ encodes both a {\em rest mass} related to the energy $\hbar \omega^{(q)}_o$ of a photon at rest in the zero-momentum $\kk_\parallel=0$ state and a {\em kinetic mass} related to growth of the energy of a photon when this is set into motion. In the bottom left panel of Fig.\ref{fig:basics}, the former can be read out as the energy gap below the dispersion, while the latter gives the curvature of the dispersion around its bottom.
Under the non-relativistic approximation of \eqref{eq:dispersion}, a parabolic form of the dispersion is obtained for each $q$ branch at small in-plane momenta, 
\begin{equation}
\omega^{(q)}(\kk_\parallel) \simeq \omega^{(q)}_o + \frac{\hbar}{2m^{(q)}}
\,\kk_\parallel^2\,.
\label{eq:parabolic}
\end{equation}

Upon quantization, the field in the cavity can be expressed in terms of quantum field operators $\Psih_q(\rr_\parallel)$ satisfying bosonic commutation rules, 
\begin{equation}
\left[\Psih_q(\rr_\parallel),\Psihd_{q'}(\rr'_\parallel)\right]=\delta_{q,q'}\,\delta^{(2)}(\rr_\parallel,\rr'_\parallel)\,. 
\end{equation}
These are related to the amplitude of the in-cavity electric field component,
\begin{equation}
 \hat{E}(\rr_\parallel,z)=\sum_{q=1}^\infty\,\mathcal{E}_q(z)\,\Psih(\rr_\parallel)+\mathcal{E}_q^*(z)\,\Psihd(\rr_\parallel) 
 \end{equation}
 where the wavefunction of mode $q$ around the bottom of the branch
 \begin{equation}
  \mathcal{E}_q(z)=\sqrt{\frac{4\pi \hbar \omega_o}{n^2 L}}\,\sin\left(\frac{\pi\,q\, z}{L}\right)
  \label{eq:Eqz}
 \end{equation}
 displays nodes at $z=0,L$ and gives an energy $\hbar \omega^{(q)}_o$ per photon.

The form \eqref{eq:parabolic} of the dispersion suggests a straightforward way to generate an external potential for photons. Replacing in \eqref{eq:parabolic} the explicit dependence of $\omega^{(q)}_o$ on the cavity parameters \eqref{eq:mass}, it is in fact immediate to see how a slow lateral dependence of the refractive index $n(\rr_\parallel)$ and/or of the cavity thickness $L(\rr_\parallel)$ can be exploited to generate an effective in-plane potential profile
\begin{equation}
 V^{(q)}(\rr_\parallel)
 = \omega_o^{(q)}(\rr_\parallel)=\frac{c \pi q}{n(\rr_\parallel)\, L(\rr_\parallel)}\,.
 \label{eq:potential}
\end{equation}
Such a potential can be used, e.g., to exert uniform forces in the case of a uniform gradient or localize the photons in suitably defined potential wells.
Of course, in order to avoid introducing undesired losses due to the abrupt in-plane variation of the field mode, the $\rr_\parallel$-dependence of $L$ and $n$ has to be sufficiently smooth.

Even though these simple formulas have been obtained for the simplest case of metallic cavities, qualitatively similar results hold for other kind of planar microcavities with translational symmetry along the  $xy$ plane, in particular semiconductor DBR microcavities enclosed by distributed Bragg reflectors~\cite{kavokin2011microcavities}.

\subsubsection{Photon-photon interactions}


A superposition principle for light fields directly follows from the linearity of the classical Maxwell equations in vacuo~\cite{jackson1999classical}: light beams propagate independently of each other, and a light beam can not be used to modify the propagation of another beam. Quite remarkably, this cornerstone of classical electrodynamics ceases to be true in quantum electrodynamics where photons can be converted into electron-positron pairs and back.

Among a plethora of other effects, this interaction vertex results into the effective photon-photon interaction process first pointed out by Heisenberg and Euler in the 1930's~\cite{heisenberg1936folgerungen} and illustrated in the bottom right panel of Fig.\ref{fig:basics}. As this process is mediated by  the creation of virtual electron-positron pairs, the effective scattering cross section is only significant for high energy photons of energies comparable to the electron/positron rest mass of around $0.5\,\textrm{MeV}$, while it is dramatically suppressed at lower energies according to
\begin{equation}
 \sigma\propto \alpha^4 \, \left(\frac{h}{m_{el} c}\right)^2\,\left(\frac{\hbar \omega}{m_{el} c^2}\right)^6\,:
 \label{eq:H-E}
\end{equation}
in addition to the four fine-structure $\alpha=e^2/\hbar c\sim 1/137$ factors corresponding to the four interaction vertices and to the electron Compton wavelength $h / (m_{el} c)\simeq 2.4\,\textrm{pm}$ factors, the main suppression factor comes from the detuning of the virtual intermediate states due to the large rest mass of the virtual electron-positron pairs compared to the incident photon energy. For visible light in the eV range, this suppression factor can be as huge as $10^{-36}$ making the observation of such processes in table-top experiments very difficult. Some experimental efforts are in progress to observe photon-photon scattering in vacuo using $\gamma$-rays and experimental results have been reported for a related effect of photons scattering off the strong electrostatic field surrounding a heavy-ion~\cite{atlas2017evidence}.

The microscopic mechanism of photon-photon scattering in vacuo suggests that a huge reinforcement of the effect would be obtained if lighter particles were available as intermediate states of the scattering process. This naturally occurs in solid-state insulating materials, whose ground state consists of a filled valence band and an empty conduction band. Promoting an electron from the valence to the conduction band can be reinterpreted~\cite{ashcroft2022solid} as the creation of an electron in the conduction band and a hole in the valence band. The energy cost of this process is roughly determined by the electronic energy gap of the material, typically in the eV range. Translating this result into the formula \eqref{eq:H-E} predicts a dramatic reinforcement of the photon-photon scattering cross section by a factor of the order $10^{36}$ and suggests the actual observability of photon-photon collision processes in material media.

This intuitive picture of photon-photon collisions can be put on solid grounds in the framework of nonlinear optics~\cite{Butcher,Boyd}: here, binary photon-photon interactions are encoded in the $\chi^{(3)}$ optical nonlinearity of the medium, which describes a polarization proportional to the cube of the electric field. Such nonlinear optical processes occur in a variety of optical media; strong efforts are presently being paid to realize new materials where the strength of the optical nonlinearities is reinforced well beyond the prediction of a naive extension of \eqref{eq:H-E}. Among the most promising platforms for this purpose, we can mention superconductor-based circuit-QED devices operating in the microwave range~\cite{carusotto2020photonic,blais_RMP2021} or optically dressed atomic gases in the so-called Rydberg-EIT confgurations~\cite{chang2014quantum,peyronel2012quantum,firstenberg2013attractive}: such exceptional media are at the heart of the recent developments towards the realization of strongly interacting photon fluids that  we are going to present later on in Sec.\ref{sec:strongly}.

For the sake of simplicity, we will restrict our quantitative discussion of the photon-photon interaction Hamiltonian to a simplest case that can be worked out in analytical terms. We model photon-photon interactions as a local quartic term of the form
\begin{equation}
H_{\rm int}=\frac{\hbar g_{\rm 3d}}{2}\,\int\,d^3\rr\,: \hat{E}(\rr)^4 :
\label{eq:H_gnl}
\end{equation}
By equating\footnote{We have assumed a spatio-temporally local nonlinearity, so the total nonlinear polarization is 
\mbox{$P_{\rm nl}^{\rm (tot)}(\rr,t)=\chi^{(3)}\,E^3(\rr,t)$~.}
The uniform-space mode normalization is
\mbox{$ \mathcal{E}=({2\pi\,\hbar \omega}/{n^2})^{1/2}$~.} The amplitude of the coherent state is normalized as \mbox{$\langle \ah_\kk\rangle=(2\pi)^3\,\delta(\kk-\kk')\,\bar{\alpha}$~.}
} the nonlinear frequency shift predicted by the Hamiltonian \eqref{eq:H_gnl} for a given plane-wave mode at $\kk$ in a spatially uniform geometry when only this mode is coherently occupied with amplitude $\bar{\alpha}$,
\begin{equation}
 \delta\omega=6 g_{\rm 3d}\,|\mathcal{E}|^4 \, |\bar{\alpha}|^2
\end{equation}
with the amplitude of the nonlinear polarization at the frequency of the mode predicted by the nonlinear optics formalism,
\begin{equation}
 P_{\rm nl}=3\chi^{(3)}\,|\mathcal{E}|^2\,|\,\mathcal{E}\,|\bar{\alpha}|^2\,\bar{\alpha},
\end{equation}
we obtain that
\begin{equation}
g_{\rm 3d}=-\frac{\chi^{(3)}}{2 \hbar}\,.
\end{equation}
Moving to a planar cavity geometry completely filled by the nonlinear medium and restricting our attention to the lowest $q=1$ mode, we can insert the explicit expression for the cavity mode profile \eqref{eq:Eqz} into \eqref{eq:H_gnl} and integrate over $z$, so to we obtain a nonlinear interaction term of the usual form
\begin{equation}
 H_{\rm int}=\frac{\hbar g_{\rm nl}}{2}\,\int d^2\rr_\parallel\, \Psihd(\rr_\parallel)\,\Psihd(\rr_\parallel)\,\Psih(\rr_\parallel)\,\Psih(\rr_\parallel)
\end{equation}
with an interaction strength $g_{\rm nl}$ that is related to the third-order nonlinear susceptibility $\chi^{(3)}$ of the medium by~\cite{Ferretti:PRB2012}
\begin{equation}
 g_{\rm nl}=-\frac{18\,\pi^2 (\hbar \omega_o)^2}{n^4\,L}\,\chi^{(3)}\,.
 \label{eq:gnl_chi3}
\end{equation}

\subsubsection{The Hamiltonian}

Putting all these effects together, the dynamics of the quantum fluid of light in the lowest energy states around the bottom of the $q=1$ branch can be summarized by the following bosonic Hamiltonian~\cite{carusotto2013quantum}:
\begin{multline}
H=\int\!d^2\rr_\parallel\,\left[\hbar \omega_o \Psihd(\rr_\parallel)\,\Psih(\rr_\parallel) + \frac{\hbar^2 }{2m^*} \nabla_\parallel\Psihd(\rr_\parallel)\,\nabla_\parallel\Psih(\rr_\parallel) +\hbar V(\rr_\parallel)\,\Psihd(\rr_\parallel)\,\Psih(\rr_\parallel) +\right.\\
+ \left.\frac{\hbar g_{\rm nl}}{2} \Psihd(\rr_\parallel)\,\Psihd(\rr_\parallel)\,\Psih(\rr_\parallel)\,\Psih(\rr_\parallel)\right]\,.
\label{eq:Hcav}
\end{multline}
Here, the first and second terms respectively describe the rest energy of the photons and their (nom-relativistic) in-plane kinetic energy, as defined in (\ref{eq:dispersion}-\ref{eq:mass}). The third term models the external potential \eqref{eq:potential} acting on the photons and the fourth one accounts for the (contact) photon-photon interactions mediated by the optical nonlinearity \eqref{eq:gnl_chi3}. All together, it is immediate to recognize this Hamiltonian exactly corresponds to the one of a bosonic gas of material particles, in particular of ultracold atoms~\cite{pitaevskii2016bose}.

\subsection{Driving and dissipation} 
\label{sec:theory_DD}

In contrast to fluids of material particles where the lifetime of the microscopic constituents is typically very long compared to the many-body dynamics of interest, photons in the typical cavity devices used for experiments on quantum fluids of light experience a significant decay rate, due to a combination of non-radiative losses by absorption processes in the cavity material and radiative ones through the cavity mirrors. 

Two philosophies are then available to deal with these processes. On one hand, a part of the community is making strong efforts to improve the cavity design to maximize the lifetime and, in this way, the time-window available for the conservative many-body dynamics. In addition to the optimized planar microcavity devices to study, e.g.,  quasi-equilibrium Bose-Einstein condensation effects in polariton fluids~\cite{Nelsen:PRX2013,Sun:PRL2017}, such a strategy is implicitly undertaken in experiments addressing the many-body physics of microwave photons in superconductor-based circuit-QED platforms~\cite{Wang:Science2024}.

On the other hand, one can try to {\em turn a bug into a feature} and exploit the driven-dissipative nature of the fluid of light to explore the new physics that arises from the interplay of driving and dissipation with the conservative many-body dynamics. Some external pump is in fact needed to overcome the effect of losses that would normally tend to dissipate the fluid: rather than considering it just as an experimental complication, the presence of the pump opens a range of new possibilities to manipulate the many-body state of the fluid. On top of this, one must not forget that radiative losses transfer all properties of the in-cavity field to the emitted light~\cite{QuantumOptics}, so that the quantum state of the fluid of light can be reconstructed in real time just by looking at the quantum optical properties of the emitted radiation. As we are going to see in what follows, these are crucial advantages of fluids of light as compared to, e.g., ultracold atomic systems where experiments often require destructive imaging techniques. 

Before dwelling into this rich physics, it is important to complete the theoretical framework by discussing how the driving and dissipation processes can be included into the theoretical model. This will be the subject of the next subsections.

\subsubsection{Losses}
\label{subsubsec:loss}

Losses in usual devices are linear in the field amplitude and have a frequency- and wavevector-independent rate, so that they can be described at the level of the Master equation for the density operator
\begin{equation}
\frac{d\rhoh}{dt}=-\frac{i}{\hbar} [H,\rhoh] +\mathcal{L}[\rhoh]
\label{eq:Lindblad}
\end{equation}
by Markovian Lindblad terms of the form
\begin{equation}
\mathcal{L}_{\rm loss}[\rhoh]=\frac{\gamma_{\rm loss}}{2}\int d^2\rr_\parallel\, \left[2 \Psih(\rr_\parallel)\,\rhoh\,\Psihd(\rr_\parallel) - \Psihd(\rr_\parallel)\,\Psih(\rr_\parallel)\,\rhoh-\rhoh\,\Psihd(\rr_\parallel)\,\Psih(\rr_\parallel) \right]\,.
\label{eq:Lind-loss}
\end{equation}
While this theory is sufficient to deal with typical experimental configurations with moderate to long photon lifetimes, one must not forget that some extra care has to be paid if the loss rate is comparable to the frequency of the cavity mode and/or one is operating in the so-called ultra-strong light-matter coupling regime where the quantum vacuum state of the cavity is strongly distorted from the usual QED vacuum: in this case, non-Markovian terms are often needed to avoid unphysical {\em perpetuum mobile} behaviours in the theory~\cite{Ciuti:PRB2005,Ciuti:2006PRA,frisk2019ultrastrong}. As the cavities that are typically used for quantum fluid experiments are far from this regime, in what follows we will safely stick to the form \eqref{eq:Lind-loss} of the loss term.

\subsubsection{Coherent pump}
\label{subsubsec:coherent}
The coherent pumping associated to the injection of photons by an coherent field incident on the cavity mirror can be described within the input-output formalism~\cite{QuantumOptics,QuantumNoise} by additional terms in the conservative Hamiltonian. In the simplest case where the transmission amplitude $\eta$ of the cavity mirror is flat in both frequency and wavevector, these have the simple form
\begin{equation}
H_{\rm coh}=\int d^2\rr_\parallel\,\left[\hbar\,\eta\,\mathcal{E}_{\rm inc}(\rr_\parallel,t)\,\Psihd(\rr_\parallel)+\hbar\,\eta^*\,\mathcal{E}^*_{\rm inc}(\rr_\parallel,t)\,\Psih(\rr_\parallel)\right]
\label{eq:coh_pump}
\end{equation}
in terms of the (fully arbitrary) space- and time-dependent amplitude of the incident field $\mathcal{E}_{\rm inc}(\rr_\parallel,t)$ at the cavity mirror location. 

While this formulation is sufficient to deal with the typical microcavity devices used in quantum fluid of light experiments, more complicated situations can be described by replacing the constant mirror transmission amplitude $\eta$ with a convolution with a spatio-temporal kernel $\eta(\rr_\parallel,t)$ to account for the frequency- and wavevector-dependence of the mirror transmission amplitude. As usual in the input-output theory, such a generalized form of the transmission amplitude must be associated to a corresponding frequency and wavevector dependence of the radiative loss rate~\cite{Ciuti:2006PRA}.

\subsubsection{Incoherent pump}
\label{subsubsec:incoh}

Incoherent pumping terms originate from an incident thermal radiation or from spontaneous emission processes by some suitably excited medium embedded in the cavity. This can consist, e.g., of population-inverted two-level systems or an electrically-injected electron-hole gas. In contrast to the coherent pumping described of \eqref{eq:coh_pump}, such incoherent processes are totally phase-insensitive and do not inherit any coherence from the electric/optical mechanism used to excite the two-level systems or inject the electron-hole gas. While the coherent pumping processes may induce reversible oscillatory Rabi-flopping dynamics, incoherent pumping is characterized by an intrinsic irreversible push towards the excited states.

At a simplest level of approximation, one may be tempted to describe the effect of incoherent pumping within a Born and Markov approximation by including  \eqref{eq:Lindblad} a Lindblad term of the form
\begin{equation}
\mathcal{L}_{\rm pump}[\rhoh]=\frac{\gamma_{\rm pump}}{2}\, \int d^2\rr_\parallel\,\left[2 \Psihd(\rr_\parallel)\,\rhoh\,\Psih(\rr_\parallel)- \Psih(\rr_\parallel)\,\Psihd(\rr_\parallel)\,\rhoh-\rhoh\,\Psih(\rr_\parallel)\,\Psihd(\rr_\parallel)\right]
\label{eq:Lind-pump}
\end{equation}
into the Master equation.
This is a good approximation as long as the gain has a broad-band nature and, more importantly, the pump rate remains well below the loss rate $\gamma_{\rm pump}<\gamma_{\rm loss}$. In this regime, the effect of pumping is to sporadically inject some extra photon into the cavity, which is then rapidly lost: in addition to spontaneous processes injecting photons (almost) uniformly into all modes of the cavity, such emission processes may also enjoy bosonic stimulation so the emission will be reinforced into those cavity modes that already contain photons.

On the other hand, a most interesting physics occurs when incoherent pumping overcomes losses and a coherent fluid of light is spontaneously generated in the cavity via a non-equilibrium condensation mechanism analogous to lasing~\cite{bloch2022non}: in this case, a naive use of \eqref{eq:Lind-pump} would lead to a divergence of the in-cavity field amplitude and an accurate description of the steady-state of the in-plane cavity field requires including gain saturation effects.   

\subsubsection{Gain saturation \& frequency selectivity}
\label{subsubsec:Hemitter}

A common strategy to account for gain saturation is to explicitly include the microscopic dynamics of the emission process into the quantum theoretical description. Such an approach dates back to early works in laser theory~\cite{Lamb:PR1964,scully1997quantum}, where the emitters are described as two-level systems driven in a population-inverted state by some external incoherent pumping.

In a simplest geometry where the field is discretized in a lattice of cavities each supporting a single optical mode and embedding a single two-level emitter, the local Hamiltonian of each site (to be supplemented by hopping terms in a full lattice model) has the form
\begin{equation}
 H_{\rm site}=\hbar \omega_0 \ahd\ah + \hbar \omega_{\rm em}\,\hat{\sigma}_z + \hbar \Omega_R \left[ \ahd \hat{\sigma}^- +  \ah \hat{\sigma}^+ \right]
 \label{eq:Hsite}
\end{equation}
where the bosonic $\ah$ and the spin $\sigma^{z,\pm}$ operators respectively describe the photonic field (of natural frequency $\omega_0$) and the two-level emitter (of natural frequency $\omega_{\rm em}$). 
The population inversion is enforced by a incoherent pumping acting on the emitter, as described by a Lindblad term of the form
\begin{equation}
\mathcal{L}_{\rm pump-em}[\rhoh]=\frac{\Gamma_{\rm P}}{2}\,\left[2 \hat{\sigma}^+ \rhoh \hat{\sigma}^- - \hat{\sigma}^-\hat{\sigma}^+ \rhoh - \rhoh \hat{\sigma}^-\hat{\sigma}^+ \right]\,.
\label{eq:Lind-pump-em}
\end{equation}
which pushes the population towards the excited state.

\paragraph{Effective incoherent pumping rate} -- 
Emission by the emitters into the cavity occurs via the light-matter coupling term in \eqref{eq:Hsite}. If the pump rate $\Gamma_{\rm P}$ well exceeds the Rabi frequency $\Omega_R$, the dynamics of the emitter can be adiabatically eliminated and the coherent Rabi oscillations typical of the Jaynes-Cummings model of cavity-QED~\cite{QuantumOptics,scully1997quantum} are replaced by an irreversible emission into the cavity mode.

If we further assume that the in-cavity field is weak enough not to affect the population inversion, gain saturation is not significant and we obtain an effective incoherent pumping of the form \eqref{eq:Lind-pump} with a rate
\begin{equation}
 \gamma_{\rm pump}^{\rm eff,0}=\frac{4\Omega_R^2}{\Gamma_{\rm P}}\,. 
 \label{eq:gamma_pump_eff}
\end{equation}
Of course, such a rate is only obtained when the emitter is exactly on resonance with the photon mode,
otherwise the pumping rate is reduced according to a Lorentzian lineshape of width $\Gamma_{\rm P}$ around the resonant point $\omega=\omega_{\rm em}$,
\begin{equation}
 \gamma_{\rm pump}^{\rm eff}(\omega)= \gamma_{\rm pump}^{\rm eff,0}\,\frac{\Gamma_{\rm P}^2/4}{(\omega-\omega_{\rm em})^2 +\Gamma_{\rm P}^2/4}\,.
\end{equation}
Including this $\omega$ dependence is straightforwardly done in a linear system: for each mode at frequency $\bar{\omega}$, one can write an incoherent pumping term the form \eqref{eq:Lind-pump} with an effective $\gamma_{\rm pump}^{\rm eff}(\omega)$ evaluated at the mode frequency $\omega=\bar{\omega}$.

Things get more complex in interacting many-mode models, where different frequencies $\omega$ have to be considered for the different transitions between many-body eigenstates.
In this case, the standard Lindblad term \eqref{eq:Lind-pump} can be replaced by a generalized term in the form~\cite{Lebreuilly:CRAS2016}
\begin{equation}
\mathcal{L}_{\rm{em}}(\rho)  = \frac{ \gamma_{\rm pump}^{\rm eff,0}}{2}\sum_{j}\left[\tilde{a}_{i}^{\dagger}\rho \ah_{i}+\ah_{i}^{\dagger}\rho\tilde{a}_{i}-\ah_{i}\tilde{a}_{i}^{\dagger}\rho-\rho\tilde{a}_{i}\ah_{i}^{\dagger}\right]. \label{eq:gainmarkov}
\end{equation}
where the sum runs over the lattice sites $j$ and the frequency dependence of the pumping process is encoded by the modified jump operator $\tilde{a}_{j}^\dagger$. 
The effective jump amplitude from the many-body state $f$ with $N$ photons to the many-body state $f'$ with $N+1$ photons under the creation of a photon in the $j$ cavity mode now includes the frequency difference $\omega_{f,f'}=\omega_{f'}-\omega_f$ of the many-body states as
\begin{equation}
\bra{f'}\tilde{a}_{j}^\dagger \ket{f} =  \frac{\Gamma_{\rm{p}}/2}{i(\omega_{f,f'}-\omega_{\rm{em}})+\Gamma_{\rm{p}}/2}\bra{f'} \ahd_{j}\ket{f}\,.
\label{eq:crea_lorent}
\end{equation}
This immediately leads to a Lorentzian dependence of the total $f\to f'$ transition rate on the frequency difference $\omega_{f,f'}$,
\begin{equation}
\mathcal{T}_{\rm pump}(f\to f')= \gamma_{\rm pump}^{\rm eff}(\omega_{f,f'})\,\sum_j|\bra{f'} \ahd_j \ket{f}|^2\,, 
\label{eq:Tpump}
\end{equation}
to be compared with the frequency-independent transition rate due to the loss processes discussed in Sec.\ref{subsubsec:loss},
\begin{equation}
 \mathcal{T}_{\rm loss}({f'\to f})= \gamma_{\rm{loss}}\,\sum_j|\bra{f} \ah_j \ket{f'}|^2\,.
\end{equation}
In addition to the frequency-dependent pump rate \eqref{eq:Tpump}, the imaginary part at the denominator of \eqref{eq:crea_lorent} can be recast in terms of a conservative Hamiltonian term giving a (typically small) energy shift of the energy levels of the photonic system. This effect can be physically understood as originating from the effective refractive index of the emitters. In the simplest case of a linear single-mode cavity, an explicit calculation leads to a frequency-shift of the cavity mode by
\begin{equation}
 \delta\omega_0=\frac{\Gamma_P\,\gamma_{\rm pump}^{\rm eff,0}/4}{(\omega_{\rm em}-\omega_0)^2+\Gamma_P^2/4}\,(\omega_{\rm em}-\omega_0)\,:
\end{equation}
note how this effect has an opposite sign compared to the usual refractive index of ground-state atoms so, in contrast to the usual polariton splitting~\cite{carusotto2013quantum}, it leads to the attraction of the cavity mode towards the emitter in a sort of {\em frequency pulling} effect. Before proceeding, it is important to note that the generalized master equation \eqref{eq:gainmarkov} is valid under a sort of perturbative expansion in the strength of the driving/dissipation processes: the jump amplitudes are in fact modified according to the energies of the conservative eigenstates. Writing master equations that go beyond this approximation and allow to deal with stronger driving/dissipation regimes is a topic of present-day research~\cite{breuer2002theory,Becker:PRE2021,Tello:PRB2024}.

\paragraph{Gain saturation} --
As mentioned above, the effective description based on \eqref{eq:gainmarkov} also assumes that the emission into the photon mode is weak enough not to deplete the population inversion in the emitters. Of course, bosonic stimulation of the emission processes makes this constraint more severe when there is a sizable population in the photonic modes, $\gamma_{\rm pump}^{\rm eff}\,n_{\rm ph}\ll \Gamma_{\rm P}$.

In order to go beyond this assumption and account for those gain-saturation effects that play a major role in many physical phenomena, one needs to 
fully include the back-reaction effect of the emission onto the emitter dynamics. At the quantum level, this requires a complete description of the emitter and field dynamics: this is a challenging many-body task and typically has to be carried out under some other simplifying hypothesis, e.g. mimicking gain saturation by means of two-photon loss processes or performing some Gutzwiller site-decoupling approximation~\cite{Caleffi:PRL2023}. 
At the semiclassical level, instead, calculations can be performed in terms of generalized mean-field equations inspired by the semiclassical theory of lasers as we are going to see in the next Section.

\paragraph{Alternative configurations} -- 
To conclude this section, it is important to also mention that alternative configurations to obtain a frequency-dependent pumping were proposed in~\cite{Kapit:2014PRX} using parametric emission processes. The key element of such schemes is to replace the two-level emitter used in \eqref{eq:Hsite} with an additional ancilla cavity displaying strong losses $\Gamma_{\rm anc}$ described by a standard Lindblad master equation. Parametric processes generate pairs of photons in the photonic mode and in the ancilla cavity according to
\begin{equation}
H = \hbar \omega_0 \ahd\ah + \hbar \omega_{\rm anc}\,\bhd\bh + \left[\hbar \Omega_{\rm par} e^{-i\omega_{\rm par} t} \ahd \bhd +  \hbar \Omega^*_{\rm par} e^{i\omega_{\rm par} t} \ah \bh \right]\,:
\end{equation}
an effective irreversibility naturally arises if the ancilla decay rate $\Gamma_{\rm anc}$ well exceeds the parametric emission rate $\Omega_{\rm par}$.
Right at the parametric resonance point, this gives an effective pumping rate,
\begin{equation}
\gamma_{\rm pump}^{\rm eff,0}=\frac{4|\Omega_{\rm par}|^2}{\Gamma_{\rm anc}} 
\end{equation}
Such a value, together with the parametric resonance condition $\omega_{\rm par}-\omega_{\rm anc}=\omega_0$ can be directly inserted in the theory of Eq.\eqref{eq:gainmarkov} and leads to closely analogous results up to the many-body level.
Such a strategy was experimentally put into practice in~\cite{ma2019dissipatively} to generate a Mott insulator state of impenetrable photons. Note how related techniques for autonomous dissipative stabilization of desired quantum states are of widespread use also beyond quantum fluids of light, for various tasks in quantum science and technology~\cite{Mamaev:Quantum2018,Harrington:NatRevPhys2022,Li:NatComm2024,Lescanne:NatPhys2020}.

\section{Weakly interacting fluid}
\label{sec:weakly}

The new possibilities opened for quantum fluids of light by the presence of the driving and dissipation processes were first addressed in the context of weakly interacting fluids for which a mean-field approximation is legitimate. Such regimes have been long studied in the context of nonlinear optics and laser physics and, more recently, in the context of polariton fluids and polariton condensates. The literature on this physics is immense and we refer the interested reader to recent review articles~\cite{carusotto2013quantum,bloch2022non}. With no hope of completeness, in what follows we will illustrate a few cases where the physics is -to our eyes- most clear and illustrative of the general principles.

\subsection{Mean-field approximation}

The key idea of the mean-field approximation is to write an approximate equation of motion for the expectation value of the field operator $\psi(\rr_\parallel,t)=\langle \Psih(\rr_\parallel,t) \rangle$. Such an equation of motion can be derived by taking the Heisenberg equation for the field operator and then factorizing all operator products. Usually, this factorization assumption is accurate as long as the fluctuations around the mean-field are small, which is generally the case if interactions between the particles are weak.

Application of this procedure to the conservative Hamiltonian \eqref{eq:Hcav} leads to the Heisenberg equation
\begin{equation}
 i\hbar \frac{d\Psih(\rr_\parallel,t)}{dt}=[\Psih,H]=\hbar \omega_o\Psih(\rr_\parallel) -\frac{\hbar^2}{2m}\nabla^2_\parallel\Psih(\rr_\parallel)+ \hbar V(\rr_\parallel)\,\Psih(\rr_\parallel)
+ \hbar g_{\rm nl} \Psihd(\rr_\parallel)\,\Psih(\rr_\parallel)\,\Psih(\rr_\parallel)
 \end{equation}
on which the factorization approximation
\begin{equation}
 \langle \Psihd(\rr_\parallel)\,\Psih(\rr_\parallel)\,\Psih(\rr_\parallel) \rangle \approx |\psi(\rr_\parallel,t)|^2\,\psi(\rr_\parallel,t)\,,
\end{equation}
gives the usual Gross-Pitaevskii equation for an interacting Bose-Einstein condensate,
\begin{equation}
 i \frac{\partial \psi}{\partial t}=\left[\omega_o -\frac{\hbar \nabla^2}{2m^*}\right]\,\psi + V\,\psi+ g_{\rm nl}\,|\psi|^2\,\psi\,.
\end{equation}
Necessary conditions for the validity of this mean-field approach can be obtained from the general many-body theory of dilute Bose-Einstein condensates at equilibrium~\cite{pitaevskii2016bose,CastinLectures}. For instance, for three-dimensional gases, the condition can be expressed in terms of a diluteness condition $n_{3d}\,a_{sc}^3\ll 1$, where $n_{3d}$ is the three-dimensional density and $a_{sc}$ is the atom-atom collisional scattering length, related to the interaction parameter by $g_{\rm nl}=4\pi\hbar^2 a_{sc}/m$. More subtle conditions have to be imposed in low-dimensional geometries, but a quite general formulation involves the characteristic kinetic and interaction energies, $\varepsilon_{kin}=\hbar^2 n^{2/d}/2m \gg \varepsilon_{int}=g_{\rm nl} n$ or, equivalently, the number of particles per healing length, $n\xi^d\gg 1$ where $n$ is the $d$-dimensional density and $\xi =\sqrt{\hbar^2/m g_{\rm nl} n}$ is the healing length characterizing the microscopic structure of the fluid. 

The same factorization procudure can be performed on the pumping and loss terms introduced in Sec.\ref{sec:theory_DD}. The coherent pump is straightforwardly included at the level of the Heisenberg equation. As the corresponding Hamiltonian term \eqref{eq:coh_pump} is linear in $\Psih$, no additional factorization approximation has to be made. As a result, one obtains a forcing term in the equation of motion for $\psi$ of the form:
\begin{equation}
\ldots + \eta\,\mathcal{E}(\rr_\parallel,t)\,.
\end{equation}
The linear loss and incoherent pumping terms \eqref{eq:Lind-loss} and \eqref{eq:Lind-pump} give linear, phase-insensitive gain and loss terms of the form
\begin{equation}
\ldots -\frac{\gamma_{\rm loss}}{2} \psi + \frac{\gamma_{\rm pump}}{2} \psi\,,
\end{equation}
from which it is immediate to see how an unlimited exponential growth may arise if $\gamma_{\rm pump}>\gamma_{\rm loss}$ and no gain saturation mechanism is included in the model.

\subsubsection{Gain saturation \& Frequency selectivity}

Gain saturation can be included starting from the model of Sec.\ref{subsubsec:Hemitter}. In analogy to the semiclassical equations of lasers~\cite{Lamb:PR1964,scully1997quantum}, we have to write the equations of motions for the single-mode field amplitude $\alpha =\langle \ah\rangle$,
\begin{equation}
  \label{eq:a}
 \frac{d\alpha}{dt}=-i \omega_0 \alpha -i \Omega_R\sigma^- -\frac{\gamma_{\rm loss}}{2}\alpha
\end{equation}
as well as for the expectation values of the emitter operators $\sigma^{\pm,z}=\langle \hat{\sigma}^{\pm,z}\rangle$. Performing all the factorization approximations of mean-field theory, we obtain
\begin{eqnarray}
 \label{eq:sigmaz}
 \frac{d\sigma^z}{dt}&=&\Gamma_{\rm P}(1-\sigma^z)  - 2 i \Omega_R ( \sigma^+ \alpha - \sigma^-  \alpha^*)   \\
  \frac{d\sigma^-}{dt} &=& - i \omega_{\rm em} \sigma^- -\frac{\Gamma_{\rm P}}{2}\,\sigma^- + i \Omega_R \, \sigma^z a  
 \label{eq:sigmapm}
\end{eqnarray}
These equations for the single-mode configuration can be straightforwardly extended to spatially continuous geometries with many uniformly distributed emitters of in-plane density $n_{\rm em}$. For this, we introduce space-dependent emitter spin variables $\sigma^{\pm,z}(\rr_\parallel)$ and we replace the localized field amplitude $\alpha$ with the continuous field $\psi(\rr_\parallel)$. 

In the so-called good-cavity limit where the emitter pump rate $\Gamma_{\rm P}$ is much faster than the field dynamics and of the light-matter coupling $\Omega_R$, one can eliminate the emitter degrees of freedom and write an effective equation of motion for the field only,
\begin{equation}
 i\frac{\partial \psi}{\partial t}=\left[\omega_o -\frac{\hbar \nabla^2}{2m^*}\right]\,\psi + V\,\psi+ g_{\rm nl}\,|\psi|^2\,\psi 
 +\frac{i}{2} \left[\frac{P-2i \,\delta\omega_0}{1+|\psi|^2/n_s}-\gamma_{\rm loss}
 \right]\psi + \eta \, E_{\rm inc}(\rr,t)\,
 \label{eq:dd_psi}
\end{equation}
where the cavity mode experiences an effective pumping
\begin{equation}
 P=\frac{n_{\rm em}\,\Omega_R^2 {\Gamma_P}}{(\omega_{\rm em}-\omega_0)^2 + {\Gamma_P^2}/{4}}
 \label{eq:freqdepP}
\end{equation}
with a saturation density
\begin{equation}
 n_s=\frac{(\omega_{\rm em}-\omega_0)^2 + {\Gamma_P^2}/{4}}{2\Omega_R^2}
 \label{eq:nsat}
\end{equation}
and a Lorenzian frequency-dependence of linewidth $\Gamma_P$ analogous to the one of \eqref{eq:Tpump}.
In analogy to what was found in \eqref{eq:pulling}, the presence of the emitters also leads to a frequency-pulling effect towards the emitter of magnitude
\begin{equation}
\delta\omega_0 =  \frac{n_{\rm em}\,\Omega_R^2}{(\omega_{\rm em}-\omega_0)^2 + {\Gamma_P^2}/{4}}\,(\omega_{\rm em}-\omega_0)\,.
\label{eq:pulling}
\end{equation}

These formulas starting from \eqref{eq:dd_psi} implicitly assume that the cavity field dynamics takes place in a very narrow frequency band around $\omega_0$. At a slightly more sophisticated level, we can Taylor-expand the frequency-dependence of the different quantities. For the most relevant case of the pumping rate $P$, this can be modeled in an effective way~\cite{Wouters:PRL2010} by replacing
\begin{equation}
P\rightarrow P \left[1-\frac{2(\omega_0-\omega_{\rm em})}{(\omega_{\rm em}-\omega_0)^2 + {\Gamma_P^2}/{4}} \, \left(i \frac{\partial}{\partial t} -\omega_0\right)\right]
\end{equation}
and letting the last time-derivative to only act on the field $\psi$. In some physical models of lasing, such a frequency-dependence of gain is indeed important to stabilize laser action at the bottom of the photon band and get a more accurate description of the experiment. Of course, if needed, similar Taylor expansions might be performed also on the other two quantities \eqref{eq:nsat} and \eqref{eq:pulling}.

As a final point, it is important to highlight that equations analogous to \eqref{eq:dd_psi} have been derived in several other contexts from different points of views. The approach of this article is inspired from the many-body theory of dilute Bose gases and the closely related idea of quantum fluid of light: in this context, it is natural to call this equation {\em generalized non-equilibrium Gross-Pitaevskii equation}. The very similar equations that are investigated in the context of nonlinear dynamics typically go under the name of  complex Ginzburg-Landau equations~\cite{Cross:RMP1993,Aranson:RMP2002}. In the specific context of nonlinear optics, the motion equation for the field in the presence of coherent driving and losses goes under the name of {\em Lugiato-Lefever equation}. This equation, originally formulated in~\cite{Lugiato:PRL1987}, is still now a powerful workhorse for the description of optical devices based on Kerr cavity solitons and frequency combs~\cite{Lugiato:Varenna}.

%

\subsubsection{Quantum fluctuations beyond mean-field}
\label{sec:Wigner}

Within a semiclassical approach, quantum and thermal fluctuations around the mean-field can be described within the so-called truncated Wigner approximation~\cite{QuantumNoise,Carusotto:PRB2005} by adding a stochastic white noise term $\xi(\rr_\parallel,t)$ on the right-hand side of the motion equation \eqref{eq:dd_psi} so to account, at least at a lowest order, for quantum fluctuations. In the case of coherent pumping, the complex-valued noise term must have a random phase and its variance is to be set to
\begin{equation}
 \langle \xi^*(\rr_\parallel,t) \,  \xi(\rr'_\parallel,t') \rangle = {\frac{\gamma_{loss}}{2}}\,\delta^{(2)}(\rr_\parallel-\rr'_\parallel)\,\delta(t-t')\,:
\end{equation} 
this is the minimum noise level as determined by the quantum fluctuations associated to the discreteness of the photon in the loss process. Of course, a higher noise level is found in incoherent pumping configurations~\cite{Wouters:PRB2009} where additional contributions are given by spontaneous events in the pumping process,
\begin{equation}
 \langle \xi^*(\rr_\parallel,t) \,  \xi(\rr'_\parallel,t') \rangle =\frac{1}{2}\,{\left(\gamma_{loss}+\frac{\gamma_{pump}}{1+|\psi|^2/n_S}\right)}\,\delta^{(2)}(\rr_\parallel-\rr'_\parallel)\, \delta(t-t')\,.
 \label{eq:semicl_noise}
\end{equation}
Further sources of quantum noise may stem from the interaction of the Bogoliubov modes of the photon fluid with phonon modes in the solid-state crystal hosting the fluid of light~\cite{Frerot:PRX2023}. Generalization of truncated Wigner methods to systems involving emitters as described e.g. in the Hamiltonian model of \eqref{eq:Hsite} are under way~\cite{Marzena_private}.

As typical of Wigner representations~\cite{QuantumNoise}, for any product of fields $\psi(\rr_\parallel)$ and $\psi^*(\rr_\parallel')$, the statistical average over the noise provides information on the corresponding {\em symmetrized} product of quantum operators $\Psih(\rr_\parallel)$ and $\Psihd(\rr_\parallel')$, so commutators have to be subtracted to obtain the more standard normal-ordered quantities. While this is a quite trivial step for single-time operator products, it poses serious challenges for calculation multi-time quantities~\cite{Berg:PRA2009}.

Interestingly, the driven-dissipative condition makes the truncated Wigner approach much more stable than in the conservative case, as the field fluctuations get to a driven-dissipative steady-state determined by the interplay of stochastic noise and dissipation well before approaching an inaccurate classical field thermal equilibrium state~\cite{Sinatra:JPhysB2002}. While this is often sufficient to get physically reliable predictions for the field correlation functions~\cite{Carusotto:PRB2005}, it is important to note that the approximation of truncating higher-order noise terms~\cite{Steel:PRA1998} that underlies \eqref{eq:semicl_noise} typically loses accuracy when strong interactions and/or very non-perturbative processes in the quantum fluctuations are considered, such as quantum tunneling in bistable systems~\cite{Drummond:PRA1986,Vogel:PRA1988}. Quantitative studies of the limitations of the truncated Wigner approach~\cite{VanRegemortel:PRA2017} and of possible avenues to include the higher-order derivative terms~\cite{Polkovnikov:PRA2003} are the topic of current research.

A most remarkable prediction of this theory~\cite{Busch:PRA2014} is that the driven-dissipative stationary state displays sizable fluctuations for the field $\psi$, well larger than the zero-point fluctuations of the Bogoliubov ground state of the corresponding equilibrium system. Within the Bogoliubov picture, the loss process for a particle at momentum $\kk$ involves in fact either the destruction of a quasi-particle at $\kk$ or the creation of a quasi-particle at $-\kk$. The overall magnitude of these fluctuations roughly corresponds to a temperature of the order of the interaction energy $g_{\rm nl}|\psi|^2$ but fluctuations may also display non trivial correlations.



%

\subsection{Coherent pumping}

The configuration where the new possibilities opened by driving and dissipations are clearest is the coherent pumping one. This configuration was exploited in the first experiments demonstrating superfluid light, as the flow speed and the density of the fluid can be directly controlled by the incidence angle (i.e. the in-plane wavevector), the frequency and the intensity of the incident field~\cite{Carusotto:PRL2004,Amo:NPhys2009}. As a key difference from equilibrium condensates, the oscillation frequency of the field is not bound to the chemical potential but is freely determined by the one of the incident coherent field. As a result, the equation of state becomes richer and new phenomena can be observed.

\subsubsection{Homogeneous fluid}

\begin{figure}
    \centering
    \includegraphics[width=0.47\textwidth]{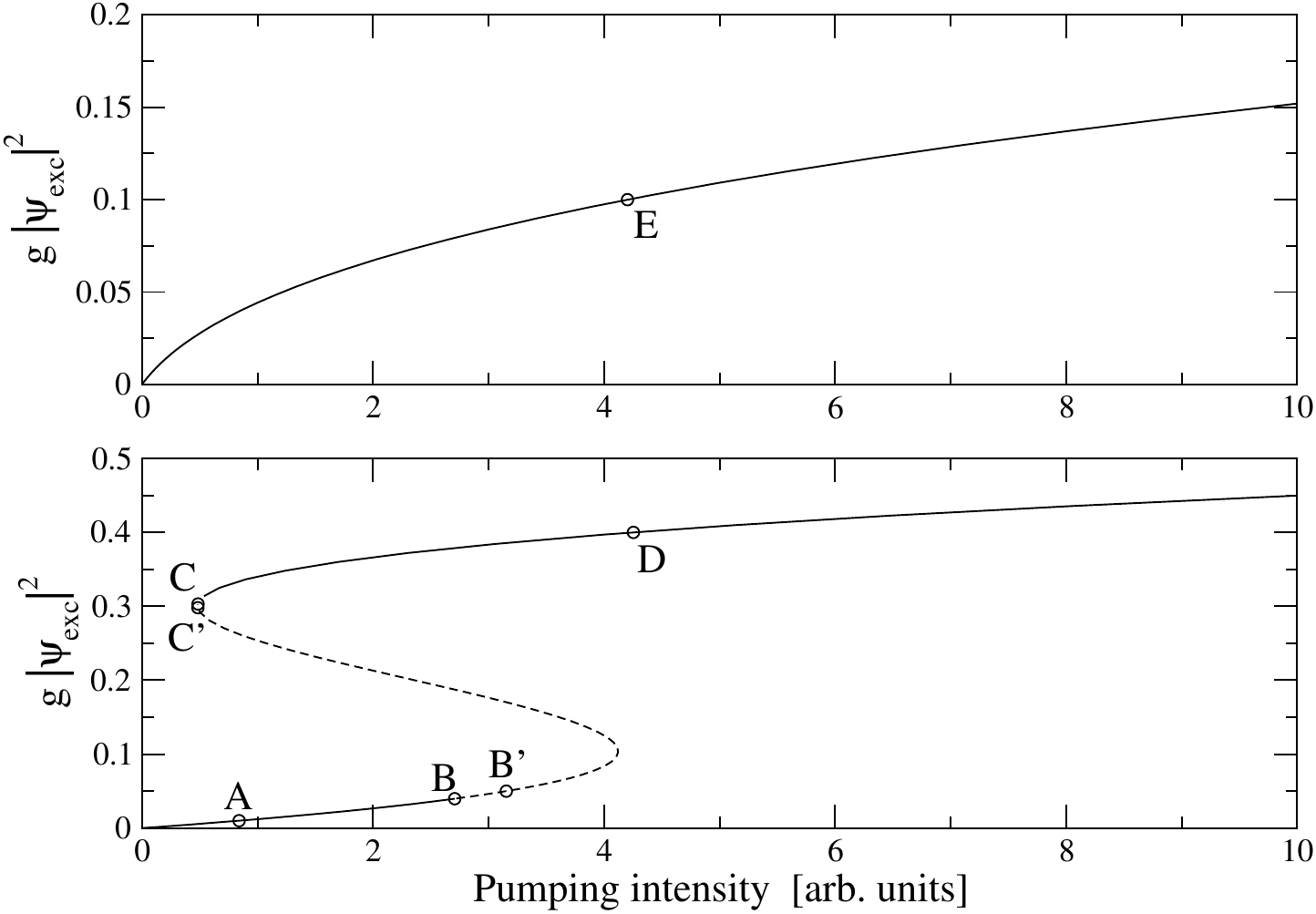}\hspace{0,05\textwidth}
    \includegraphics[width=0.47\textwidth]{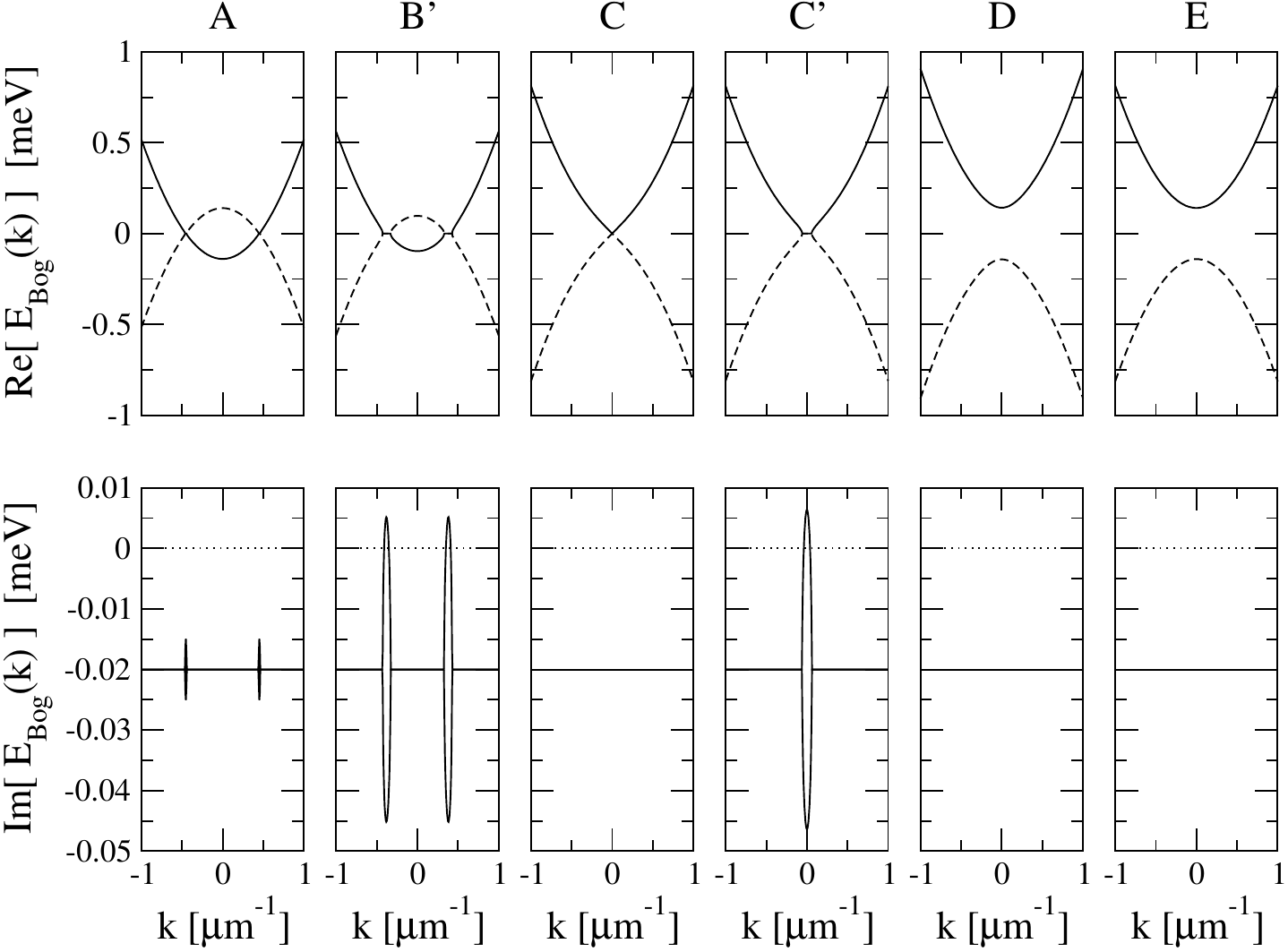}
    \caption{Left panels: plot of the equation of state of the fluid of light, expressed as fluid density as a function of pump intensity for two different incident frequencies in respectively the optical limiter regime (top) and the bistable regime (bottom). The dashed line indicates the dynamically unstable regions. Right panel: Real part (top) and imaginary part (bottom) of the excitation frequencies for the nonequilibrium Bogoliubov modes corresponding to the points indicated as A, B', C, C', D, E in the left panel.  \label{fig:coherent}}
\end{figure}

In the simplest case of a monochromatic incident field at $\omega_{\rm inc}$ with a spatially homogeneous profile, the fluid is at rest and the equation of state
\begin{equation}
\left[\left(\omega_{\rm inc}-\omega_0+g_{\rm nl}|\psi|^2\right)^2 + \frac{\gamma_{\rm loss}^2}{4} \right]\,|\psi|^2=|\eta\,E_{\rm inc}|^2\,,
\label{eq:eos}
\end{equation}
gives a complex nonlinear dependence of the density $|\psi|^2$ on the different parameters of the incident field. 
Some of the peculiar features of these non-equilibrium fluids are visually illustrated in Fig.\ref{fig:coherent}: the prediction of the equation of state \eqref{eq:eos} is displayed in the two left panels for the two cases of a pump frequency respectively below $\omega_{\rm inc}<\omega_0$ (top) and above $\omega_{\rm inc}>\omega_0$ (bottom) the bare resonance frequency (we are assuming $g_{\rm nl}>0$): in the former case, the in-cavity intensity (proxied by $g|\psi_{\rm exc}|^2$ in the figure) grows monotonically but sublinearly with the pump intensity $|E_{\rm inc}|^2$; in the latter case, bistability effects are apparent, with two values of the density $|\psi|^2$ being available for the same value of the pump intensity $|E_{\rm inc}|^2$ (the intermediate branch is dynamically unstable).

Further consequences of the non-equilibrium condition are visible in the right panels, where the dispersion of the Bogoliubov modes on top of the homogeneous fluid~\cite{Carusotto:PRL2004,carusotto2013quantum} are plotted for the different operating points indicated by the capital letters in the left panels. As the oscillation frequency of the field $\psi$ is freely tunable via the incident frequency $\omega_{\rm inc}$, a wide range of behaviours can be observed depending on the pumping conditions. In particular, the Bogoliubov modes display a sonic long-wavelength behaviour $\omega=c_s\,|\kk|$ only for a very specific operating point (C) at the end of the upper branch of the bistability loop. Away from this point, the dispersion may display finite gaps (D,E) and, even, precursors (A) or fully fledged (B,C') dynamical instabilities. Detailed experimental studies of the Bogoliubov dispersion in the different pumping regimes have been carried out by pump-probe spectroscopy in~\cite{Claude:PRL2022}.

\subsubsection{Acoustic horizons and analog Hawking radiation}

This tunability of the Bogoliubov modes can be exploited to observe exciting new behaviours in spatially inhomogeneous configurations when the spatial profile of the coherent pump is suitably structured. In this perspective, a variety of configurations have been investigated in the literature, for instance vortices and vortex lattices that appear under a coherent pump carrying orbital angular momentum~\cite{Boulier:CRAS2016}. 

As a specific most exciting example, we focus here on configurations featuring a sonic horizon separating an upstream region of sub-sonic flow $c_u>v_u$ from a downstream region of super-sonic flow $v_d>c_d$ where $c_{u,d}$ and $v_{u,d}$ are respectively the sound and flow speeds in the two $u,d$ regions. From the point of view of sound propagation, this configuration displays strong analogies with light propagation in the neighborhood of a black hole in gravitational physics and, upon quantization, sonic analogs of Hawking radiation has been anticipated to be emitted by the acoustic horizon~\cite{barcelo2011analogue}. 

With a careful design of the coherent pump, such analog black holes naturally arise as the driven-dissipative steady-state of a photon fluid. A natural possibility to achieve this~\cite{Gerace:PRB2012} is to use a monochromatic pump and to design its phase profile so to have a small in-plane wavevector in the upstream region and a larger one in the downstream region. Alternatively, the downstream region can be left unpumped so that the photon fluid is able to ballistically expand into this region. This latter configuration was experimentally realized in~\cite{Nguyen:PRL2015}. A spatially-resolved experimental characterization of the Bogoliubov modes on the two sides of the horizon has been performed in~\cite{Falque:arXiv2023}, explicitly showing the transition from a sub- to a super-sonic flow as the horizon is crossed: in the experiment, a super-sonic flow is signalled by the positive-norm branch of the Bogoliubov dispersion being pushed down to negative frequencies.

The on-going challenge is then to obtain experimental evidence of the analog Hawking emission by the acoustic horizon: to this purpose, numerical studies based on the Wigner formalism have shown that a clear signature of the Hawking radiation is present in the correlation function of density fluctuations on opposite sides of the horizon~\cite{Gerace:PRB2012,Nguyen:PRL2015}. Physically, this correlation feature reflects the (quantum) correlations between the Hawking phonon emitted outside the horizon into the sub-sonic region and its partner that propagates inside the horizon into the super-sonic region.
Most interestingly, in contrast to atomic fluids where destructive measurements of atomic density profile are typically needed~\cite{Munoz:Nature2019}, intensity fluctuations are of direct experimental access in driven-dissipative photon fluids: radiative losses directly transfer the correlation functions of the in-cavity field into the ones of the emitted light, where the Hawking signal can be observed as peculiar intensity correlations of the emitted light.

\subsubsection{Integer quantum Hall effect}

One of the most intriguing family of phenomena of quantum condensed matter physics is the quantum Hall effect~\cite{Tong:QHbook}. This effect was originally discovered in two-dimensional electron gases subject to strong magnetic fields and cooled to ultralow temperatures. While in room-temperature metals the transverse Hall resistivity $\rho_{xy}$ shows a structureless linear dependence on the applied magnetic field $B$, the {\em quantum} Hall effect is visible as a series of plateaus in $\rho_{xy}$ when the so-called filling factor
\begin{equation}
 \nu=\frac{2\pi\hbar n_{2D}}{e B}
\end{equation}
is in the neighborhood of integer (the so-called {\em integer quantum Hall (IQH)} effect) or fractional ({\em fractional quantum Hall (FQH)} effect) values. In this section, we are going to focus on the IQH effect, while a brief discussion of the FQH in an optical context will be given later on in Sec.\ref{sec:strongly}.

\begin{figure}
    \centering
    \includegraphics[width=0.8\textwidth,trim=1.75cm 1.5cm 6cm 1.5cm, clip]{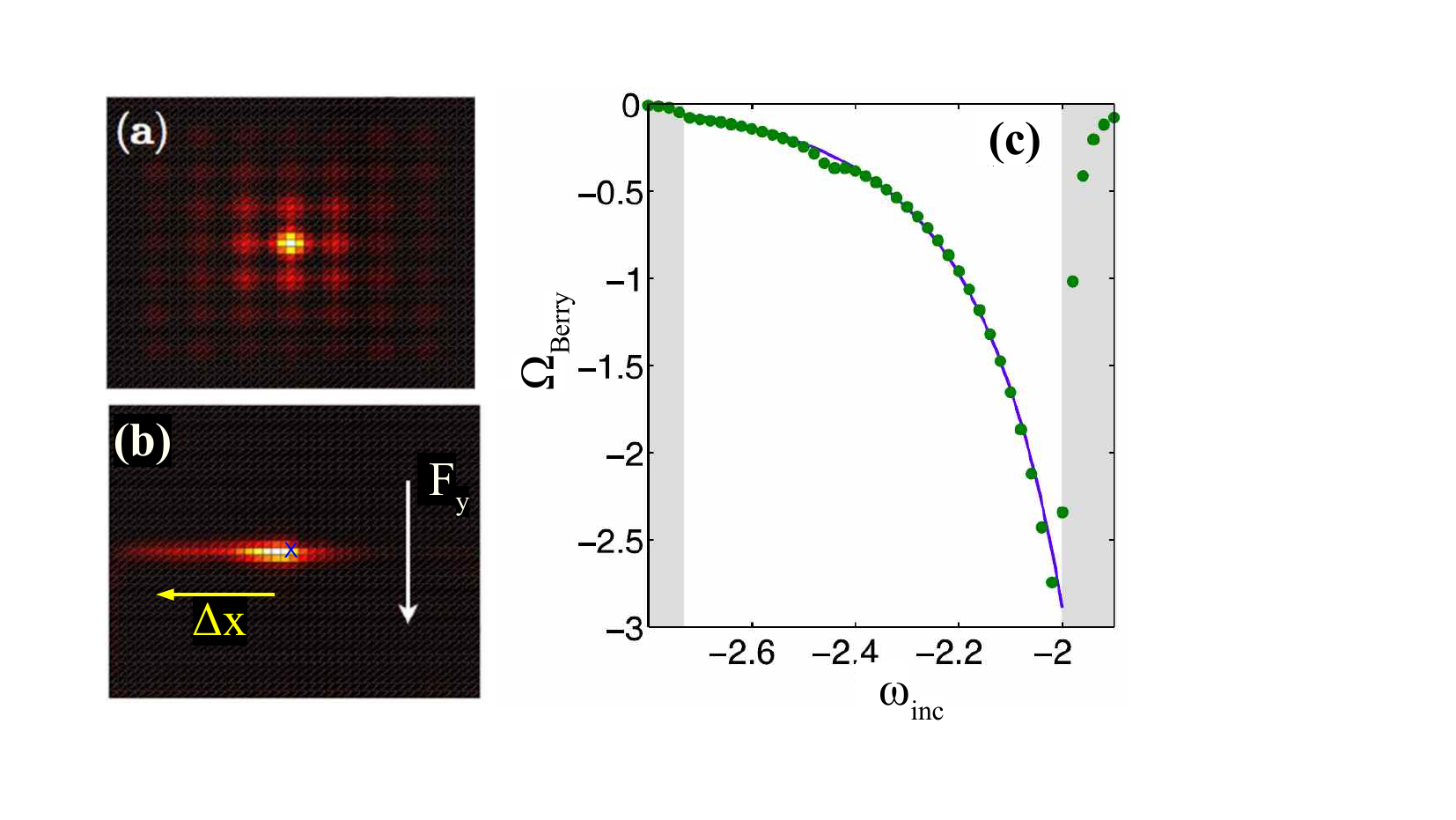}
    \caption{  \label{fig:topo} Left (a-b) panels: color plots of the numerically calculated steady-state intensity distribution in a coherently pumped $q=5$ photonic Harper-Hofstadter model. The pump is localized on the central site and is resonant with the lowest photonic band. The loss rate is taken to be larger than the bandwidth $\gamma_{\rm loss}/ \Delta\omega_{\rm band}=2$, so to be in the IQH regime.
    In the upper (a) panel, no external force is present and the distribution has a $\pi/2$ rotational symmetry. In the lower (b) panel, a force along the negative $y$ direction is applied, and the intensity distribution moves leftwards along the transverse $x$ direction. The magnitude of this displacement agrees with the prediction of the driven-dissipative IQH formula \eqref{eq:transverse_shift_quantized}. Right panel: for a loss rate smaller than the bandwidth $\gamma_{\rm loss}/\Delta\omega_{\rm band}\simeq 1/30$ of the lowest band of a $q=3$ Harper-Hofstadter, we are in the anomalous quantum Hall regime and the displacement is determined by the average of the Berry curvature over all states resonant with the pump frequency $\omega_{\rm inc}$. Within a photonic band (white region), the value of the Berry curvature extracted from the numerically calculated transverse shift (green points) are successfully compared to the value of the Berry curvature extracted from the Harper-Hofstadter band structure.
    Panels adapted from~\cite{Ozawa:PRL2014}.}
\end{figure}

Physically, 
the origin of the IQH can be traced to the complete and homogeneous filling of all states in the $\nu$ lowest Landau levels in the magnetic field $B$. In electronic systems where the number of electrons is conserved, the value of $\nu$ can be fixed by tuning the externally applied magnetic field $B$ or the electronic density $n_{2D}$ with suitable gates, so to fix the Fermi energy between two Landau levels. While this simple argument explains the quantized value of the conductivity, the finite width of the plateaux results instead from a more subtle interplay between localized and delocalized electronic states in the unavoidable disorder of any realistic system~\cite{Tong:QHbook}.

In analogy to these advances in quantum condensed matter systems, a strong interest is being devoted to the so-called topological photonic systems where new phenomena stemming from the non-trivial band topology of the photonic states have been observed, e.g. unidirectionally propagating chiral edge states~\cite{ozawaRMP2019topological}. In this context, a basic ingredient is the so-called synthetic magnetic field acting on photons: in the presence of a strong and uniform synthetic magnetic field, photonic modes closely resemble Landau levels.
Even though in driven-dissipative photonic systems particles are continuously lost and replenished and no Fermi-Dirac statistics is available to impose a uniform population, a recent work~\cite{Ozawa:PRL2014} has pointed out that a spatially very localized coherent pump can be used to uniformly drive all states within a given Landau level. Even though the occupation is not pinned to $1$ as it happens in the electronic case, this uniform population is sufficient to observe a quantized response under the effect of a uniform force. In our photonic case, such a force --a sort of synthetic electric field-- is obtained with a uniform spatial gradient of the cavity resonance frequency as described by a linearly varying potential term $V(\rr_\parallel)=-\mathbf{F}\cdot \rr_\parallel$ in \eqref{eq:dd_psi}. 

Focusing for concreteness on the square-lattice Harper-Hofstadter model at $1/q$ flux per plaquette with a large $q$, the lateral spatial shift of the center of mass of the photon intensity distribution was in fact predicted~\cite{Ozawa:PRL2014} to be
\begin{equation}
 \Delta x=\frac{q\mathcal{C} F}{\pi\gamma_{\rm loss}}
 \label{eq:transverse_shift_quantized}
\end{equation}
where a force of magnitude $F$ is applied along the direction $y$ and $\mathcal{C}=1$ is the Chern number of the photonic band under examination: the driven-dissipative nature results in the transverse susceptibility being proportional to the lifetime $\gamma_{\rm loss}^{-1}$ times a quantized value. 

While Landau levels in continuous space are perfectly flat in energy, the energy bands of discrete topological models on a lattice (e.g. the Harper-Hofstadter or the Haldane models) may display a significant bandwidth $\Delta\omega_{\rm band}$ in energy. In our driven-dissipative systems, this offers a further knob to be exploited to stabilize and manipulate different states of the quantum fluid: depending on the relative value of the energy bandwidth $\Delta\omega_{\rm band}$ and of the loss rate $\gamma_{\rm loss}$, one can in fact transition from a IQH regime where $\gamma_{\rm loss}\gg\Delta\omega_{\rm band}$ and the coherent pumping uniformly drives all states within the band, to an anomalous Hall regime where $\gamma_{\rm loss}<\Delta\omega_{\rm band}$ and the frequency of the coherent drive specifically selects a set of resonant states. In this latter case, the lateral shift is no longer quantized and is determined by the averaged value of the Berry curvature over the set of resonant states. This provides a further example where the driven-dissipative nature of the photon fluid can be exploited to stabilize interesting many-body states and probe their physical properties.

An experiment that exploits driving and dissipation for a comprehensive study of anomalous and integer quantum Hall effect in a photonic Haldane model using a frequency-encode synthetic dimension platform based on an optical fiber loop was recently reported in ~\cite{chenier2024quantized}.

\subsection{Incoherent pumping}
\label{sec:incoherent_weak}


In the previous section we have discussed coherent driving schemes, where the phase of the in-cavity field is locked to the one of the incident field.
A different range of physical phenomena are observed when an incoherent pump is used to inject photons in the system: as a most significant example, we will focus on the spontaneous generation of coherence via a non-equilibrium phase transition, a phenomenon that opens the way towards the study of critical phenomena in a novel non-equilibrium setting.
While the underlying physics is very similar in the different systems, this phenomenon takes distinct names in distinct communities, e.g. laser oscillation in the optical literature or non-equilibrium Bose-Einstein condensation in the polaritonic one. An interdisciplinary review of this physics can be found in~\cite{bloch2022non}. 

\begin{figure}
    \centering
    \parbox[c]{0.28\textwidth}{\includegraphics[width=0.27\textwidth,clip]{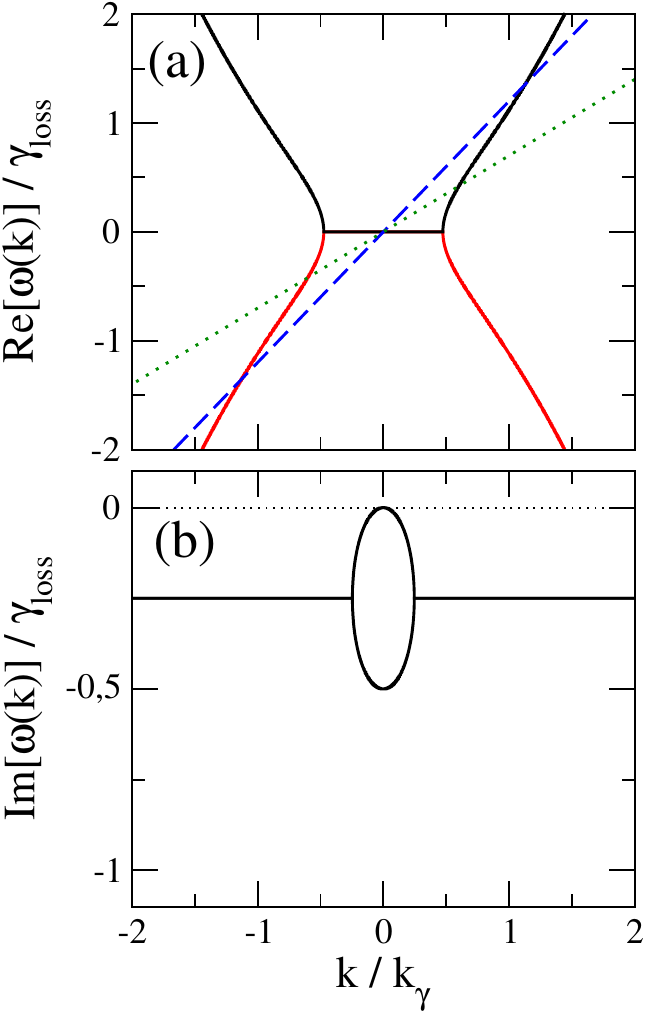}}
\parbox[c]{0.71\textwidth}{\includegraphics[width=0.7\textwidth,clip]{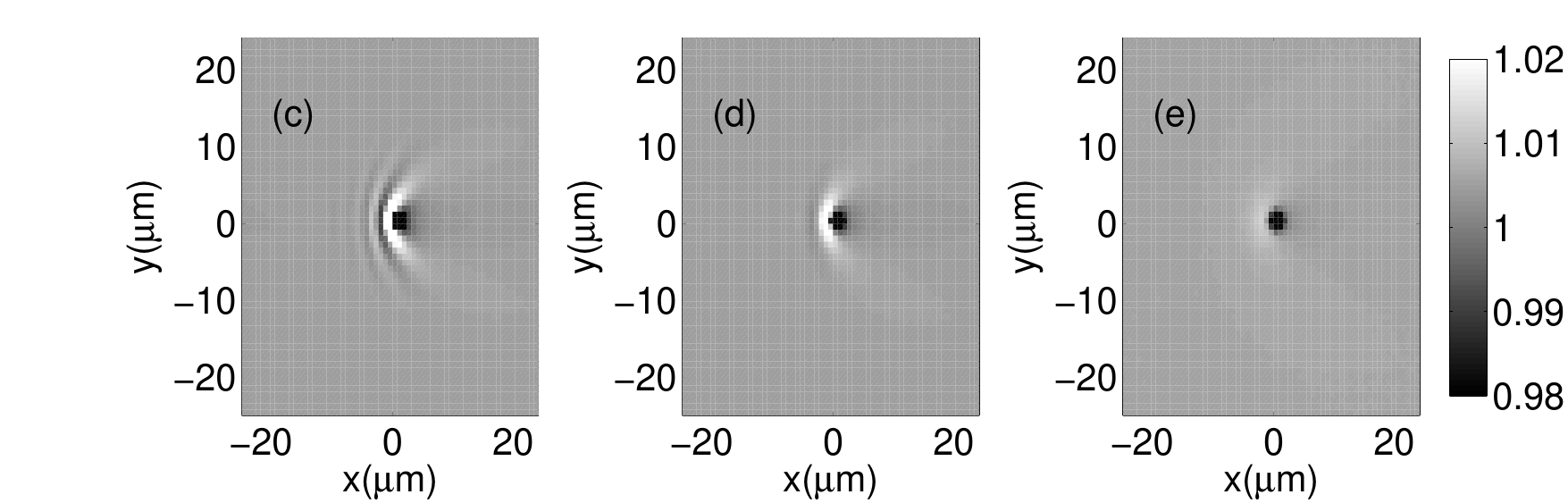}\\
    \includegraphics[width=0.7\textwidth,clip]{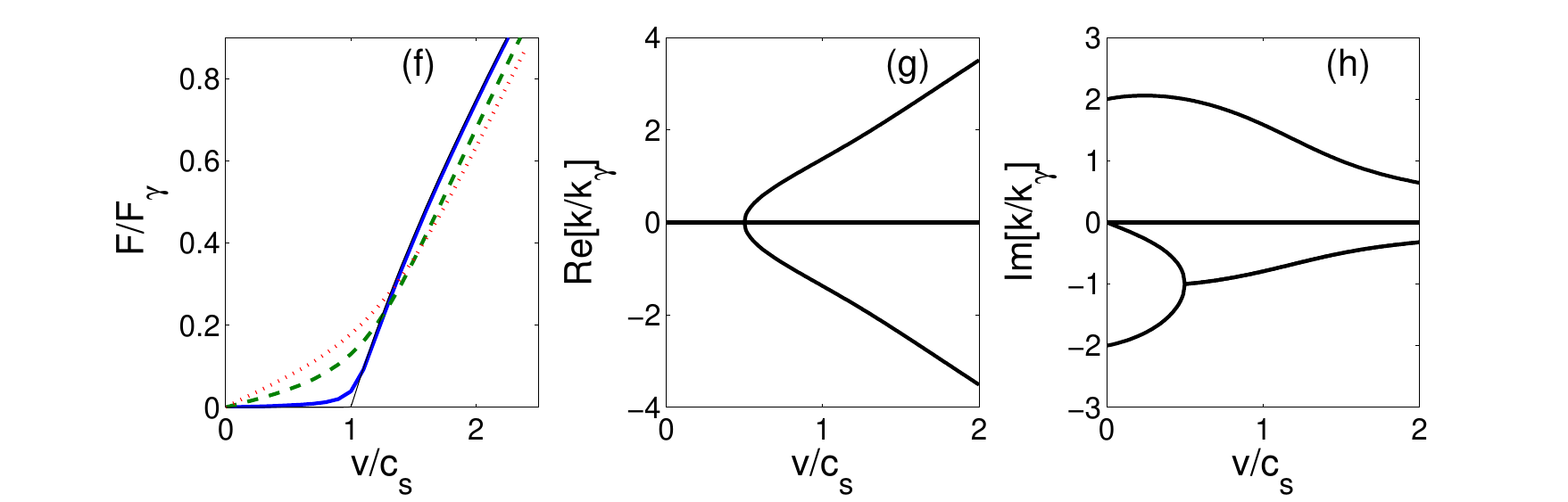}}
    \caption{ Left panels: plot of the real (a) and imaginary (b) parts of the dispersion of the collective excitations of a non-equilibrium condensate under incoherent pumping at $P/\gamma_{\rm loss}=2$. For small $k$ values, the dispersion displays a diffusive behaviour with a zero real part and a quadratically growing imaginary part. Right (c-h) panels: generalized Landau criterion for non-equilibrium condensates. Panels (c-e) show the density perturbation induced in a fluid moving in the positive $x$ direction by an impurity at rest. The different panels refer to different values of the condensate velocity $v/c_s =1.5, 1, 0.4$ across the (equilibrium) speed of sound $c_s=\sqrt{g_{\rm nl}\,|\psi|^2/m}$. Panel (f) shows the (normalized) force exerted by the fluid on the defect as a function of the condensate velocity $v/c_s$ for different values of the non-equilibrium parameter $\gamma_{\rm loss}/g_{\rm nl}\,|\psi|^2=0,\,0.1,\,1,\,2$. For the smallest $\gamma_{\rm loss}$ values, the sudden onset of friction that is visible in the vicinity of the critical speed $v/c_s= 1$  is a non-equilibrium counterpart of the Landau critical velocity of equilibrium superfluids~\cite{pitaevskii2016bose}.
    Panels (g-h) show a cut of the real and the imaginary part of the wavevector of the collective excitations emitted in the negative $x$ direction as a function of the speed for an intermediate loss case with $\gamma_{\rm loss}/g_{\rm nl}\,|\psi|^2=1$. The sudden onset of friction in (f) corresponds to the singular point that is visible in (g-h) slightly below $v/c_s= 1$. Panels (c-h) are adapted from~\cite{Wouters:PRL2010}. \label{fig:incoh_mean-field}}
\end{figure}

\subsubsection{Mean-field effects}

Restricting again to a weak interaction regime, a theoretical description of the field dynamics can be obtained at the mean-field level starting from the generalized Gross-Pitaevskii equation \eqref{eq:dd_psi}. In the absence of any coherent incident field $E_{\rm inc}=0$, the system transitions from a trivial $\psi=0$ steady-state solution for a weak pump $P<\gamma_{\rm loss}$ below threshold to a condensed (or lasing) solution for a strong pump $P>\gamma_{\rm loss}$ above threshold at $P_c=\gamma_{\rm loss}$. Here, the steady-state intensity of the field is stabilzed by nonlinear gain saturation effects to
\begin{equation}
 |\psi|^2= n_s \left(\frac{P}{\gamma_{\rm loss}}-1\right)
\end{equation}
but the phase of $\psi$ remains free and is randomly chosen at every instance of the experiment. The equation of motion \eqref{eq:dd_psi} displays in fact a global $U(1)$ symmetry $\psi\rightarrow \psi\,e^{i\varphi}$ for arbitrary $\varphi$ and this symmetry is spontaneously broken above threshold where the field gets a specific, uniform phase across the whole system. According to \eqref{eq:dd_psi}, the time-evolution of the field then displays a monochromatic oscillation at an intensity-dependent frequency $\omega$. The combination of a spatially uniform phase and a monochromatic oscillation in time showcases the interest of this configuration as a source of spatio-temporally coherent light.

\paragraph{Diffusive Goldstone mode} --
Even though the driven-dissipative steady-state shares many properties with a standard Bose-Einstein condensate at equilibrium, one must not forget that the non-equilibrium condition is responsible for important deviations in the collective properties of the fluid, in particular for what concerns the Goldstone mode that is associated to the spontaneously broken $U(1)$ symmetry. As it is illustrated in Fig.\ref{fig:incoh_mean-field}(a-b), the dispersion of the long-wavelength, low-frequency collective excitations does not show the usual sonic behaviour $\omega=c_s\,|\kk|$ of equilibrium condensates with a speed of sound $c_s$, but rather displays a diffusive form $\omega=-i \alpha k^2$ with a real and positive $\alpha$.
The usual behaviour with propagating collective excitations is recovered at higher frequencies and shorter wavelengths. The transition between the two regimes is typically associated to an exceptional point in the dispersion and its position is determined by the strength of the departure from equilibrium, namely by the value of the loss rates: the larger $\gamma_{\rm loss}$, the wider the non-equilibrium region of the dispersion. On top of this general rule, critical slowing down phenomena are visible for $P\gtrsim \gamma_{\rm loss}$ in the vicinity of the phase transition point. 

The theoretical prediction of a diffusive behaviour of the Goldstone mode of a non-equilibrium condensate has been recently confirmed in a pump-and-probe experimental measurement of the dispersion of the collective modes of an exciton-polariton condensate~\cite{claude2023observation}.

\paragraph{Generalized Landau criterion for superfluidity} --
This diffusive behaviour has been anticipated in~\cite{Wouters:PRL2010} to impact the conceptual meaning of the Landau criterion for superfluidity in the driven-dissipative context. In stark contrast to the equilibrium case, it is shown in Fig.\ref{fig:incoh_mean-field}(a) how the real part of the dispersion has an intersection with the $\omega=k\,v$ straight line for any non-zero value of $v$. A naive application of the Landau criterion for superfluidity~\cite{pitaevskii2016bose} would then predict that an impurity moving across a fluid will experience a significant friction force for any value of the flow speed $v$, with no specific feature at the speed of sound $c_s=\sqrt{g_{\rm nl}\,|\psi|^2/m}$.

This prediction is shown in Fig.\ref{fig:incoh_mean-field}(f) to disagree with a full numerical solution of the generalized Gross-Pitaevskii equation \eqref{eq:dd_psi}: for small or moderate values of the non-equilibrium parameter $\gamma_{\rm loss}/g_{\rm nl}\,|\psi|^2$, the friction force exerted by an impurity at rest onto a moving fluid displays in fact a sudden upwards kink at a value of $v/c_s$ close to the (equilibrium) critical velocity $v_{\rm cr}=c_s$. Such a change in behaviour can be associated to the singularity in the wavevector of the collective excitations emitted by the impurity shown in  Fig.\ref{fig:incoh_mean-field}(g-h): at low speed, this wavevector is purely imaginary, so the collective excitation generated by the impurity does not transport momentum; at large speeds, instead, this wavevector acquires a finite real part, which is directly associated to the momentum dissipated by the impurity into the fluid. 

\paragraph{Spatial structure of the condensate and supersolid behaviours} --
While thermodynamical arguments impose that equilibrium condensates form in the lowest-energy single-particle orbital~\cite{Huang,pitaevskii2016bose}, exciting features are observed in the spatial structure of finite-size non-equilibrium condensates. For instance, restricting the incoherent pump to a limited region of space allows to control the shape of the condensate via the so-called volcano effect: condensation mostly occurs at the center of the pump spot, and then the coherent particles get expelled by the repulsive interactions and form a macroscopic outward-expanding radial flow. Originally observed in~\cite{Richard:PRL2005}, this effect was theoretically interpreted in~\cite{Wouters:PRB2008} and fully confirmed experimentally in~\cite{Wertz:NatPhys2010}.

While this is the typical behaviour of spatially homogeneous systems, even more intriguing phenomena are observed in spatially periodic geometries. Here, condensation into an excited state at the top of a photonic band is favoured by the interplay of a positive interaction energy and a negative effective mass which result in effectively attractive interactions. These keep the condensate localized in space with a soliton-like shape~\cite{Tanese:NatComm2013,Baboux:Optica2018} and, in this way, reduce the condensation threshold with respect to condensation into the expanding positive-mass states at the bottom of the band.

A presently very active direction of research concerns the realization of quantum fluids of light that simultaneously display a long-range coherence and a  spatially periodic density modulation. This physics has been recently investigated in both theory and experiments~\cite{Nigro:PRL2025,trypogeorgos2025emerging} using spatially periodic devices displaying several photonic bands in the same energy range. While the main condensation process occurs at the top of a photonic band, a spatial modulation of the condensate density spontaneously appears as the result of coherent parametric scattering processes into finite wavevector modes in the side photonic bands.
Such configurations can be seen as photonic analogs of the so-called supersolid states of matter as recently observed in atomic condensates~\cite{recati2023supersolidity}. 

With a proper design of the device geometry, e.g. in ring resonators, related parametric processes can be induced into modes at different frequencies: in this case, the density modulation continuously moves in space and gives temporally periodic oscillations in the emitted light intensity, while preserving a long-time phase coherence of the emission. The importance of such {\em optical comb generators} for a number of opto-electronic applications~\cite{diddams2020optical,chang2022integrated} suggests that the new insight on the stabilization of complex states of matter offered by the use of many-body physics concepts in the novel driven-dissipative context will open interesting perspectives also for opto-electronic technologies.


\subsubsection{Fluctuation and quasi-condensation effects}

Going beyond the mean-field approximation, it is interesting to look at the process of condensation from the point of view of statistical mechanics. Equilibrium Bose-Einstein condensation in a spatially uniform geometry is typically associated~\cite{Huang,pitaevskii2016bose}  to the appearance of a sharp and macroscopic peak in the momentum distribution $n(\kk)=\langle \hat{\tilde{\Psi}}^\dagger(\kk)\,\hat{\tilde{\Psi}}(\kk)\rangle$ where $\tilde{\Psi}(\kk)$ is the Fourier transforn of the quantum field $\Psih(\rr)$. From a real-space perspective, condensation is instead signalled by the onset of long-range order in the correlation function
\begin{equation}
 G^{(1)}(\rr,\rr')=\langle \Psihd(\rr)\,\Psih(\rr') \rangle
\end{equation}
which tends to a finite value at long distances $|\rr-\rr'|\to \infty$~\cite{CCT:CdF}.

This physics can be understood in analogy with ferromagnetism in the $xy$ model
where the ferromagnetic transition is characterized in terms of the magnetization correlation function
\begin{equation}
 G^{(1)}(\rr,\rr')=\langle \vec{M}(\rr)\,\vec{M}(\rr') \rangle\,.
\end{equation}
In the high-temperature disordered state, the correlation function only displays short-range correlations and quickly tends to zero at long distances. On the other hand, long-range order is visible in the low-temperature ferromagnetic state where the correlation function tends to a finite value at long distances signalling a uniform magnetization throughout the whole system\footnote{It is interesting to note that a three-dimensional easy-plane (in the $xy$ plane) model of ferromagnetism subject to an orthogonal magnetic field (along $z$) is expected to displays a steady rotation of the  magnetization rotation along $z$. As a result, the $x$ and $y$ components of the magnetization display monochromatic oscillations as it happens to the field $\psi$ of the optical model.}.

In particular, the $SO(2)$ rotational symmetry that is spontaneously broken by the ferromagnetic transition of the $xy$ model is mathematically equivalent to the $U(1)$ group describing the $\Psi\rightarrow \Psi\,e^{i\phi}$ symmetry that is spontaneously broken at the Bose-Einstein condensation transition.
As the specific direction of the magnetization is randomly selected at every instance of the experiment, the average value of the magnetization $\vec{M}$ is zero by symmetry unless some (even weak) external magnetic field is present to pin its direction. In this case, a gap opens in the dispersion of the Goldstone mode associated to the spontaneous symmetry breaking, namely the magnon branch.
The optical analog of this idea was investigated in~\cite{Wouters:PRA2007,claude2023observation}: the phase of a condensate can be pinned by an additional incident coherent field. Also in this case, the explicit breaking of the symmetry leads to the opening of a gap in the collective excitation spectrum. As a consequence of the non-equilibrium condition, however, the gap may open in both the imaginary and the real parts of the Goldstone mode dispersion.

While three-dimensional geometries are the most natural ones in condensed matter systems, standard optical systems are typically limited to a maximum of two dimensions. In analogy to the bounds imposed on phase transitions by the Hohenberg-Mermin-Wagner theorem of equilibrium statistical mechanics~\cite{Huang}, it is then natural to wonder whether the reduced dimensionality has a similar impact also on the non-equilibrium phase transition. Even though the general theorem can not be directly applied to the non-equilibrium context, specific calculations have predicted~\cite{Graham:ZPhys1970,Wouters:PRB2006} that in one dimension the long range order does not survive the long-wavelength fluctuations and is replaced by a exponential decay of the same-time coherence, yet on relatively long distance scales. Even more subtle features have been pointed out for the intermediate two-dimensional case where the non-equilibrium condition may have a strong impact on the nature of the Berezinskii-Kosterlitz-Thouless transition~\cite{Altman:PRX2015,Sieberer:RPP2016,Zamora:PRX2017}. A common tool for the theoretical study of these phenomena is offered by the stochastic Gross-Pitaevskii equation discussed in Sec.\ref{sec:Wigner}.


Once again, a very important asset for the experimental investigation of the critical properties and of the spatio-temporal correlation function in the different regimes is provided by the radiative loss channel: as the coherence properties of the in-cavity field are imprinted into the one of the emitted radiation, their measurement can be straightforwardly performed with standard optical techniques. This feature has been exploited in experimental studies of the correlation properties across the phase transition and, in particular, for the recent observation of Kardar-Parisi-Zhang (KPZ) universal behaviours in a one-dimensional configuration~\cite{fontaine2022kardar}.

While moving up to two-dimensional systems is not expected to pose additional conceptual challenges to the experiment, three- and higher-dimensional configurations require the use of synthetic dimensions: extra spatial dimension can in fact be realized by exploiting some degrees of freedom of the photon, e.g. the different cavity modes of a ring resonator~\cite{Ozawa:PRA2016,Ehrhardt:LPR2023,Yuan:Optica2018}. 
By introducing gain into these platforms, we anticipate the possibility of studying condensation in novel high-dimensional contexts and, in this way, of unveiling those peculiar statistical behaviours that result from the interplay of KPZ physics with the roughening transition~\cite{TAKEUCHI201877,wiese1998perturbation,marinari2000critical,Canet:PRL2010}.

%
%
%
%
%

\section{Strongly correlated systems}
\label{sec:strongly}
%

In the previous Sections we have discussed the physics of dilute driven-dissipative systems: for sufficiently weak interactions, a mean-field theory based on (stochastic) partial differential equations for classical fields provides an accurate description of the steady-state and of the dynamics of the system. In this last Section, we briefly summarize the state-of-the-art and some future developments in the direction of strongly quantum correlated states of photonic matter that can be observed in the presence of strong optical nonlinearities.

\subsection{Coherent drive}

\subsubsection{Photon blockade in a single-mode cavity}

The first proposal in this direction~\cite{Imamoglu:PRL97} was the so-called {\em photon blockade} effect. The dynamics of the quantum electromagnetic field in a nonlinear single-mode cavity can be described by a nonlinear oscillator Hamiltonian of the form 
\begin{equation}
H=\hbar \omega_o \hat{a}^\dagger \hat{a} + \frac{\hbar \omega_{\rm nl}}{2} \hat{a}^\dagger \hat{a}^\dagger \hat{a} \hat{a}
\label{eq:anharmonic}
\end{equation}
where the anharmonicity is quantified by the parameter $\omega_{\rm nl}$ proportional to the $\chi^{(3)}$ optical nonlinearity of the nonlinear medium embedded in the cavity. In a planar geometry, we can make use of \eqref{eq:Hcav} to get
\begin{equation}
 \omega_{\rm nl}=\frac{g_{\rm nl}}{\mathcal{S}}
\end{equation}
where the effective mode area $\mathcal{S}$ is defined as 
\begin{equation}
  \mathcal{S}=\frac{\left(\int\!d^2\rr_\parallel\,|\mathcal{E}_\parallel(\rr_\parallel)|^2\right)^2}{\int\!d^2\rr_\parallel\,|\mathcal{E}_\parallel(\rr_\parallel)|^4} 
\end{equation}
in terms of the mode profile
$\mathcal{E}_\parallel(\rr_\parallel)$ along the cavity plane.

\begin{figure}
    \centering
    \parbox[c]{0.23\textwidth}{\frame{\includegraphics[width=0.23\textwidth,trim=2cm 0cm 19cm 3.cm, clip]{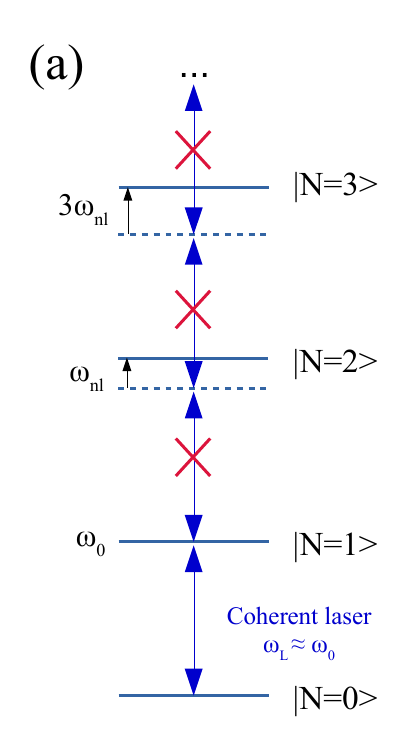}}}\hspace{0.01\textwidth}
    \parbox[c]{0.24\textwidth}{ \frame{\includegraphics[width=0.24\textwidth,trim=0cm 0cm 19cm 0cm, clip]{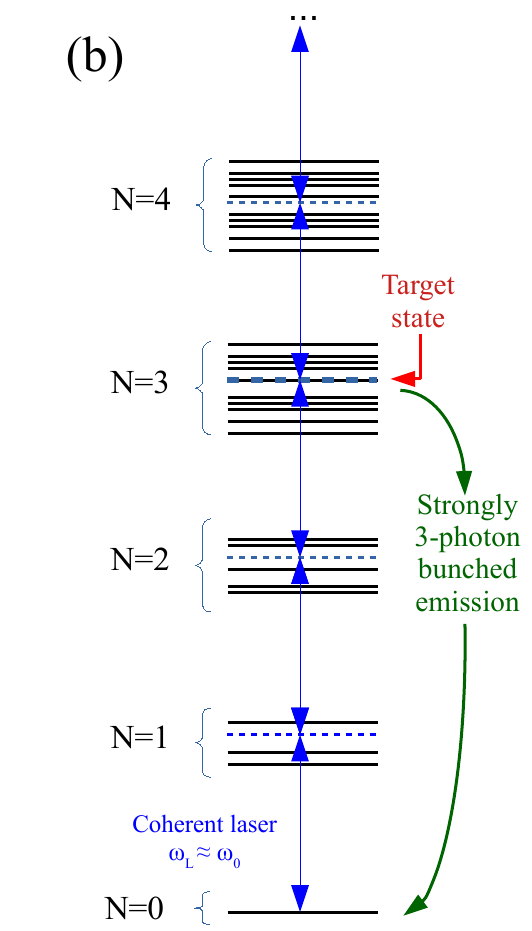}}}\hspace{0.03\textwidth}
    \parbox[c]{0.4\textwidth}{\includegraphics[width=0.4\textwidth]{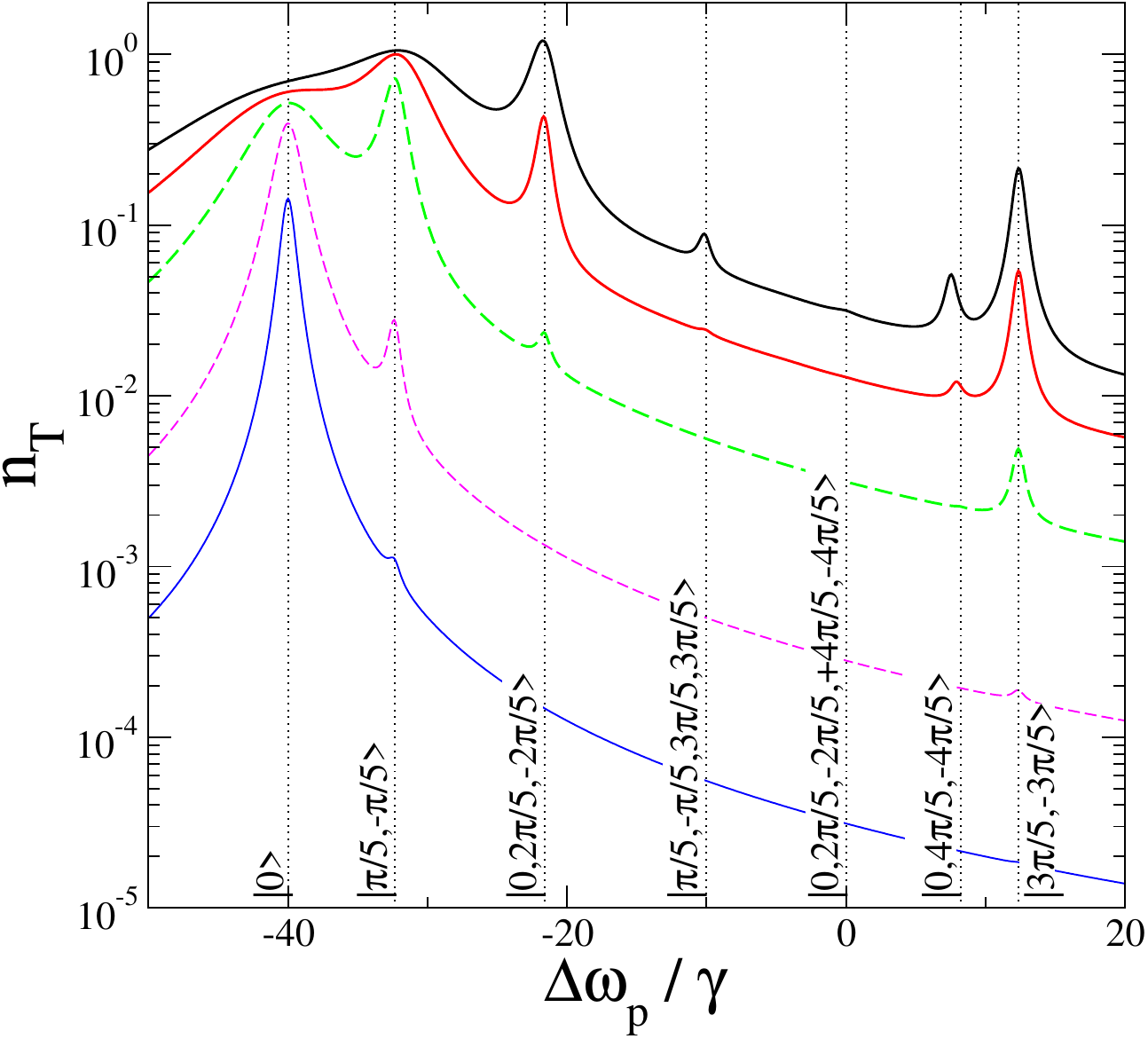} \\
    \includegraphics[width=0.44\textwidth]{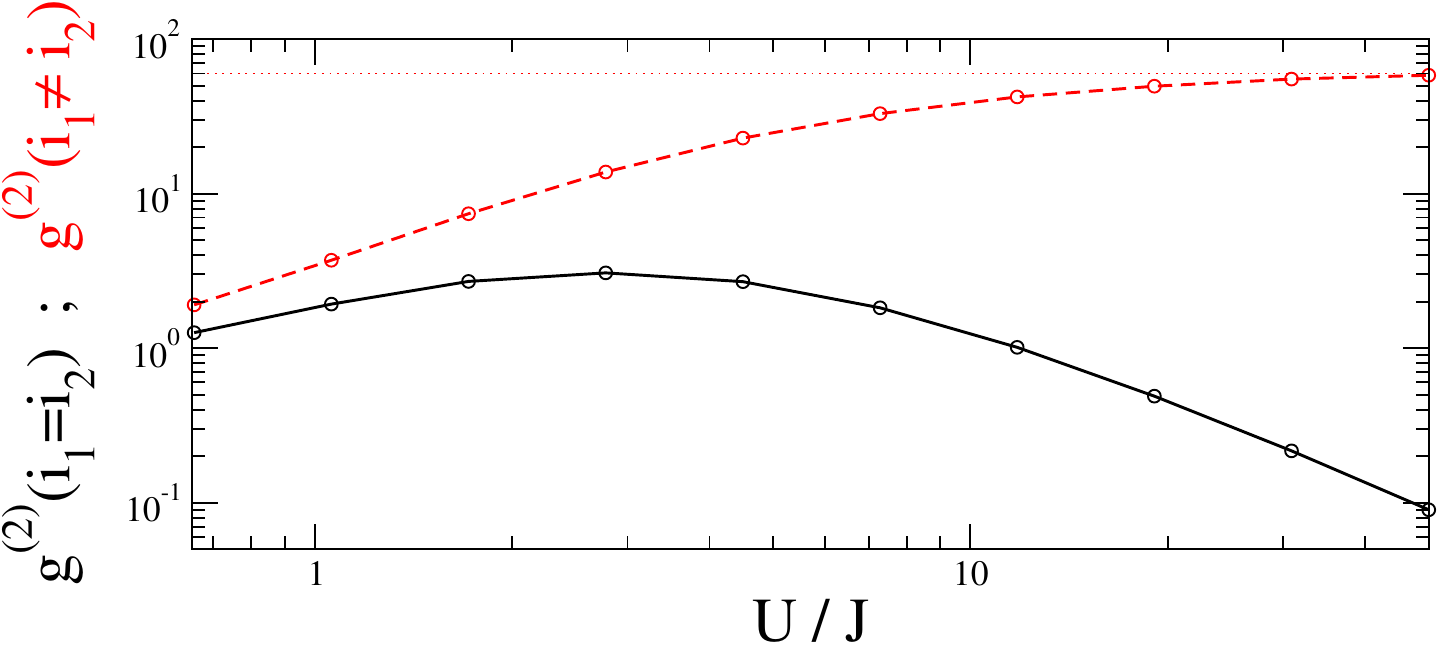}}
    \caption{Left panel (a): sketch of the photon blockade mechanism for a single-mode cavity under a coherent pumping exactly on resonance with the bare cavity frequency, $\omega_L=\omega_o$. Because of the strong nonlinearity $\omega_{\rm nl}\gg \gamma_{\rm loss}$,  transitions to all $N>1$ states are non-resonant and the dynamics is confined to the $N=0,1$ states, giving effectively impenetrable photons. Central panel (b): sketch of the coherent pumping scheme to selectively excite a $N=3$-photon state via a three-photon transition.
    Top-right panel: generation of a Tonks-Girardeau gas of fermionized photons under coherent pumping. Spectrum of total photon number $n_T$ as a function of the incident frequency $\Delta\omega_p=\omega_{\rm inc}-\omega_o$ for different values of the pump intensity, $F/\gamma_{\rm loss}=0.1,\,0.3,\,1,\,2,\,3$. Photons are assumed to be perfectly impenetrable $U/\gamma_{\rm loss}=\infty$ and $J/\gamma_{\rm loss}=20$. For each many-body state, the vertical dotted lines indicate the theoretical prediction for the position of the $N$-photon resonance peak; the label indicates the orbitals filled in the corresponding fermionic wavefunction.
    Bottom-right panel: second order photon correlations in the emission from the same (black) and from different (red) sites as a function of the interaction constant $\omega_{\rm nl}/\gamma_{\rm loss}$. At each point, the coherent drive is set on resonance with the lowest two-photon state of a 3-sites chain. The right panels are adapted from~\cite{Carusotto:PRL2009}.  \label{fig:TG}}
\end{figure}

Independently of the specific geometry, the eigenstates of \eqref{eq:anharmonic} are labelled by the photon number and have an energy
\begin{equation}
E_n=n\hbar\omega_o+ \hbar \omega_{\rm nl}\frac{n(n-1)}{2}\,.
\end{equation}
As it is sketched in the left-most panel of Fig.\ref{fig:TG}, the energy separation $E_{n+1}-E_{n}=\hbar\omega_o+n\hbar\omega_{\rm nl}$ is a growing function of $n$. As such, a coherent incident field resonant with the $0\rightarrow 1$ transition will not be able to further excite the system to the $n\geq 2$ states as the corresponding transitions are no longer resonant. For instance, the $1\rightarrow 2$ transition is detuned by an effective interaction energy $\omega_{\rm nl}$: as soon as this exceeds the cavity linewidth $\omega_{\rm nl}\gg \gamma_{\rm loss}$, the coherent field is not able to efficiently drive this second transition, so the cavity system ends up behaving as an effective two-level system. In physical terms, this can be understood as photons becoming effectively impenetrable objects, a situation that is in stark contrast with the prediction of the linear Maxwell equations where photons are rigorously non-interacting particles.

\subsubsection{Many-body states}

A straightforward generalization of this pumping scheme can be exploited to generate strongly correlated many-photon states in spatially extended geometries such as cavity arrays. The key idea, sketched in the central panel of Fig.\ref{fig:TG}, is to resonantly drive a $N$-photon transition to bring the system from its zero-photon vacuum state to the desired $N$-photon eigenstate of the conservative many-body Hamiltonian: the crucial requirement for an efficient selectivity of this preparation scheme is that the energy gap between the desired $N$-photon eigenstate and its neighbors exceeds the linewidth of the $N$-photon state, which typically grows as $N\,\gamma_{\rm loss}$.

As a first example, this idea was theoretically explored in view of realizing a Tonks-Girardeau gas of impenetrable photons in a one-dimensional chain of coupled cavities~\cite{Carusotto:PRL2009}. This is described by a Hamiltonian in the form
\begin{equation}
 H=\sum_j\hbar \omega_o \hat{a}_j^\dagger \hat{a}_j + \frac{\hbar \omega_{\rm nl}}{2} \hat{a}_j^\dagger \hat{a}_j^\dagger \hat{a}_j \hat{a}_j - \hbar J \sum_{\langle j,j'\rangle} \ahd_j \ah_{j'}
 \label{eq:H_lattice}
\end{equation}
where the indices $j,j'$ run over the sites of the chain, assumed to have periodic boundary conditions, and the hopping terms of amplitude $J>0$ are restricted to next-neighbor sites. The impenetrability condition is enforced by assuming $\omega_{\rm nl}\gg J$.
A coherent pump is assumed to drive all sites according to the discrete version of \eqref{eq:coh_pump}, 
\begin{equation}
H_{\rm coh}=\hbar \sum_j \left[ F_j(t)\, \ahd_j + \hbar F_j(t)^*\, \ah_j\right]\,.
\end{equation}
In what follows, we assume  that all sites are driven in a monochromatic with the same amplitude $F$ and frequency $\omega_{\rm inc}$, that is $F_j(t)=F\,\,e^{-i\omega_{\rm inc} t}$. As the master equation is in the Lindblad form, the steady-state and the dynamics of the system can be calculated by solving it with standard numerical tools, e.g. by projecting it on the basis of number states. Note that efficient high-level numerical libraries are nowadays available for these tasks~\cite{qutip}.

Examples of spectra of the steady-state total population $n_T= \sum_j \langle \ahd_j\ah_j \rangle$ as a function of the incident frequency $\omega_{\rm inc}$ are shown in the upper-right panel of Fig.\ref{fig:TG} for growing values of the incident intensity $|F|^2$: at low intensity (blue line), the spectrum is characterized by a single main peak corresponding to the $1$-photon transition to the $k=0$ orbital. At growing intensities, additional peaks at frequencies corresponding to the higher $N=2,3,\ldots$ photon transitions appear. As expected, the position of each of these peaks matches the frequency of the $N$-photon resonance from the vacuum state (at zero energy) to the some $N$-photon state of energy $\omega_{N}^\alpha$ within the $k=0$ sector,
\begin{equation}
\omega_{\rm inc}= \omega_{N}^\alpha/N\,.
\label{eq:coh_resonance}
\end{equation}
In the specific Tonks-Girardeau case, the different excited states are labeled by the quasi-momenta of the (fermionized) particles and their energy can be analytically calculated~\cite{CastinLectures2004}. The  expected position of the resonances is indicated in the Figure by the vertical lines, which indeed match well with the position of the numerically calculated resonance peaks. 

In order to be able to selectively drive the desired many-body state, the frequency separation from the nighboring states has to exceed the linewidth: as the separation is typically set by the hopping $J$ and interaction $\omega_{\rm int}$ energies and the broadening grows as $N\gamma_{\rm loss}$, the requirement on the linewidth $\gamma_{\rm loss}$ becomes more and more stringent as larger $N$ states are targeted. On top of this, one must not forget that for growing values of the drive amplitude $F$, each peak displays a sizable power-broadening on top of its natural linewidth.

As it is well known from atomic physics~\cite{CCT4}, a coherent illumination leads to a reversible oscillatory dynamics between the different states, with absorption processes followed by stimulated emission ones until a steady-state is enforced via spontaneous emission processes. In particular, for very strong incident intensities, the steady-state tends to an equidistribution of the population among the different states. As such, a continuous-wave excitation can hardly be used to transfer all the population to a given excited state, but more complex time-dependent schemes have to be adopted to generate, e.g., a state with a well-defined $N$.

Nonetheless, as the population in the steady-state is typically a decreasing function of the photon number $N$, this difficulty can be circumvented by specifically looking at $N$-photon observables. Such observables are in fact zero on all lower $N'<N$ photon states and, up to moderate values of the incident intensity, only receive a sub-dominant contribution from the higher $N'>N$ ones.

A concrete example of application of this idea is illustrated in the bottom-right panel of Fig.\ref{fig:TG}, where we specifically look at $N=2$ observables such as the second-order correlation function of the emission
\begin{equation}
g^{(2)}(i_1,i_2)=\frac{\langle \ahd_{i_1} \ahd_{i_2} \ah_{i_2}\ah_{i_1}\rangle}{\langle \ahd_{i_1} \ah_{i_1}\rangle \langle \ahd_{i_2} \ah_{i_2}\rangle} \, .
\end{equation}
Depending on the value of the interaction parameter $\omega_{\rm int}/J$, the $N=2$-photon eigenstates go from being delocalized, approximately non-interacting-photon states to fermionized, impenetrable-photon states. This key difference directly translate to the shape of $g^{(2)}(i_1,i_2)$: for non-interacting photons it has a position-independent value equal to $1$; for impenetrable photons, its same-point value is much suppressed with respect to the different-point value, $g^{(2)}(i_1=i_2)\ll g^{(2)}(i_1\neq i_2)$. This illustrates how a coherent drive followed by a measurement of a $N$-photon correlation function allows to obtain interesting information on the many-body eigenstates of a strongly interacting system. An conceptually related experiment was reported using a superconducting circuit-QED platform in~\cite{Fedorov:PRL2021}.

\begin{figure}[htbp]
    \centering
    {\includegraphics[width=0.9\textwidth,trim=0cm 0cm 8cm 0cm, clip]{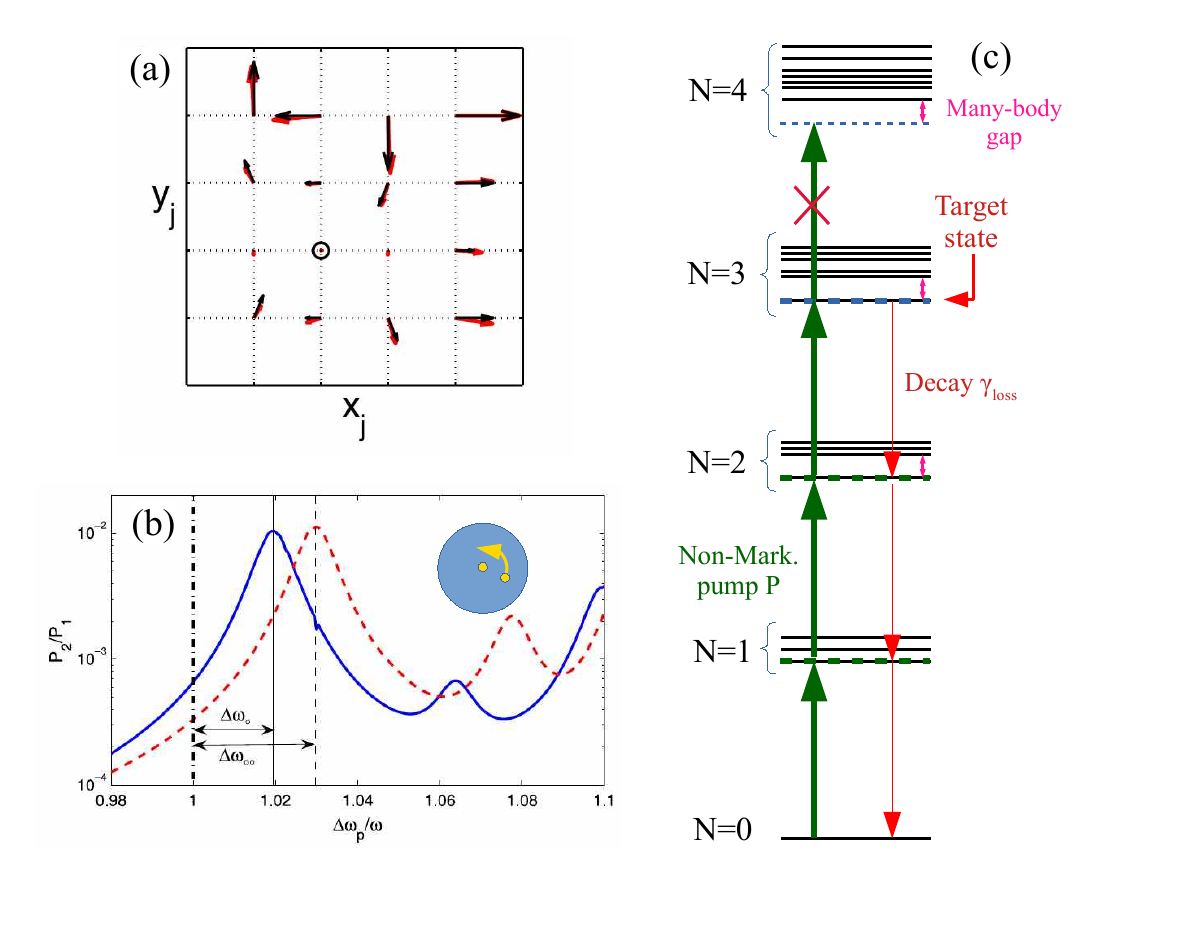}}
    \caption{ Top-left panel (a): reconstruction of the two-particle wavefunction of a Laughlin state from the two-photon correlations of the light emitted by the two-photon FQH state generated under coherent pumping. The red arrows indicate the two-photon emission amplitude, the black ones indicate the value of the complex-valued Laughlin wavefunction discretized in the lattice. Bottom-left panel (b): frequency-shift of the two-photon resonance peak under the effect of the braiding of a quasi-hole. The continuous blue line is the spectrum for a single quasi-hole braided around the cloud at a radius $r/\ell=0.4$; the dashed red line spectrum includes also a stationary second hole located at $r=0$, as indicated in the sketch in the inset. 
    Right panel (c): sketch of the frequency-selective incoherent pumping scheme to stabilize a $N=3$ FQH state. The non-Markovian nature of the pump prevents excitation across the many-body gap. The (Markovian) photon losses can not lead to states above the many-body gap. All together, this restricts the dynamics to states with $N\leq 3$ photons below the many-body gap.
    Panel (a) is adapted from~\cite{Umucalilar:PRL2012}. Panels (b) is adapted from~\cite{Umucalilar:PLA2013}. \label{fig:FQH}}
\end{figure}

Further theoretical work~\cite{Umucalilar:PRL2012} has extended this idea to a two-dimensional lattice geometry where impenetrable photons are also subject to a synthetic magnetic field. In this geometry, the optical transitions occur from the vacuum state to strongly correlated $N$-photon states analogous to the ones of fractional quantum Hall systems. An example of the peculiar physics of this regime is illustrated in the top-left panel (a) of Fig.\ref{fig:FQH}. Here, we show how tuning the coherent pump frequency on the lowest $N=2$-photon resonance peak, one can extract information on the complex microscopic structure of the Laughlin states from the two-photon correlations of the emitted light: even in a small $4\times 4$ lattice, we recognize in the wavefunction a clear antibunching feature between pairs of strongly interacting photons as well as a complex phase profile as a function of the relative position.

A closely related idea has been experimentally explored in a recent experiment~\cite{clark2020observation}: strong interactions between photons are obtained by coupling light to optically dressed atoms in a Rydberg-EIT configuration~\cite{Gorshkov:PRL2011,peyronel2012quantum}; a synthetic magnetic field in a cylindrically symmetric geometry is obtained by using a twisted optical ring-cavity~\cite{schine2016synthetic}. Combining these elements, a two-photon Laughlin state is selectively created using a coherent laser pump with finite angular momentum and its microscopic structure is probed by looking at the correlation functions of the emitted light at different angles and in different orbital angular momentum channels.

Scaling up the idea to larger photon numbers is naturally the next goal, but several challenges are still present along this way. As it was mentioned above, an obvious difficulty comes from the decreasing value of the spacing between many-body resonances which imposes stricter constraints on their linewidth.
An even more serious challenge comes from the quickly decreasing value of the matrix element for driving a $N$-photon transition from the trivial vacuum state to a topologically non-trivial FQH state with a coherent light beam formed by uncorrelated photons.

\subsubsection{Signatures of the braiding phase}

In spite of these difficulties, an intriguing new possibility offered by the driven-dissipative condition is illustrated in the bottom-left panel Fig.\ref{fig:FQH}(b). One of the key feature of a topological state of matter is encoded in the so-called {\em braiding phase} that the many-body wavefunction acquires when two quasi-holes are made to adiabatically move through the fluid and encircle each other as sketched in the inset. While the braiding phase is a mathematically well defined concept~\cite{Tong:QHbook,stern2008anyons}, an experimental observation of the overall phase of a quantum wavefunction requires to design some interference process between different quantum states~\cite{Paredes:PRL2001,Grusdt:NatComm2016,de2020anyonic}. Quantum superpositions between the vacuum state and the desired many-body state are naturally generated by a coherent pumping process, so any extra phase acquired by the many-body wavefunction may be observed as a shift in the resonance frequency of the corresponding transition.

A concrete implementation of this general idea to measure the braiding phase in driven-dissipative fractional quantum Hall fluids of light was proposed in~\cite{Umucalilar:PLA2013}. In the presence of one or more localized repulsive potentials, the lowest energy two-photon eigenstate consists of a Laughlin state pierced by quasi-holes pinned at the position of the potentials. When a single quasi-hole is braided around the fluid at a radius $r_o$ from the center with a constant angular velocity $\Omega$, the many-body wavefunction acquires at each roundtrip a phase given by the flux of synthetic magnetic field enclosed by the trajectory. This phase, of geometric origin, is responsible for the shift $\Delta\omega_o$ of the main $N=2$-photon peak visible on the blue spectrum of Fig.\ref{fig:FQH}(b). When a second hole is present at the origin, the shift $\Delta\omega_{oo}$ acquires an additional contribution due to the braiding of one quasi-hole around the other,
\begin{equation}
 \Delta\omega_{oo}-\Delta\omega_{o}=\varphi_{Br} \frac{\Omega}{2\pi}\,.
 \label{eq:bradingOmega}
\end{equation}
where $\varphi_{Br}$ is the braiding phase accumulated at each round trip and $2\pi/\Omega$ is the time needed for a round-trip.
The frequency shift observed in the numerical simulation in Fig.\ref{fig:FQH}(b) is in excellent agreement with the value theoretically extracted from a microscopic calculation of the many-body braiding phase for this configuration. As it is discussed in detail in~\cite{Umucalilar:PLA2013}, the value of the braiding phase tends to the topological value $\pi$ as soon as the cloud has a macroscopic size and edge effects can be neglected. In principle this requires a number $N\gg 1$ of particles; interestingly in view of experiments, a qualitative effect is already visible for $N=3$ and the asymptotic value is closely approached for $N\sim 10$.

This theoretical prediction shows how the quantum superposition nature of the state that is naturally generated in a coherent pumping process is not only a limitation of the preparation process, but can be exploited as a new asset to observe the peculiar properties of the many-body wavefunction of topological states of matter.


\subsection{Incoherent pumping}

In the previous Subsection we have seen how the reversible nature of the coherent pumping process makes the typical form of the driven-dissipative steady-state to consist of a superposition of different number states. If one wishes to concentrate the population on a single excited state, many-photon generalizations of the $\pi$-pulses used in the coherent manipulation of the internal state of atoms~\cite{CCT4} can be envisaged. As the many-body states are very close in energy, sophisticated quantum control techniques may be required to achieve a good selectivity, which furthermore rely on some preliminary theoretical knowledge of the system.

An alternative way, inspired to population inversion in laser physics, is to exploit a incoherent pumping scheme to irreversibly push the population upwards towards higher-$N$ states.
In its simplest formulation with a broad-band incoherent pump as described within the Markovian setting of Eqs.\eqref{eq:Lind-pump}, this incoherent pump scheme allows to efficiently push the population upwards but the absence of energy-selectivity makes the population to spread among the different many-body states forming each $N$-photon manifold, typically in a uniform way~\cite{Lebreuilly:CRAS2016}. As a result, the fluid ends up displaying an effectively infinite temperature which prevents observation of interesting quantum many-body effects.

A promising way-out is to introduce a frequency-selectivity of the pumping scheme which allows for non-Markovian features in the irreversible coupling between the system and the external bath used for pumping. The sketch of a simple implementation of this idea is shown in the right panel of Fig.\ref{fig:FQH} in the context of generating a FQH fluid of light: provided the spectrum of the incoherent pump is narrower than the many-body gap protecting the target FQH state, the pumping can not lead to states above the many-body gap. As states above the many-body gap can not be reached by loss processes either\footnote{This is easy to see on the analytical form of the Laughlin wavefunctions of FQH states~\cite{Tong:QHbook},
$$\psi(z_1,\ldots,z_N)=P(z_1,\ldots,z_N)\,\Pi_{i>j}(z_i-z_j)^p\,e^{-\sum_i |z_i|^2}\,, $$ 
where $P$ is an arbitrary symmetric polynomial in the normalized complex $z_i=(x_i+i\,y_i)/\ell_B$ coordinates of the $i=1,\ldots,N$ particles and $p$ is an integer (even for bosonic particles like photons and odd for fermionic particles like electrons) describing the specific FQH state under consideration. The non-interacting nature of this state is enforced by the presence of factors $(z_i-z_j)$ for any pair $i,j$ of particles, which give zeros in the wavefunction for overlapping particles. This impenetrability condition is preserved when a particle is removed from the system.}, the population remains concentrated within the states below the many-body gap. If the geometry of the fluid has periodic boundary conditions, the incompressibility of the bulk puts an upper bound to the number of particles that can be injected before one has to cross the many-body energy gap, e.g. a maximum of $N=3$ particles can be accommodated in a non-interacting state for the configuration illustrated in the sketch. If the incoherent pumping rate $\gamma_{\rm pump}$ exceeds the loss rate $\gamma_{\rm loss}$, the population is irreversibly pushed up along the ladder of $N$ states and gets eventually concentrated in the upmost available state which is indeed our target FQH state.

The situation is slightly more complicate if the periodic boundary conditions are replaced by a finite geometry with physical edges, e.g., when the fluid is subject to a confining potential~\cite{Macaluso:PRA2017} or has open boundary conditions~\cite{Nardin:PRL2024,Binanti:PRR2024}. In these cases, the energy gap to the lowest edge excitation is typically much smaller than the many-body gap in the bulk and scales inversely to the cloud perimeter. This puts more and more severe constraints on the energy selectivity of the pumping process that is required to suppress spurious edge excitations. However, since these excitations are confined on the edge of the cloud, they do not create any harm to the topological properties of the bulk which remains well stabilized in a very pure fractional quantum Hall state.

In order to put these qualitative arguments onto solid and quantitative grounds one needs to develop a theoretical description that is able to account for the non-Markovian features associated to the frequency-dependence of the pumping process. In~\cite{Biella:PRA2017,Caleffi:PRL2023}, this was done by explicitly introducing into the model two-level emitters that provide a frequency-dependent injection of photons as discussed at the beginning of Sec.\ref{subsubsec:Hemitter}. This allows to reformulate the frequency-dependent incoherent pumping in terms of a master equation in the Lindblad form, which can then be solved with standard methods, e.g. using one of the numerical libraries that are available for these tasks~\cite{qutip}. In the fast repumping regime where the emitters can be adiabatically eliminated, a simplified photon-only description based on generalized Lindblad master equation can be used, as presented in Sec.\ref{subsubsec:Hemitter}. This has provided interesting insight both in lattice and in continuum geometries~\cite{Kapit:2014PRX,Umucalilar:PRA2017}.
On top of these numerical works, some analytical results have also been obtained in specific limiting cases~\cite{umucalilar2021autonomous}, further confirming the promise of this framework for the autonomous stabilization of FQH states.
On top of these feasibility studies, interesting consequences of the driven-dissipative nature of the fluid have also been pointed out in theoretical works. While isolated quasi-holes in the bulk of a standard equilibrium FQH fluid move along deterministic trajectories, typically with very low speeds, a recent work~\cite{kurilovich2022stabilizing} has anticipated that repeated loss and repumping processes in the neighborhood of a quasi-hole give rise to an effectively stochastic motion of quasi-holes.

The conceptual facility of this mechanism for the {\em autonomous stabilization} of a fractional quantum Hall fluid of light is a consequence of the peculiar structure of the many-body spectrum and of the many-body eigenstates. Something similar occurs for the deep Mott insulator state, whose stabilization was experimentally demonstrated~\cite{ma2019dissipatively} using a superconductor-based circuit-QED platform operating in the $\omega_{\rm nl}\gg J$ regime (in the notation of the lattice Hamiltonian \eqref{eq:H_lattice}). In this limit, the thickness of the band of hole states (on the order of $J$) is much smaller than the separation from extra-particle states featuring some doubly occupied site (on the order of $\omega_{\rm nl}$): choosing a repumping rate $J\ll \Gamma_P \ll \omega_{\rm nl}$ then guarantees an effective replenishing of any hole, while suppressing the creation of extra particles.

An even richer physics occurs when $J$ and $\omega_{\rm nl}$ have comparable values, so that the driven-dissipative condition interplays with the superfluid-insulator quantum phase transition occurring at a finite critical value of $J/\omega_{\rm nl}$~\cite{bloch2008many}. In~\cite{Hafezi:PRB2015,Lebreuilly:PRA2017}, a square-like form the the incoherent emission spectrum, arising e.g. from a inhomogeneously broadened distribution of emitters, was considered, that leads to an equilibrium-like superfluid-insulator transition. The case of a Lorentzian spectrum was instead considered in~\cite{Biella:PRA2017} and specific non-equilibrium features pointed out for the superfluid-insulator transition. Beyond this, the spectrum of the collective excitations on top of the driven-dissipative steady state was studied across the different regimes in~\cite{Caleffi:PRL2023}: similarly to the equilibrium case, gapped particle-hole excitations are found in the insulating state and a gapless Goldstone mode in the superfluid state. As a typical feature of driven-dissipative systems, this latter was predicted to have the same diffusive properties as in the weakly-interacting case discussed in~Sec.\ref{sec:incoherent_weak}.

\section{Conclusions and perspectives}
\label{sec:conclusions}

In this article we have reviewed the basic concepts of the driven-dissipative many-body physics of quantum fluids of light and we have summarized the main theoeretical tools used for their description. A special emphasis has been given to those new effects and those new tools that the driven-dissipative nature introduces for the stabilization and the manipulation of interesting many-body states. 

As illustrative examples, in the weakly interacting regime, a suitable coherent pump can be used to generate inhomogeneous flows displaying acoustic horizons; in topological photonic systems, the interplay between the bandwidth of energy bands and the loss rate can be exploited to smoothly transition from an anomalous Hall to a quantum integer Hall regime in the transverse drift under crossed synthetic magnetic and electric fields. In the strongly interacting regime, spectroscopical signatures of the anyonic braiding phase in fractional quantum Hall fluids of light have been anticipated under coherent pumping and promising incoherent pumping schemes to stabilize macroscopic strongly correlated fluids are being actively investigated.

These advances are opening exciting perspectives in a numer of interdisciplinary directions, at the crossroad of many-body physics, nonlinear statistical mechanics, and quantum simulation of gravitational problems. As we have mentioned along the article, acoustic horizons in inhomogeneous flows have been predicted to give acoustic analogs of Hawking radiation, which could be experimentally detected in the intensity correlations of the emitted light. The flexibility in the generation and the manipulation of different flow geometries is of utmost interest in view of studying more complex phenomena where, e.g., Hawking radiation interplays with superradiance and localized quasi-normal modes~\cite{Jacquet:PRL2023} of the underlying space-time. The coherence properties of condensates and quasi-condensates display universality features that fall in the framework of the Kardar-Parisi-Zhang framework: going beyond the one-dimensional geometries considered so far in experiments, the combination of condensation with synthetic dimension platforms is expected to open the way to studies of non-equilibrium statistical mechanics and, more generally, many-body effects in high dimensions, possibly larger than the three dimensions of our physical space~\cite{ozawaprice2019topological}. Finally, application of the frequency-dependent incoherent pumping techniques to strongly interacting photons in complex geometries and/or in the presence of non-trivial band topologies offers new promising avenues towards the driven-dissipative stabilization of new states of photonic matter and the observation of their peculiar collective excitations, possibly with new anyonic statistics of interest for topological quantum computing~\cite{Kapit:2014PRX}.

On top of these research directions in the context of quantum fluids of light, driven-dissipative schemes are attracting a growing interest also in the context of ultracold atomic gases. A well-established direction in this context is to design dissipation channels that tend to push the gas towards a desired many-body state~\cite{Diehl:NatPhys2008,Bardyn:PRL2012,Budich:PRA2015,Mivehvar:AdP2021,Ferri:PRX2021,Marsh:PRX2024}. A more recent research avenue aims at investigating configurations where a net particle flux is established by letting the system exchange atoms with several suitably designed reservoirs, either in an incoherent way via inter-atomic collisions or in a coherent way via internal transitions driven by incident electromagnetic fields.
An example of such strategy is the continuously loaded Bose Einstein condensate that was generated in the driven-dissipative steady-state of a complex atom cooling apparatus  in~\cite{Chen:Nature2022}; taking inspiration from the accumulated expertise in condensation effects in fluids of light~\cite{claude2023observation,fontaine2022kardar}, a natural question is therefore to explore the consequences of the driven-dissipative nature on the coherence and on the collective modes of the condensate. At the same time, macroscopic reservoirs involving an atomic condensate have been used to realize atomic analogs of the coherent pumping schemes: first experiments have demonstrated bistability effects in single mode geometries~\cite{Labouvie:PRL2016,Benary:NJP2022}; taking inspiration from the recent experiments demonstrating baby fractional quantum Hall fluids of light~\cite{clark2020observation}, analogous work is in progress in the direction of stabilizing strongly correlated states.

\section*{Acknowledgments}

We acknowledge financial support by: Provincia Autonoma di Trento (PAT); from Q@TN, the joint lab between University of Trento, FBK-Fondazione Bruno Kessler, INFN-National Institute for Nuclear Physics and CNR-National Research Council; the National Quantum Science and Technology Institute through the PNRR MUR Project under Grant PE0000023-NQSTI, co-funded by the European Union - NextGeneration EU; the Deutsche Forschungsgemeinschaft (DFG, German Research Foundation) via the Research Unit FOR 5688 ``Driven-dissipative many-body systems of ultracold atoms'', project number 521530974.



\bibliography{biblio}

\begin{thebibliography}{137}%
\makeatletter
\providecommand \@ifxundefined [1]{%
 \@ifx{#1\undefined}
}%
\providecommand \@ifnum [1]{%
 \ifnum #1\expandafter \@firstoftwo
 \else \expandafter \@secondoftwo
 \fi
}%
\providecommand \@ifx [1]{%
 \ifx #1\expandafter \@firstoftwo
 \else \expandafter \@secondoftwo
 \fi
}%
\providecommand \natexlab [1]{#1}%
\providecommand \enquote  [1]{``#1''}%
\providecommand \bibnamefont  [1]{#1}%
\providecommand \bibfnamefont [1]{#1}%
\providecommand \citenamefont [1]{#1}%
\providecommand \href@noop [0]{\@secondoftwo}%
\providecommand \href [0]{\begingroup \@sanitize@url \@href}%
\providecommand \@href[1]{\@@startlink{#1}\@@href}%
\providecommand \@@href[1]{\endgroup#1\@@endlink}%
\providecommand \@sanitize@url [0]{\catcode `\\12\catcode `\$12\catcode
  `\&12\catcode `\#12\catcode `\^12\catcode `\_12\catcode `\%12\relax}%
\providecommand \@@startlink[1]{}%
\providecommand \@@endlink[0]{}%
\providecommand \url  [0]{\begingroup\@sanitize@url \@url }%
\providecommand \@url [1]{\endgroup\@href {#1}{\urlprefix }}%
\providecommand \urlprefix  [0]{URL }%
\providecommand \Eprint [0]{\href }%
\providecommand \doibase [0]{https://doi.org/}%
\providecommand \selectlanguage [0]{\@gobble}%
\providecommand \bibinfo  [0]{\@secondoftwo}%
\providecommand \bibfield  [0]{\@secondoftwo}%
\providecommand \translation [1]{[#1]}%
\providecommand \BibitemOpen [0]{}%
\providecommand \bibitemStop [0]{}%
\providecommand \bibitemNoStop [0]{.\EOS\space}%
\providecommand \EOS [0]{\spacefactor3000\relax}%
\providecommand \BibitemShut  [1]{\csname bibitem#1\endcsname}%
\let\auto@bib@innerbib\@empty
\bibitem [{\citenamefont {Pitaevski{\u\i}}\ and\ \citenamefont
  {Stringari}(2016)}]{pitaevskii2016bose}%
  \BibitemOpen
  \bibfield  {author} {\bibinfo {author} {\bibfnamefont {L.}~\bibnamefont
  {Pitaevski{\u\i}}}\ and\ \bibinfo {author} {\bibfnamefont {S.}~\bibnamefont
  {Stringari}},\ }\href {https://books.google.it/books?id=k\_ZGCwAAQBAJ} {\emph
  {\bibinfo {title} {Bose-Einstein Condensation and Superfluidity}}},\
  International series of monographs on physics\ (\bibinfo  {publisher} {Oxford
  University Press},\ \bibinfo {year} {2016})\BibitemShut {NoStop}%
\bibitem [{\citenamefont {Gross}\ and\ \citenamefont
  {Bloch}(2017)}]{gross2017quantum}%
  \BibitemOpen
  \bibfield  {author} {\bibinfo {author} {\bibfnamefont {C.}~\bibnamefont
  {Gross}}\ and\ \bibinfo {author} {\bibfnamefont {I.}~\bibnamefont {Bloch}},\
  }\bibfield  {title} {\bibinfo {title} {Quantum simulations with ultracold
  atoms in optical lattices},\ }\href@noop {} {\bibfield  {journal} {\bibinfo
  {journal} {Science}\ }\textbf {\bibinfo {volume} {357}},\ \bibinfo {pages}
  {995} (\bibinfo {year} {2017})}\BibitemShut {NoStop}%
\bibitem [{\citenamefont {Bloch}\ \emph {et~al.}(2008)\citenamefont {Bloch},
  \citenamefont {Dalibard},\ and\ \citenamefont {Zwerger}}]{bloch2008many}%
  \BibitemOpen
  \bibfield  {author} {\bibinfo {author} {\bibfnamefont {I.}~\bibnamefont
  {Bloch}}, \bibinfo {author} {\bibfnamefont {J.}~\bibnamefont {Dalibard}},\
  and\ \bibinfo {author} {\bibfnamefont {W.}~\bibnamefont {Zwerger}},\
  }\bibfield  {title} {\bibinfo {title} {Many-body physics with ultracold
  gases},\ }\href@noop {} {\bibfield  {journal} {\bibinfo  {journal} {Reviews
  of modern physics}\ }\textbf {\bibinfo {volume} {80}},\ \bibinfo {pages}
  {885} (\bibinfo {year} {2008})}\BibitemShut {NoStop}%
\bibitem [{\citenamefont {Cooper}\ \emph {et~al.}(2019)\citenamefont {Cooper},
  \citenamefont {Dalibard},\ and\ \citenamefont
  {Spielman}}]{cooper2019topological}%
  \BibitemOpen
  \bibfield  {author} {\bibinfo {author} {\bibfnamefont {N.}~\bibnamefont
  {Cooper}}, \bibinfo {author} {\bibfnamefont {J.}~\bibnamefont {Dalibard}},\
  and\ \bibinfo {author} {\bibfnamefont {I.}~\bibnamefont {Spielman}},\
  }\bibfield  {title} {\bibinfo {title} {Topological bands for ultracold
  atoms},\ }\href@noop {} {\bibfield  {journal} {\bibinfo  {journal} {Reviews
  of modern physics}\ }\textbf {\bibinfo {volume} {91}},\ \bibinfo {pages}
  {015005} (\bibinfo {year} {2019})}\BibitemShut {NoStop}%
\bibitem [{\citenamefont {Carusotto}\ and\ \citenamefont
  {Ciuti}(2013)}]{carusotto2013quantum}%
  \BibitemOpen
  \bibfield  {author} {\bibinfo {author} {\bibfnamefont {I.}~\bibnamefont
  {Carusotto}}\ and\ \bibinfo {author} {\bibfnamefont {C.}~\bibnamefont
  {Ciuti}},\ }\bibfield  {title} {\bibinfo {title} {Quantum fluids of light},\
  }\href@noop {} {\bibfield  {journal} {\bibinfo  {journal} {Reviews of Modern
  Physics}\ }\textbf {\bibinfo {volume} {85}},\ \bibinfo {pages} {299}
  (\bibinfo {year} {2013})}\BibitemShut {NoStop}%
\bibitem [{\citenamefont {Chang}\ \emph {et~al.}(2014)\citenamefont {Chang},
  \citenamefont {Vuleti{\'c}},\ and\ \citenamefont {Lukin}}]{chang2014quantum}%
  \BibitemOpen
  \bibfield  {author} {\bibinfo {author} {\bibfnamefont {D.~E.}\ \bibnamefont
  {Chang}}, \bibinfo {author} {\bibfnamefont {V.}~\bibnamefont {Vuleti{\'c}}},\
  and\ \bibinfo {author} {\bibfnamefont {M.~D.}\ \bibnamefont {Lukin}},\
  }\bibfield  {title} {\bibinfo {title} {Quantum nonlinear optics—photon by
  photon},\ }\href@noop {} {\bibfield  {journal} {\bibinfo  {journal} {Nature
  Photonics}\ }\textbf {\bibinfo {volume} {8}},\ \bibinfo {pages} {685}
  (\bibinfo {year} {2014})}\BibitemShut {NoStop}%
\bibitem [{\citenamefont {Carusotto}\ \emph {et~al.}(2020)\citenamefont
  {Carusotto}, \citenamefont {Houck}, \citenamefont {Koll{\'a}r}, \citenamefont
  {Roushan}, \citenamefont {Schuster},\ and\ \citenamefont
  {Simon}}]{carusotto2020photonic}%
  \BibitemOpen
  \bibfield  {author} {\bibinfo {author} {\bibfnamefont {I.}~\bibnamefont
  {Carusotto}}, \bibinfo {author} {\bibfnamefont {A.~A.}\ \bibnamefont
  {Houck}}, \bibinfo {author} {\bibfnamefont {A.~J.}\ \bibnamefont
  {Koll{\'a}r}}, \bibinfo {author} {\bibfnamefont {P.}~\bibnamefont {Roushan}},
  \bibinfo {author} {\bibfnamefont {D.~I.}\ \bibnamefont {Schuster}},\ and\
  \bibinfo {author} {\bibfnamefont {J.}~\bibnamefont {Simon}},\ }\bibfield
  {title} {\bibinfo {title} {Photonic materials in circuit quantum
  electrodynamics},\ }\href@noop {} {\bibfield  {journal} {\bibinfo  {journal}
  {Nature Physics}\ }\textbf {\bibinfo {volume} {16}},\ \bibinfo {pages} {268}
  (\bibinfo {year} {2020})}\BibitemShut {NoStop}%
\bibitem [{\citenamefont {Bloch}\ \emph {et~al.}(2022)\citenamefont {Bloch},
  \citenamefont {Carusotto},\ and\ \citenamefont {Wouters}}]{bloch2022non}%
  \BibitemOpen
  \bibfield  {author} {\bibinfo {author} {\bibfnamefont {J.}~\bibnamefont
  {Bloch}}, \bibinfo {author} {\bibfnamefont {I.}~\bibnamefont {Carusotto}},\
  and\ \bibinfo {author} {\bibfnamefont {M.}~\bibnamefont {Wouters}},\
  }\bibfield  {title} {\bibinfo {title} {Non-equilibrium bose--einstein
  condensation in photonic systems},\ }\href@noop {} {\bibfield  {journal}
  {\bibinfo  {journal} {Nature Reviews Physics}\ }\textbf {\bibinfo {volume}
  {4}},\ \bibinfo {pages} {470} (\bibinfo {year} {2022})}\BibitemShut {NoStop}%
\bibitem [{\citenamefont {Sieberer}\ \emph
  {et~al.}(2016{\natexlab{a}})\citenamefont {Sieberer}, \citenamefont
  {Buchhold},\ and\ \citenamefont {Diehl}}]{sieberer2016keldysh}%
  \BibitemOpen
  \bibfield  {author} {\bibinfo {author} {\bibfnamefont {L.~M.}\ \bibnamefont
  {Sieberer}}, \bibinfo {author} {\bibfnamefont {M.}~\bibnamefont {Buchhold}},\
  and\ \bibinfo {author} {\bibfnamefont {S.}~\bibnamefont {Diehl}},\ }\bibfield
   {title} {\bibinfo {title} {Keldysh field theory for driven open quantum
  systems},\ }\href@noop {} {\bibfield  {journal} {\bibinfo  {journal} {Reports
  on Progress in Physics}\ }\textbf {\bibinfo {volume} {79}},\ \bibinfo {pages}
  {096001} (\bibinfo {year} {2016}{\natexlab{a}})}\BibitemShut {NoStop}%
\bibitem [{\citenamefont {Diehl}\ \emph {et~al.}(2008)\citenamefont {Diehl},
  \citenamefont {Micheli}, \citenamefont {Kantian}, \citenamefont {Kraus},
  \citenamefont {B{\"u}chler},\ and\ \citenamefont
  {Zoller}}]{Diehl:NatPhys2008}%
  \BibitemOpen
  \bibfield  {author} {\bibinfo {author} {\bibfnamefont {S.}~\bibnamefont
  {Diehl}}, \bibinfo {author} {\bibfnamefont {A.}~\bibnamefont {Micheli}},
  \bibinfo {author} {\bibfnamefont {A.}~\bibnamefont {Kantian}}, \bibinfo
  {author} {\bibfnamefont {B.}~\bibnamefont {Kraus}}, \bibinfo {author}
  {\bibfnamefont {H.}~\bibnamefont {B{\"u}chler}},\ and\ \bibinfo {author}
  {\bibfnamefont {P.}~\bibnamefont {Zoller}},\ }\bibfield  {title} {\bibinfo
  {title} {Quantum states and phases in driven open quantum systems with cold
  atoms},\ }\href@noop {} {\bibfield  {journal} {\bibinfo  {journal} {Nature
  Physics}\ }\textbf {\bibinfo {volume} {4}},\ \bibinfo {pages} {878} (\bibinfo
  {year} {2008})}\BibitemShut {NoStop}%
\bibitem [{\citenamefont {Bardyn}\ \emph {et~al.}(2012)\citenamefont {Bardyn},
  \citenamefont {Baranov}, \citenamefont {Rico}, \citenamefont {\ifmmode
  \dot{I}\else \.{I}\fi{}mamo\ifmmode~\breve{g}\else \u{g}\fi{}lu},
  \citenamefont {Zoller},\ and\ \citenamefont {Diehl}}]{Bardyn:PRL2012}%
  \BibitemOpen
  \bibfield  {author} {\bibinfo {author} {\bibfnamefont {C.-E.}\ \bibnamefont
  {Bardyn}}, \bibinfo {author} {\bibfnamefont {M.~A.}\ \bibnamefont {Baranov}},
  \bibinfo {author} {\bibfnamefont {E.}~\bibnamefont {Rico}}, \bibinfo {author}
  {\bibfnamefont {A.}~\bibnamefont {\ifmmode \dot{I}\else
  \.{I}\fi{}mamo\ifmmode~\breve{g}\else \u{g}\fi{}lu}}, \bibinfo {author}
  {\bibfnamefont {P.}~\bibnamefont {Zoller}},\ and\ \bibinfo {author}
  {\bibfnamefont {S.}~\bibnamefont {Diehl}},\ }\bibfield  {title} {\bibinfo
  {title} {Majorana modes in driven-dissipative atomic superfluids with a zero
  chern number},\ }\href {https://doi.org/10.1103/PhysRevLett.109.130402}
  {\bibfield  {journal} {\bibinfo  {journal} {Phys. Rev. Lett.}\ }\textbf
  {\bibinfo {volume} {109}},\ \bibinfo {pages} {130402} (\bibinfo {year}
  {2012})}\BibitemShut {NoStop}%
\bibitem [{\citenamefont {Budich}\ \emph {et~al.}(2015)\citenamefont {Budich},
  \citenamefont {Zoller},\ and\ \citenamefont {Diehl}}]{Budich:PRA2015}%
  \BibitemOpen
  \bibfield  {author} {\bibinfo {author} {\bibfnamefont {J.~C.}\ \bibnamefont
  {Budich}}, \bibinfo {author} {\bibfnamefont {P.}~\bibnamefont {Zoller}},\
  and\ \bibinfo {author} {\bibfnamefont {S.}~\bibnamefont {Diehl}},\ }\bibfield
   {title} {\bibinfo {title} {Dissipative preparation of chern insulators},\
  }\href {https://doi.org/10.1103/PhysRevA.91.042117} {\bibfield  {journal}
  {\bibinfo  {journal} {Phys. Rev. A}\ }\textbf {\bibinfo {volume} {91}},\
  \bibinfo {pages} {042117} (\bibinfo {year} {2015})}\BibitemShut {NoStop}%
\bibitem [{\citenamefont {Mivehvar}\ \emph {et~al.}(2021)\citenamefont
  {Mivehvar}, \citenamefont {Piazza}, \citenamefont {Donner},\ and\
  \citenamefont {Ritsch}}]{Mivehvar:AdP2021}%
  \BibitemOpen
  \bibfield  {author} {\bibinfo {author} {\bibfnamefont {F.}~\bibnamefont
  {Mivehvar}}, \bibinfo {author} {\bibfnamefont {F.}~\bibnamefont {Piazza}},
  \bibinfo {author} {\bibfnamefont {T.}~\bibnamefont {Donner}},\ and\ \bibinfo
  {author} {\bibfnamefont {H.}~\bibnamefont {Ritsch}},\ }\bibfield  {title}
  {\bibinfo {title} {Cavity qed with quantum gases: new paradigms in many-body
  physics},\ }\href@noop {} {\bibfield  {journal} {\bibinfo  {journal}
  {Advances in Physics}\ }\textbf {\bibinfo {volume} {70}},\ \bibinfo {pages}
  {1} (\bibinfo {year} {2021})}\BibitemShut {NoStop}%
\bibitem [{\citenamefont {Ferri}\ \emph {et~al.}(2021)\citenamefont {Ferri},
  \citenamefont {Rosa-Medina}, \citenamefont {Finger}, \citenamefont {Dogra},
  \citenamefont {Soriente}, \citenamefont {Zilberberg}, \citenamefont
  {Donner},\ and\ \citenamefont {Esslinger}}]{Ferri:PRX2021}%
  \BibitemOpen
  \bibfield  {author} {\bibinfo {author} {\bibfnamefont {F.}~\bibnamefont
  {Ferri}}, \bibinfo {author} {\bibfnamefont {R.}~\bibnamefont {Rosa-Medina}},
  \bibinfo {author} {\bibfnamefont {F.}~\bibnamefont {Finger}}, \bibinfo
  {author} {\bibfnamefont {N.}~\bibnamefont {Dogra}}, \bibinfo {author}
  {\bibfnamefont {M.}~\bibnamefont {Soriente}}, \bibinfo {author}
  {\bibfnamefont {O.}~\bibnamefont {Zilberberg}}, \bibinfo {author}
  {\bibfnamefont {T.}~\bibnamefont {Donner}},\ and\ \bibinfo {author}
  {\bibfnamefont {T.}~\bibnamefont {Esslinger}},\ }\bibfield  {title} {\bibinfo
  {title} {Emerging dissipative phases in a superradiant quantum gas with
  tunable decay},\ }\href {https://doi.org/10.1103/PhysRevX.11.041046}
  {\bibfield  {journal} {\bibinfo  {journal} {Phys. Rev. X}\ }\textbf {\bibinfo
  {volume} {11}},\ \bibinfo {pages} {041046} (\bibinfo {year}
  {2021})}\BibitemShut {NoStop}%
\bibitem [{\citenamefont {Marsh}\ \emph {et~al.}(2024)\citenamefont {Marsh},
  \citenamefont {Kroeze}, \citenamefont {Ganguli}, \citenamefont
  {Gopalakrishnan}, \citenamefont {Keeling},\ and\ \citenamefont
  {Lev}}]{Marsh:PRX2024}%
  \BibitemOpen
  \bibfield  {author} {\bibinfo {author} {\bibfnamefont {B.~P.}\ \bibnamefont
  {Marsh}}, \bibinfo {author} {\bibfnamefont {R.~M.}\ \bibnamefont {Kroeze}},
  \bibinfo {author} {\bibfnamefont {S.}~\bibnamefont {Ganguli}}, \bibinfo
  {author} {\bibfnamefont {S.}~\bibnamefont {Gopalakrishnan}}, \bibinfo
  {author} {\bibfnamefont {J.}~\bibnamefont {Keeling}},\ and\ \bibinfo {author}
  {\bibfnamefont {B.~L.}\ \bibnamefont {Lev}},\ }\bibfield  {title} {\bibinfo
  {title} {Entanglement and replica symmetry breaking in a driven-dissipative
  quantum spin glass},\ }\href@noop {} {\bibfield  {journal} {\bibinfo
  {journal} {Physical Review X}\ }\textbf {\bibinfo {volume} {14}},\ \bibinfo
  {pages} {011026} (\bibinfo {year} {2024})}\BibitemShut {NoStop}%
\bibitem [{\citenamefont {Chen}\ \emph {et~al.}(2022)\citenamefont {Chen},
  \citenamefont {Gonz{\'a}lez~Escudero}, \citenamefont {Min{\'a}{\v{r}}},
  \citenamefont {Pasquiou}, \citenamefont {Bennetts},\ and\ \citenamefont
  {Schreck}}]{Chen:Nature2022}%
  \BibitemOpen
  \bibfield  {author} {\bibinfo {author} {\bibfnamefont {C.-C.}\ \bibnamefont
  {Chen}}, \bibinfo {author} {\bibfnamefont {R.}~\bibnamefont
  {Gonz{\'a}lez~Escudero}}, \bibinfo {author} {\bibfnamefont {J.}~\bibnamefont
  {Min{\'a}{\v{r}}}}, \bibinfo {author} {\bibfnamefont {B.}~\bibnamefont
  {Pasquiou}}, \bibinfo {author} {\bibfnamefont {S.}~\bibnamefont {Bennetts}},\
  and\ \bibinfo {author} {\bibfnamefont {F.}~\bibnamefont {Schreck}},\
  }\bibfield  {title} {\bibinfo {title} {Continuous bose--einstein
  condensation},\ }\href@noop {} {\bibfield  {journal} {\bibinfo  {journal}
  {Nature}\ }\textbf {\bibinfo {volume} {606}},\ \bibinfo {pages} {683}
  (\bibinfo {year} {2022})}\BibitemShut {NoStop}%
\bibitem [{\citenamefont {Labouvie}\ \emph {et~al.}(2016)\citenamefont
  {Labouvie}, \citenamefont {Santra}, \citenamefont {Heun},\ and\ \citenamefont
  {Ott}}]{Labouvie:PRL2016}%
  \BibitemOpen
  \bibfield  {author} {\bibinfo {author} {\bibfnamefont {R.}~\bibnamefont
  {Labouvie}}, \bibinfo {author} {\bibfnamefont {B.}~\bibnamefont {Santra}},
  \bibinfo {author} {\bibfnamefont {S.}~\bibnamefont {Heun}},\ and\ \bibinfo
  {author} {\bibfnamefont {H.}~\bibnamefont {Ott}},\ }\bibfield  {title}
  {\bibinfo {title} {Bistability in a driven-dissipative superfluid},\ }\href
  {https://doi.org/10.1103/PhysRevLett.116.235302} {\bibfield  {journal}
  {\bibinfo  {journal} {Phys. Rev. Lett.}\ }\textbf {\bibinfo {volume} {116}},\
  \bibinfo {pages} {235302} (\bibinfo {year} {2016})}\BibitemShut {NoStop}%
\bibitem [{\citenamefont {Benary}\ \emph {et~al.}(2022)\citenamefont {Benary},
  \citenamefont {Baals}, \citenamefont {Bernhart}, \citenamefont {Jiang},
  \citenamefont {R{\"o}hrle},\ and\ \citenamefont {Ott}}]{Benary:NJP2022}%
  \BibitemOpen
  \bibfield  {author} {\bibinfo {author} {\bibfnamefont {J.}~\bibnamefont
  {Benary}}, \bibinfo {author} {\bibfnamefont {C.}~\bibnamefont {Baals}},
  \bibinfo {author} {\bibfnamefont {E.}~\bibnamefont {Bernhart}}, \bibinfo
  {author} {\bibfnamefont {J.}~\bibnamefont {Jiang}}, \bibinfo {author}
  {\bibfnamefont {M.}~\bibnamefont {R{\"o}hrle}},\ and\ \bibinfo {author}
  {\bibfnamefont {H.}~\bibnamefont {Ott}},\ }\bibfield  {title} {\bibinfo
  {title} {Experimental observation of a dissipative phase transition in a
  multi-mode many-body quantum system},\ }\href@noop {} {\bibfield  {journal}
  {\bibinfo  {journal} {New Journal of Physics}\ }\textbf {\bibinfo {volume}
  {24}},\ \bibinfo {pages} {103034} (\bibinfo {year} {2022})}\BibitemShut
  {NoStop}%
\bibitem [{\citenamefont {Kavokin}\ \emph {et~al.}(2011)\citenamefont
  {Kavokin}, \citenamefont {Baumberg}, \citenamefont {Malpuech},\ and\
  \citenamefont {Laussy}}]{kavokin2011microcavities}%
  \BibitemOpen
  \bibfield  {author} {\bibinfo {author} {\bibfnamefont {A.}~\bibnamefont
  {Kavokin}}, \bibinfo {author} {\bibfnamefont {J.}~\bibnamefont {Baumberg}},
  \bibinfo {author} {\bibfnamefont {G.}~\bibnamefont {Malpuech}},\ and\
  \bibinfo {author} {\bibfnamefont {F.}~\bibnamefont {Laussy}},\ }\href
  {https://books.google.it/books?id=2g7wHcMcaJ0C} {\emph {\bibinfo {title}
  {Microcavities}}},\ Oxford science publications\ (\bibinfo  {publisher} {OUP
  Oxford},\ \bibinfo {year} {2011})\BibitemShut {NoStop}%
\bibitem [{\citenamefont {Jackson}(1999)}]{jackson1999classical}%
  \BibitemOpen
  \bibfield  {author} {\bibinfo {author} {\bibfnamefont {J.~D.}\ \bibnamefont
  {Jackson}},\ }\href@noop {} {\bibinfo {title} {Classical electrodynamics}}
  (\bibinfo {year} {1999})\BibitemShut {NoStop}%
\bibitem [{\citenamefont {Heisenberg}\ and\ \citenamefont
  {Euler}(1936)}]{heisenberg1936folgerungen}%
  \BibitemOpen
  \bibfield  {author} {\bibinfo {author} {\bibfnamefont {W.}~\bibnamefont
  {Heisenberg}}\ and\ \bibinfo {author} {\bibfnamefont {H.}~\bibnamefont
  {Euler}},\ }\bibfield  {title} {\bibinfo {title} {Folgerungen aus der
  diracschen theorie des positrons},\ }\href@noop {} {\bibfield  {journal}
  {\bibinfo  {journal} {Zeitschrift f{\"u}r Physik}\ }\textbf {\bibinfo
  {volume} {98}},\ \bibinfo {pages} {714} (\bibinfo {year} {1936})}\BibitemShut
  {NoStop}%
\bibitem [{\citenamefont {ATLAS~Collaboration}(2017)}]{atlas2017evidence}%
  \BibitemOpen
  \bibfield  {author} {\bibinfo {author} {\bibfnamefont {.~.}\ \bibnamefont
  {ATLAS~Collaboration}},\ }\bibfield  {title} {\bibinfo {title} {Evidence for
  light-by-light scattering in heavy-ion collisions with the atlas detector at
  the lhc},\ }\href@noop {} {\bibfield  {journal} {\bibinfo  {journal} {Nature
  physics}\ }\textbf {\bibinfo {volume} {13}},\ \bibinfo {pages} {852}
  (\bibinfo {year} {2017})}\BibitemShut {NoStop}%
\bibitem [{\citenamefont {Ashcroft}\ and\ \citenamefont
  {Mermin}(2022)}]{ashcroft2022solid}%
  \BibitemOpen
  \bibfield  {author} {\bibinfo {author} {\bibfnamefont {N.~W.}\ \bibnamefont
  {Ashcroft}}\ and\ \bibinfo {author} {\bibfnamefont {N.~D.}\ \bibnamefont
  {Mermin}},\ }\href@noop {} {\emph {\bibinfo {title} {Solid state physics}}}\
  (\bibinfo  {publisher} {Cengage Learning},\ \bibinfo {year}
  {2022})\BibitemShut {NoStop}%
\bibitem [{\citenamefont {Butcher}\ and\ \citenamefont
  {Cotter}(2008)}]{Butcher}%
  \BibitemOpen
  \bibfield  {author} {\bibinfo {author} {\bibfnamefont {P.~N.}\ \bibnamefont
  {Butcher}}\ and\ \bibinfo {author} {\bibfnamefont {D.}~\bibnamefont
  {Cotter}},\ }\href@noop {} {\emph {\bibinfo {title} {The elements of
  nonlinear optics}}},\ Cambridge Studies in Modern Optics\ (\bibinfo
  {publisher} {Cambridge University Press},\ \bibinfo {year}
  {2008})\BibitemShut {NoStop}%
\bibitem [{\citenamefont {Boyd}(2008)}]{Boyd}%
  \BibitemOpen
  \bibfield  {author} {\bibinfo {author} {\bibfnamefont {R.~W.}\ \bibnamefont
  {Boyd}},\ }\href@noop {} {\emph {\bibinfo {title} {Nonlinear Optics}}}\
  (\bibinfo  {publisher} {Academic Press},\ \bibinfo {year} {2008})\BibitemShut
  {NoStop}%
\bibitem [{\citenamefont {Blais}\ \emph {et~al.}(2021)\citenamefont {Blais},
  \citenamefont {Grimsmo}, \citenamefont {Girvin},\ and\ \citenamefont
  {Wallraff}}]{blais_RMP2021}%
  \BibitemOpen
  \bibfield  {author} {\bibinfo {author} {\bibfnamefont {A.}~\bibnamefont
  {Blais}}, \bibinfo {author} {\bibfnamefont {A.~L.}\ \bibnamefont {Grimsmo}},
  \bibinfo {author} {\bibfnamefont {S.~M.}\ \bibnamefont {Girvin}},\ and\
  \bibinfo {author} {\bibfnamefont {A.}~\bibnamefont {Wallraff}},\ }\bibfield
  {title} {\bibinfo {title} {Circuit quantum electrodynamics},\ }\href
  {https://doi.org/10.1103/RevModPhys.93.025005} {\bibfield  {journal}
  {\bibinfo  {journal} {Rev. Mod. Phys.}\ }\textbf {\bibinfo {volume} {93}},\
  \bibinfo {pages} {025005} (\bibinfo {year} {2021})}\BibitemShut {NoStop}%
\bibitem [{\citenamefont {Peyronel}\ \emph {et~al.}(2012)\citenamefont
  {Peyronel}, \citenamefont {Firstenberg}, \citenamefont {Liang}, \citenamefont
  {Hofferberth}, \citenamefont {Gorshkov}, \citenamefont {Pohl}, \citenamefont
  {Lukin},\ and\ \citenamefont {Vuleti{\'c}}}]{peyronel2012quantum}%
  \BibitemOpen
  \bibfield  {author} {\bibinfo {author} {\bibfnamefont {T.}~\bibnamefont
  {Peyronel}}, \bibinfo {author} {\bibfnamefont {O.}~\bibnamefont
  {Firstenberg}}, \bibinfo {author} {\bibfnamefont {Q.-Y.}\ \bibnamefont
  {Liang}}, \bibinfo {author} {\bibfnamefont {S.}~\bibnamefont {Hofferberth}},
  \bibinfo {author} {\bibfnamefont {A.~V.}\ \bibnamefont {Gorshkov}}, \bibinfo
  {author} {\bibfnamefont {T.}~\bibnamefont {Pohl}}, \bibinfo {author}
  {\bibfnamefont {M.~D.}\ \bibnamefont {Lukin}},\ and\ \bibinfo {author}
  {\bibfnamefont {V.}~\bibnamefont {Vuleti{\'c}}},\ }\bibfield  {title}
  {\bibinfo {title} {Quantum nonlinear optics with single photons enabled by
  strongly interacting atoms},\ }\href@noop {} {\bibfield  {journal} {\bibinfo
  {journal} {Nature}\ }\textbf {\bibinfo {volume} {488}},\ \bibinfo {pages}
  {57} (\bibinfo {year} {2012})}\BibitemShut {NoStop}%
\bibitem [{\citenamefont {Firstenberg}\ \emph {et~al.}(2013)\citenamefont
  {Firstenberg}, \citenamefont {Peyronel}, \citenamefont {Liang}, \citenamefont
  {Gorshkov}, \citenamefont {Lukin},\ and\ \citenamefont
  {Vuleti{\'c}}}]{firstenberg2013attractive}%
  \BibitemOpen
  \bibfield  {author} {\bibinfo {author} {\bibfnamefont {O.}~\bibnamefont
  {Firstenberg}}, \bibinfo {author} {\bibfnamefont {T.}~\bibnamefont
  {Peyronel}}, \bibinfo {author} {\bibfnamefont {Q.-Y.}\ \bibnamefont {Liang}},
  \bibinfo {author} {\bibfnamefont {A.~V.}\ \bibnamefont {Gorshkov}}, \bibinfo
  {author} {\bibfnamefont {M.~D.}\ \bibnamefont {Lukin}},\ and\ \bibinfo
  {author} {\bibfnamefont {V.}~\bibnamefont {Vuleti{\'c}}},\ }\bibfield
  {title} {\bibinfo {title} {Attractive photons in a quantum nonlinear
  medium},\ }\href@noop {} {\bibfield  {journal} {\bibinfo  {journal} {Nature}\
  }\textbf {\bibinfo {volume} {502}},\ \bibinfo {pages} {71} (\bibinfo {year}
  {2013})}\BibitemShut {NoStop}%
\bibitem [{Note1()}]{Note1}%
  \BibitemOpen
  \bibinfo {note} {We have assumed a spatio-temporally local nonlinearity, so
  the total nonlinear polarization is \protect \mbox {$P_{\protect \rm
  nl}^{\protect \rm (tot)}(\protect \mathbf {r},t)=\chi ^{(3)}\protect
  \,E^3(\protect \mathbf {r},t)$~.} The uniform-space mode normalization is
  \protect \mbox {$ \protect \mathcal {E}=({2\pi \protect \,\hbar \omega
  }/{n^2})^{1/2}$~.} The amplitude of the coherent state is normalized as
  \protect \mbox {$\langle \protect \hat {a}_\protect \mathbf {k}\rangle =(2\pi
  )^3\protect \,\delta (\protect \mathbf {k}-\protect \mathbf {k}')\protect
  \,\protect \bar {\alpha }$~.}}\BibitemShut {Stop}%
\bibitem [{\citenamefont {Ferretti}\ and\ \citenamefont
  {Gerace}(2012)}]{Ferretti:PRB2012}%
  \BibitemOpen
  \bibfield  {author} {\bibinfo {author} {\bibfnamefont {S.}~\bibnamefont
  {Ferretti}}\ and\ \bibinfo {author} {\bibfnamefont {D.}~\bibnamefont
  {Gerace}},\ }\bibfield  {title} {\bibinfo {title} {Single-photon nonlinear
  optics with kerr-type nanostructured materials},\ }\href
  {https://doi.org/10.1103/PhysRevB.85.033303} {\bibfield  {journal} {\bibinfo
  {journal} {Phys. Rev. B}\ }\textbf {\bibinfo {volume} {85}},\ \bibinfo
  {pages} {033303} (\bibinfo {year} {2012})}\BibitemShut {NoStop}%
\bibitem [{\citenamefont {Nelsen}\ \emph {et~al.}(2013)\citenamefont {Nelsen},
  \citenamefont {Liu}, \citenamefont {Steger}, \citenamefont {Snoke},
  \citenamefont {Balili}, \citenamefont {West},\ and\ \citenamefont
  {Pfeiffer}}]{Nelsen:PRX2013}%
  \BibitemOpen
  \bibfield  {author} {\bibinfo {author} {\bibfnamefont {B.}~\bibnamefont
  {Nelsen}}, \bibinfo {author} {\bibfnamefont {G.}~\bibnamefont {Liu}},
  \bibinfo {author} {\bibfnamefont {M.}~\bibnamefont {Steger}}, \bibinfo
  {author} {\bibfnamefont {D.~W.}\ \bibnamefont {Snoke}}, \bibinfo {author}
  {\bibfnamefont {R.}~\bibnamefont {Balili}}, \bibinfo {author} {\bibfnamefont
  {K.}~\bibnamefont {West}},\ and\ \bibinfo {author} {\bibfnamefont
  {L.}~\bibnamefont {Pfeiffer}},\ }\bibfield  {title} {\bibinfo {title}
  {Dissipationless flow and sharp threshold of a polariton condensate with long
  lifetime},\ }\href@noop {} {\bibfield  {journal} {\bibinfo  {journal}
  {Physical Review X}\ }\textbf {\bibinfo {volume} {3}},\ \bibinfo {pages}
  {041015} (\bibinfo {year} {2013})}\BibitemShut {NoStop}%
\bibitem [{\citenamefont {Sun}\ \emph {et~al.}(2017)\citenamefont {Sun},
  \citenamefont {Wen}, \citenamefont {Yoon}, \citenamefont {Liu}, \citenamefont
  {Steger}, \citenamefont {Pfeiffer}, \citenamefont {West}, \citenamefont
  {Snoke},\ and\ \citenamefont {Nelson}}]{Sun:PRL2017}%
  \BibitemOpen
  \bibfield  {author} {\bibinfo {author} {\bibfnamefont {Y.}~\bibnamefont
  {Sun}}, \bibinfo {author} {\bibfnamefont {P.}~\bibnamefont {Wen}}, \bibinfo
  {author} {\bibfnamefont {Y.}~\bibnamefont {Yoon}}, \bibinfo {author}
  {\bibfnamefont {G.}~\bibnamefont {Liu}}, \bibinfo {author} {\bibfnamefont
  {M.}~\bibnamefont {Steger}}, \bibinfo {author} {\bibfnamefont {L.~N.}\
  \bibnamefont {Pfeiffer}}, \bibinfo {author} {\bibfnamefont {K.}~\bibnamefont
  {West}}, \bibinfo {author} {\bibfnamefont {D.~W.}\ \bibnamefont {Snoke}},\
  and\ \bibinfo {author} {\bibfnamefont {K.~A.}\ \bibnamefont {Nelson}},\
  }\bibfield  {title} {\bibinfo {title} {Bose-einstein condensation of
  long-lifetime polaritons in thermal equilibrium},\ }\href@noop {} {\bibfield
  {journal} {\bibinfo  {journal} {Physical review letters}\ }\textbf {\bibinfo
  {volume} {118}},\ \bibinfo {pages} {016602} (\bibinfo {year}
  {2017})}\BibitemShut {NoStop}%
\bibitem [{\citenamefont {Wang}\ \emph {et~al.}(2024)\citenamefont {Wang},
  \citenamefont {Liu}, \citenamefont {Chen}, \citenamefont {Chen},
  \citenamefont {Zhao}, \citenamefont {Ying}, \citenamefont {Shang},
  \citenamefont {Wang}, \citenamefont {Huo}, \citenamefont {Peng} \emph
  {et~al.}}]{Wang:Science2024}%
  \BibitemOpen
  \bibfield  {author} {\bibinfo {author} {\bibfnamefont {C.}~\bibnamefont
  {Wang}}, \bibinfo {author} {\bibfnamefont {F.-M.}\ \bibnamefont {Liu}},
  \bibinfo {author} {\bibfnamefont {M.-C.}\ \bibnamefont {Chen}}, \bibinfo
  {author} {\bibfnamefont {H.}~\bibnamefont {Chen}}, \bibinfo {author}
  {\bibfnamefont {X.-H.}\ \bibnamefont {Zhao}}, \bibinfo {author}
  {\bibfnamefont {C.}~\bibnamefont {Ying}}, \bibinfo {author} {\bibfnamefont
  {Z.-X.}\ \bibnamefont {Shang}}, \bibinfo {author} {\bibfnamefont {J.-W.}\
  \bibnamefont {Wang}}, \bibinfo {author} {\bibfnamefont {Y.-H.}\ \bibnamefont
  {Huo}}, \bibinfo {author} {\bibfnamefont {C.-Z.}\ \bibnamefont {Peng}}, \emph
  {et~al.},\ }\bibfield  {title} {\bibinfo {title} {Realization of fractional
  quantum hall state with interacting photons},\ }\href@noop {} {\bibfield
  {journal} {\bibinfo  {journal} {Science}\ }\textbf {\bibinfo {volume}
  {384}},\ \bibinfo {pages} {579} (\bibinfo {year} {2024})}\BibitemShut
  {NoStop}%
\bibitem [{\citenamefont {Walls}\ and\ \citenamefont
  {Milburn}(2006)}]{QuantumOptics}%
  \BibitemOpen
  \bibfield  {author} {\bibinfo {author} {\bibfnamefont {D.~F.}\ \bibnamefont
  {Walls}}\ and\ \bibinfo {author} {\bibfnamefont {G.}~\bibnamefont
  {Milburn}},\ }\href@noop {} {\emph {\bibinfo {title} {Quantum Optics}}}\
  (\bibinfo  {publisher} {Springer Verlag, Berlin},\ \bibinfo {year}
  {2006})\BibitemShut {NoStop}%
\bibitem [{\citenamefont {Ciuti}\ \emph {et~al.}(2005)\citenamefont {Ciuti},
  \citenamefont {Bastard},\ and\ \citenamefont {Carusotto}}]{Ciuti:PRB2005}%
  \BibitemOpen
  \bibfield  {author} {\bibinfo {author} {\bibfnamefont {C.}~\bibnamefont
  {Ciuti}}, \bibinfo {author} {\bibfnamefont {G.}~\bibnamefont {Bastard}},\
  and\ \bibinfo {author} {\bibfnamefont {I.}~\bibnamefont {Carusotto}},\
  }\bibfield  {title} {\bibinfo {title} {Quantum vacuum properties of the
  intersubband cavity polariton field},\ }\href
  {https://doi.org/10.1103/PhysRevB.72.115303} {\bibfield  {journal} {\bibinfo
  {journal} {Phys. Rev. B}\ }\textbf {\bibinfo {volume} {72}},\ \bibinfo
  {pages} {115303} (\bibinfo {year} {2005})}\BibitemShut {NoStop}%
\bibitem [{\citenamefont {Ciuti}\ and\ \citenamefont
  {Carusotto}(2006)}]{Ciuti:2006PRA}%
  \BibitemOpen
  \bibfield  {author} {\bibinfo {author} {\bibfnamefont {C.}~\bibnamefont
  {Ciuti}}\ and\ \bibinfo {author} {\bibfnamefont {I.}~\bibnamefont
  {Carusotto}},\ }\bibfield  {title} {\bibinfo {title} {Input-output theory of
  cavities in the ultrastrong coupling regime: The case of time-independent
  cavity parameters},\ }\href {https://doi.org/10.1103/PhysRevA.74.033811}
  {\bibfield  {journal} {\bibinfo  {journal} {Phys. Rev. A}\ }\textbf {\bibinfo
  {volume} {74}},\ \bibinfo {pages} {033811} (\bibinfo {year}
  {2006})}\BibitemShut {NoStop}%
\bibitem [{\citenamefont {Frisk~Kockum}\ \emph {et~al.}(2019)\citenamefont
  {Frisk~Kockum}, \citenamefont {Miranowicz}, \citenamefont {De~Liberato},
  \citenamefont {Savasta},\ and\ \citenamefont {Nori}}]{frisk2019ultrastrong}%
  \BibitemOpen
  \bibfield  {author} {\bibinfo {author} {\bibfnamefont {A.}~\bibnamefont
  {Frisk~Kockum}}, \bibinfo {author} {\bibfnamefont {A.}~\bibnamefont
  {Miranowicz}}, \bibinfo {author} {\bibfnamefont {S.}~\bibnamefont
  {De~Liberato}}, \bibinfo {author} {\bibfnamefont {S.}~\bibnamefont
  {Savasta}},\ and\ \bibinfo {author} {\bibfnamefont {F.}~\bibnamefont
  {Nori}},\ }\bibfield  {title} {\bibinfo {title} {Ultrastrong coupling between
  light and matter},\ }\href@noop {} {\bibfield  {journal} {\bibinfo  {journal}
  {Nature Reviews Physics}\ }\textbf {\bibinfo {volume} {1}},\ \bibinfo {pages}
  {19} (\bibinfo {year} {2019})}\BibitemShut {NoStop}%
\bibitem [{\citenamefont {Gardiner}\ and\ \citenamefont
  {Zoller}(2004)}]{QuantumNoise}%
  \BibitemOpen
  \bibfield  {author} {\bibinfo {author} {\bibfnamefont {C.~W.}\ \bibnamefont
  {Gardiner}}\ and\ \bibinfo {author} {\bibfnamefont {P.}~\bibnamefont
  {Zoller}},\ }\href@noop {} {\emph {\bibinfo {title} {Quantum Noise}}}\
  (\bibinfo  {publisher} {Springer Verlag, Berlin},\ \bibinfo {year}
  {2004})\BibitemShut {NoStop}%
\bibitem [{\citenamefont {Lamb}(1964)}]{Lamb:PR1964}%
  \BibitemOpen
  \bibfield  {author} {\bibinfo {author} {\bibfnamefont {W.~E.}\ \bibnamefont
  {Lamb}},\ }\bibfield  {title} {\bibinfo {title} {Theory of an optical
  maser},\ }\href {https://doi.org/10.1103/PhysRev.134.A1429} {\bibfield
  {journal} {\bibinfo  {journal} {Phys. Rev.}\ }\textbf {\bibinfo {volume}
  {134}},\ \bibinfo {pages} {A1429} (\bibinfo {year} {1964})}\BibitemShut
  {NoStop}%
\bibitem [{\citenamefont {Scully}\ and\ \citenamefont
  {Zubairy}(1997)}]{scully1997quantum}%
  \BibitemOpen
  \bibfield  {author} {\bibinfo {author} {\bibfnamefont {M.~O.}\ \bibnamefont
  {Scully}}\ and\ \bibinfo {author} {\bibfnamefont {M.~S.}\ \bibnamefont
  {Zubairy}},\ }\href@noop {} {\emph {\bibinfo {title} {Quantum optics}}}\
  (\bibinfo  {publisher} {Cambridge university press},\ \bibinfo {year}
  {1997})\BibitemShut {NoStop}%
\bibitem [{\citenamefont {Lebreuilly}\ \emph {et~al.}(2016)\citenamefont
  {Lebreuilly}, \citenamefont {Wouters},\ and\ \citenamefont
  {Carusotto}}]{Lebreuilly:CRAS2016}%
  \BibitemOpen
  \bibfield  {author} {\bibinfo {author} {\bibfnamefont {J.}~\bibnamefont
  {Lebreuilly}}, \bibinfo {author} {\bibfnamefont {M.}~\bibnamefont
  {Wouters}},\ and\ \bibinfo {author} {\bibfnamefont {I.}~\bibnamefont
  {Carusotto}},\ }\bibfield  {title} {\bibinfo {title} {Towards strongly
  correlated photons in arrays of dissipative nonlinear cavities under a
  frequency-dependent incoherent pumping},\ }\href@noop {} {\bibfield
  {journal} {\bibinfo  {journal} {Comptes Rendus Physique}\ }\textbf {\bibinfo
  {volume} {17}},\ \bibinfo {pages} {836} (\bibinfo {year} {2016})}\BibitemShut
  {NoStop}%
\bibitem [{\citenamefont {Breuer}\ and\ \citenamefont
  {Petruccione}(2002)}]{breuer2002theory}%
  \BibitemOpen
  \bibfield  {author} {\bibinfo {author} {\bibfnamefont {H.-P.}\ \bibnamefont
  {Breuer}}\ and\ \bibinfo {author} {\bibfnamefont {F.}~\bibnamefont
  {Petruccione}},\ }\href@noop {} {\emph {\bibinfo {title} {The theory of open
  quantum systems}}}\ (\bibinfo  {publisher} {OUP Oxford},\ \bibinfo {year}
  {2002})\BibitemShut {NoStop}%
\bibitem [{\citenamefont {Becker}\ \emph {et~al.}(2021)\citenamefont {Becker},
  \citenamefont {Wu},\ and\ \citenamefont {Eckardt}}]{Becker:PRE2021}%
  \BibitemOpen
  \bibfield  {author} {\bibinfo {author} {\bibfnamefont {T.}~\bibnamefont
  {Becker}}, \bibinfo {author} {\bibfnamefont {L.-N.}\ \bibnamefont {Wu}},\
  and\ \bibinfo {author} {\bibfnamefont {A.}~\bibnamefont {Eckardt}},\
  }\bibfield  {title} {\bibinfo {title} {Lindbladian approximation beyond
  ultraweak coupling},\ }\href {https://doi.org/10.1103/PhysRevE.104.014110}
  {\bibfield  {journal} {\bibinfo  {journal} {Phys. Rev. E}\ }\textbf {\bibinfo
  {volume} {104}},\ \bibinfo {pages} {014110} (\bibinfo {year}
  {2021})}\BibitemShut {NoStop}%
\bibitem [{\citenamefont {Tello~Breuer}\ \emph {et~al.}(2024)\citenamefont
  {Tello~Breuer}, \citenamefont {Becker},\ and\ \citenamefont
  {Eckardt}}]{Tello:PRB2024}%
  \BibitemOpen
  \bibfield  {author} {\bibinfo {author} {\bibfnamefont {C.~S.}\ \bibnamefont
  {Tello~Breuer}}, \bibinfo {author} {\bibfnamefont {T.}~\bibnamefont
  {Becker}},\ and\ \bibinfo {author} {\bibfnamefont {A.}~\bibnamefont
  {Eckardt}},\ }\bibfield  {title} {\bibinfo {title} {Benchmarking quantum
  master equations beyond ultraweak coupling},\ }\href
  {https://doi.org/10.1103/PhysRevB.110.064319} {\bibfield  {journal} {\bibinfo
   {journal} {Phys. Rev. B}\ }\textbf {\bibinfo {volume} {110}},\ \bibinfo
  {pages} {064319} (\bibinfo {year} {2024})}\BibitemShut {NoStop}%
\bibitem [{\citenamefont {Caleffi}\ \emph {et~al.}(2023)\citenamefont
  {Caleffi}, \citenamefont {Capone},\ and\ \citenamefont
  {Carusotto}}]{Caleffi:PRL2023}%
  \BibitemOpen
  \bibfield  {author} {\bibinfo {author} {\bibfnamefont {F.}~\bibnamefont
  {Caleffi}}, \bibinfo {author} {\bibfnamefont {M.}~\bibnamefont {Capone}},\
  and\ \bibinfo {author} {\bibfnamefont {I.}~\bibnamefont {Carusotto}},\
  }\bibfield  {title} {\bibinfo {title} {Collective excitations of a strongly
  correlated nonequilibrium photon fluid across the insulator-superfluid phase
  transition},\ }\href@noop {} {\bibfield  {journal} {\bibinfo  {journal}
  {Physical Review Letters}\ }\textbf {\bibinfo {volume} {131}},\ \bibinfo
  {pages} {193604} (\bibinfo {year} {2023})}\BibitemShut {NoStop}%
\bibitem [{\citenamefont {Kapit}\ \emph {et~al.}(2014)\citenamefont {Kapit},
  \citenamefont {Hafezi},\ and\ \citenamefont {Simon}}]{Kapit:2014PRX}%
  \BibitemOpen
  \bibfield  {author} {\bibinfo {author} {\bibfnamefont {E.}~\bibnamefont
  {Kapit}}, \bibinfo {author} {\bibfnamefont {M.}~\bibnamefont {Hafezi}},\ and\
  \bibinfo {author} {\bibfnamefont {S.~H.}\ \bibnamefont {Simon}},\ }\bibfield
  {title} {\bibinfo {title} {Induced self-stabilization in fractional quantum
  hall states of light},\ }\href
  {http://link.aps.org/doi/10.1103/PhysRevX.4.031039} {\bibfield  {journal}
  {\bibinfo  {journal} {Phys. Rev. X}\ }\textbf {\bibinfo {volume} {4}},\
  \bibinfo {pages} {031039} (\bibinfo {year} {2014})}\BibitemShut {NoStop}%
\bibitem [{\citenamefont {Ma}\ \emph {et~al.}(2019)\citenamefont {Ma},
  \citenamefont {Saxberg}, \citenamefont {Owens}, \citenamefont {Leung},
  \citenamefont {Lu}, \citenamefont {Simon},\ and\ \citenamefont
  {Schuster}}]{ma2019dissipatively}%
  \BibitemOpen
  \bibfield  {author} {\bibinfo {author} {\bibfnamefont {R.}~\bibnamefont
  {Ma}}, \bibinfo {author} {\bibfnamefont {B.}~\bibnamefont {Saxberg}},
  \bibinfo {author} {\bibfnamefont {C.}~\bibnamefont {Owens}}, \bibinfo
  {author} {\bibfnamefont {N.}~\bibnamefont {Leung}}, \bibinfo {author}
  {\bibfnamefont {Y.}~\bibnamefont {Lu}}, \bibinfo {author} {\bibfnamefont
  {J.}~\bibnamefont {Simon}},\ and\ \bibinfo {author} {\bibfnamefont {D.~I.}\
  \bibnamefont {Schuster}},\ }\bibfield  {title} {\bibinfo {title} {A
  dissipatively stabilized mott insulator of photons},\ }\href@noop {}
  {\bibfield  {journal} {\bibinfo  {journal} {Nature}\ }\textbf {\bibinfo
  {volume} {566}},\ \bibinfo {pages} {51} (\bibinfo {year} {2019})}\BibitemShut
  {NoStop}%
\bibitem [{\citenamefont {Mamaev}\ \emph {et~al.}(2018)\citenamefont {Mamaev},
  \citenamefont {Govia},\ and\ \citenamefont {Clerk}}]{Mamaev:Quantum2018}%
  \BibitemOpen
  \bibfield  {author} {\bibinfo {author} {\bibfnamefont {M.}~\bibnamefont
  {Mamaev}}, \bibinfo {author} {\bibfnamefont {L.}~\bibnamefont {Govia}},\ and\
  \bibinfo {author} {\bibfnamefont {A.}~\bibnamefont {Clerk}},\ }\bibfield
  {title} {\bibinfo {title} {Dissipative stabilization of entangled cat states
  using a driven bose-hubbard dimer},\ }\href@noop {} {\bibfield  {journal}
  {\bibinfo  {journal} {Quantum}\ }\textbf {\bibinfo {volume} {2}},\ \bibinfo
  {pages} {58} (\bibinfo {year} {2018})}\BibitemShut {NoStop}%
\bibitem [{\citenamefont {Harrington}\ \emph {et~al.}(2022)\citenamefont
  {Harrington}, \citenamefont {Mueller},\ and\ \citenamefont
  {Murch}}]{Harrington:NatRevPhys2022}%
  \BibitemOpen
  \bibfield  {author} {\bibinfo {author} {\bibfnamefont {P.~M.}\ \bibnamefont
  {Harrington}}, \bibinfo {author} {\bibfnamefont {E.~J.}\ \bibnamefont
  {Mueller}},\ and\ \bibinfo {author} {\bibfnamefont {K.~W.}\ \bibnamefont
  {Murch}},\ }\bibfield  {title} {\bibinfo {title} {Engineered dissipation for
  quantum information science},\ }\href@noop {} {\bibfield  {journal} {\bibinfo
   {journal} {Nature Reviews Physics}\ }\textbf {\bibinfo {volume} {4}},\
  \bibinfo {pages} {660} (\bibinfo {year} {2022})}\BibitemShut {NoStop}%
\bibitem [{\citenamefont {Li}\ \emph {et~al.}(2024)\citenamefont {Li},
  \citenamefont {Roy}, \citenamefont {Lu}, \citenamefont {Kapit},\ and\
  \citenamefont {Schuster}}]{Li:NatComm2024}%
  \BibitemOpen
  \bibfield  {author} {\bibinfo {author} {\bibfnamefont {Z.}~\bibnamefont
  {Li}}, \bibinfo {author} {\bibfnamefont {T.}~\bibnamefont {Roy}}, \bibinfo
  {author} {\bibfnamefont {Y.}~\bibnamefont {Lu}}, \bibinfo {author}
  {\bibfnamefont {E.}~\bibnamefont {Kapit}},\ and\ \bibinfo {author}
  {\bibfnamefont {D.~I.}\ \bibnamefont {Schuster}},\ }\bibfield  {title}
  {\bibinfo {title} {Autonomous stabilization with programmable stabilized
  state},\ }\href@noop {} {\bibfield  {journal} {\bibinfo  {journal} {Nature
  Communications}\ }\textbf {\bibinfo {volume} {15}},\ \bibinfo {pages} {6978}
  (\bibinfo {year} {2024})}\BibitemShut {NoStop}%
\bibitem [{\citenamefont {Lescanne}\ \emph {et~al.}(2020)\citenamefont
  {Lescanne}, \citenamefont {Villiers}, \citenamefont {Peronnin}, \citenamefont
  {Sarlette}, \citenamefont {Delbecq}, \citenamefont {Huard}, \citenamefont
  {Kontos}, \citenamefont {Mirrahimi},\ and\ \citenamefont
  {Leghtas}}]{Lescanne:NatPhys2020}%
  \BibitemOpen
  \bibfield  {author} {\bibinfo {author} {\bibfnamefont {R.}~\bibnamefont
  {Lescanne}}, \bibinfo {author} {\bibfnamefont {M.}~\bibnamefont {Villiers}},
  \bibinfo {author} {\bibfnamefont {T.}~\bibnamefont {Peronnin}}, \bibinfo
  {author} {\bibfnamefont {A.}~\bibnamefont {Sarlette}}, \bibinfo {author}
  {\bibfnamefont {M.}~\bibnamefont {Delbecq}}, \bibinfo {author} {\bibfnamefont
  {B.}~\bibnamefont {Huard}}, \bibinfo {author} {\bibfnamefont
  {T.}~\bibnamefont {Kontos}}, \bibinfo {author} {\bibfnamefont
  {M.}~\bibnamefont {Mirrahimi}},\ and\ \bibinfo {author} {\bibfnamefont
  {Z.}~\bibnamefont {Leghtas}},\ }\bibfield  {title} {\bibinfo {title}
  {Exponential suppression of bit-flips in a qubit encoded in an oscillator},\
  }\href@noop {} {\bibfield  {journal} {\bibinfo  {journal} {Nature Physics}\
  }\textbf {\bibinfo {volume} {16}},\ \bibinfo {pages} {509} (\bibinfo {year}
  {2020})}\BibitemShut {NoStop}%
\bibitem [{\citenamefont {Castin}(2001)}]{CastinLectures}%
  \BibitemOpen
  \bibfield  {author} {\bibinfo {author} {\bibfnamefont {Y.}~\bibnamefont
  {Castin}},\ }\bibfield  {title} {\bibinfo {title} {Bose-einstein condensates
  in atomic gases: simple theoretical results},\ }in\ \href@noop {} {\emph
  {\bibinfo {booktitle} {``Coherent atomic matter waves'', Lecture Notes of Les
  Houches Summer School}}},\ \bibinfo {editor} {edited by\ \bibinfo {editor}
  {\bibfnamefont {R.}~\bibnamefont {Kaiser}}, \bibinfo {editor} {\bibfnamefont
  {C.}~\bibnamefont {Westbrook}},\ and\ \bibinfo {editor} {\bibfnamefont
  {F.}~\bibnamefont {David}}}\ (\bibinfo  {publisher} {EDP Sciences and
  Springer-Verlag},\ \bibinfo {year} {2001})\ pp.\ \bibinfo {pages} {1--136},\
  \bibinfo {note} {available as arXiv:cond-mat/0105058}\BibitemShut {NoStop}%
\bibitem [{\citenamefont {Wouters}\ and\ \citenamefont
  {Carusotto}(2010)}]{Wouters:PRL2010}%
  \BibitemOpen
  \bibfield  {author} {\bibinfo {author} {\bibfnamefont {M.}~\bibnamefont
  {Wouters}}\ and\ \bibinfo {author} {\bibfnamefont {I.}~\bibnamefont
  {Carusotto}},\ }\bibfield  {title} {\bibinfo {title} {Superfluidity and
  critical velocities in nonequilibrium bose-einstein condensates},\ }\href
  {https://doi.org/10.1103/PhysRevLett.105.020602} {\bibfield  {journal}
  {\bibinfo  {journal} {Phys. Rev. Lett.}\ }\textbf {\bibinfo {volume} {105}},\
  \bibinfo {pages} {020602} (\bibinfo {year} {2010})}\BibitemShut {NoStop}%
\bibitem [{\citenamefont {Cross}\ and\ \citenamefont
  {Hohenberg}(1993)}]{Cross:RMP1993}%
  \BibitemOpen
  \bibfield  {author} {\bibinfo {author} {\bibfnamefont {M.~C.}\ \bibnamefont
  {Cross}}\ and\ \bibinfo {author} {\bibfnamefont {P.~C.}\ \bibnamefont
  {Hohenberg}},\ }\bibfield  {title} {\bibinfo {title} {Pattern formation
  outside of equilibrium},\ }\href@noop {} {\bibfield  {journal} {\bibinfo
  {journal} {Reviews of modern physics}\ }\textbf {\bibinfo {volume} {65}},\
  \bibinfo {pages} {851} (\bibinfo {year} {1993})}\BibitemShut {NoStop}%
\bibitem [{\citenamefont {Aranson}\ and\ \citenamefont
  {Kramer}(2002)}]{Aranson:RMP2002}%
  \BibitemOpen
  \bibfield  {author} {\bibinfo {author} {\bibfnamefont {I.~S.}\ \bibnamefont
  {Aranson}}\ and\ \bibinfo {author} {\bibfnamefont {L.}~\bibnamefont
  {Kramer}},\ }\bibfield  {title} {\bibinfo {title} {The world of the complex
  ginzburg-landau equation},\ }\href@noop {} {\bibfield  {journal} {\bibinfo
  {journal} {Reviews of modern physics}\ }\textbf {\bibinfo {volume} {74}},\
  \bibinfo {pages} {99} (\bibinfo {year} {2002})}\BibitemShut {NoStop}%
\bibitem [{\citenamefont {Lugiato}\ and\ \citenamefont
  {Lefever}(1987)}]{Lugiato:PRL1987}%
  \BibitemOpen
  \bibfield  {author} {\bibinfo {author} {\bibfnamefont {L.~A.}\ \bibnamefont
  {Lugiato}}\ and\ \bibinfo {author} {\bibfnamefont {R.}~\bibnamefont
  {Lefever}},\ }\bibfield  {title} {\bibinfo {title} {Spatial dissipative
  structures in passive optical systems},\ }\href@noop {} {\bibfield  {journal}
  {\bibinfo  {journal} {Physical review letters}\ }\textbf {\bibinfo {volume}
  {58}},\ \bibinfo {pages} {2209} (\bibinfo {year} {1987})}\BibitemShut
  {NoStop}%
\bibitem [{\citenamefont {Lugiato}\ \emph {et~al.}(2025)\citenamefont
  {Lugiato}, \citenamefont {Prati}, \citenamefont {Brambilla},\ and\
  \citenamefont {Gatti}}]{Lugiato:Varenna}%
  \BibitemOpen
  \bibfield  {author} {\bibinfo {author} {\bibfnamefont {L.~A.}\ \bibnamefont
  {Lugiato}}, \bibinfo {author} {\bibfnamefont {F.}~\bibnamefont {Prati}},
  \bibinfo {author} {\bibfnamefont {E.}~\bibnamefont {Brambilla}},\ and\
  \bibinfo {author} {\bibfnamefont {A.}~\bibnamefont {Gatti}},\ }\bibfield
  {title} {\bibinfo {title} {The cavity kerr medium model and the surprising
  history around it},\ }in\ \href@noop {} {\emph {\bibinfo {booktitle} {Quantum
  Fluids of Light and Matter}}}\ (\bibinfo  {publisher} {IOS Press},\ \bibinfo
  {year} {2025})\ pp.\ \bibinfo {pages} {5--21}\BibitemShut {NoStop}%
\bibitem [{\citenamefont {Carusotto}\ and\ \citenamefont
  {Ciuti}(2005)}]{Carusotto:PRB2005}%
  \BibitemOpen
  \bibfield  {author} {\bibinfo {author} {\bibfnamefont {I.}~\bibnamefont
  {Carusotto}}\ and\ \bibinfo {author} {\bibfnamefont {C.}~\bibnamefont
  {Ciuti}},\ }\bibfield  {title} {\bibinfo {title} {Spontaneous
  microcavity-polariton coherence across the parametric threshold: Quantum
  monte carlo studies},\ }\href@noop {} {\bibfield  {journal} {\bibinfo
  {journal} {Physical Review B—Condensed Matter and Materials Physics}\
  }\textbf {\bibinfo {volume} {72}},\ \bibinfo {pages} {125335} (\bibinfo
  {year} {2005})}\BibitemShut {NoStop}%
\bibitem [{\citenamefont {Wouters}\ and\ \citenamefont
  {Savona}(2009)}]{Wouters:PRB2009}%
  \BibitemOpen
  \bibfield  {author} {\bibinfo {author} {\bibfnamefont {M.}~\bibnamefont
  {Wouters}}\ and\ \bibinfo {author} {\bibfnamefont {V.}~\bibnamefont
  {Savona}},\ }\bibfield  {title} {\bibinfo {title} {Stochastic classical field
  model for polariton condensates},\ }\href
  {https://doi.org/10.1103/PhysRevB.79.165302} {\bibfield  {journal} {\bibinfo
  {journal} {Phys. Rev. B}\ }\textbf {\bibinfo {volume} {79}},\ \bibinfo
  {pages} {165302} (\bibinfo {year} {2009})}\BibitemShut {NoStop}%
\bibitem [{\citenamefont {Fr{\'e}rot}\ \emph {et~al.}(2023)\citenamefont
  {Fr{\'e}rot}, \citenamefont {Vashisht}, \citenamefont {Morassi},
  \citenamefont {Lema{\^\i}tre}, \citenamefont {Ravets}, \citenamefont {Bloch},
  \citenamefont {Minguzzi},\ and\ \citenamefont {Richard}}]{Frerot:PRX2023}%
  \BibitemOpen
  \bibfield  {author} {\bibinfo {author} {\bibfnamefont {I.}~\bibnamefont
  {Fr{\'e}rot}}, \bibinfo {author} {\bibfnamefont {A.}~\bibnamefont
  {Vashisht}}, \bibinfo {author} {\bibfnamefont {M.}~\bibnamefont {Morassi}},
  \bibinfo {author} {\bibfnamefont {A.}~\bibnamefont {Lema{\^\i}tre}}, \bibinfo
  {author} {\bibfnamefont {S.}~\bibnamefont {Ravets}}, \bibinfo {author}
  {\bibfnamefont {J.}~\bibnamefont {Bloch}}, \bibinfo {author} {\bibfnamefont
  {A.}~\bibnamefont {Minguzzi}},\ and\ \bibinfo {author} {\bibfnamefont
  {M.}~\bibnamefont {Richard}},\ }\bibfield  {title} {\bibinfo {title}
  {Bogoliubov excitations driven by thermal lattice phonons in a quantum fluid
  of light},\ }\href@noop {} {\bibfield  {journal} {\bibinfo  {journal}
  {Physical Review X}\ }\textbf {\bibinfo {volume} {13}},\ \bibinfo {pages}
  {041058} (\bibinfo {year} {2023})}\BibitemShut {NoStop}%
\bibitem [{\citenamefont {Szymanska}(2025)}]{Marzena_private}%
  \BibitemOpen
  \bibfield  {author} {\bibinfo {author} {\bibfnamefont {M.~H.}\ \bibnamefont
  {Szymanska}},\ }\bibfield  {title} {\bibinfo {title} {Stochastic and tensor
  network methods for open dissipative quantum lattice models}} (\bibinfo
  {year} {2025}),\ \bibinfo {note} {talk at the ``MEOQS2025 - Methods for
  many-body open quantum systems'' workshop, Trento (2025)}\BibitemShut
  {NoStop}%
\bibitem [{\citenamefont {Berg}\ \emph {et~al.}(2009)\citenamefont {Berg},
  \citenamefont {Plimak}, \citenamefont {Polkovnikov}, \citenamefont {Olsen},
  \citenamefont {Fleischhauer},\ and\ \citenamefont {Schleich}}]{Berg:PRA2009}%
  \BibitemOpen
  \bibfield  {author} {\bibinfo {author} {\bibfnamefont {B.}~\bibnamefont
  {Berg}}, \bibinfo {author} {\bibfnamefont {L.~I.}\ \bibnamefont {Plimak}},
  \bibinfo {author} {\bibfnamefont {A.}~\bibnamefont {Polkovnikov}}, \bibinfo
  {author} {\bibfnamefont {M.~K.}\ \bibnamefont {Olsen}}, \bibinfo {author}
  {\bibfnamefont {M.}~\bibnamefont {Fleischhauer}},\ and\ \bibinfo {author}
  {\bibfnamefont {W.~P.}\ \bibnamefont {Schleich}},\ }\bibfield  {title}
  {\bibinfo {title} {Commuting heisenberg operators as the quantum response
  problem: Time-normal averages in the truncated wigner representation},\
  }\href {https://doi.org/10.1103/PhysRevA.80.033624} {\bibfield  {journal}
  {\bibinfo  {journal} {Phys. Rev. A}\ }\textbf {\bibinfo {volume} {80}},\
  \bibinfo {pages} {033624} (\bibinfo {year} {2009})}\BibitemShut {NoStop}%
\bibitem [{\citenamefont {Sinatra}\ \emph {et~al.}(2002)\citenamefont
  {Sinatra}, \citenamefont {Lobo},\ and\ \citenamefont
  {Castin}}]{Sinatra:JPhysB2002}%
  \BibitemOpen
  \bibfield  {author} {\bibinfo {author} {\bibfnamefont {A.}~\bibnamefont
  {Sinatra}}, \bibinfo {author} {\bibfnamefont {C.}~\bibnamefont {Lobo}},\ and\
  \bibinfo {author} {\bibfnamefont {Y.}~\bibnamefont {Castin}},\ }\bibfield
  {title} {\bibinfo {title} {The truncated wigner method for bose-condensed
  gases: limits of validity and applications},\ }\href
  {http://stacks.iop.org/0953-4075/35/i=17/a=301} {\bibfield  {journal}
  {\bibinfo  {journal} {Journal of Physics B: Atomic, Molecular and Optical
  Physics}\ }\textbf {\bibinfo {volume} {35}},\ \bibinfo {pages} {3599}
  (\bibinfo {year} {2002})}\BibitemShut {NoStop}%
\bibitem [{\citenamefont {Steel}\ \emph {et~al.}(1998)\citenamefont {Steel},
  \citenamefont {Olsen}, \citenamefont {Plimak}, \citenamefont {Drummond},
  \citenamefont {Tan}, \citenamefont {Collett}, \citenamefont {Walls},\ and\
  \citenamefont {Graham}}]{Steel:PRA1998}%
  \BibitemOpen
  \bibfield  {author} {\bibinfo {author} {\bibfnamefont {M.~J.}\ \bibnamefont
  {Steel}}, \bibinfo {author} {\bibfnamefont {M.~K.}\ \bibnamefont {Olsen}},
  \bibinfo {author} {\bibfnamefont {L.~I.}\ \bibnamefont {Plimak}}, \bibinfo
  {author} {\bibfnamefont {P.~D.}\ \bibnamefont {Drummond}}, \bibinfo {author}
  {\bibfnamefont {S.~M.}\ \bibnamefont {Tan}}, \bibinfo {author} {\bibfnamefont
  {M.~J.}\ \bibnamefont {Collett}}, \bibinfo {author} {\bibfnamefont {D.~F.}\
  \bibnamefont {Walls}},\ and\ \bibinfo {author} {\bibfnamefont
  {R.}~\bibnamefont {Graham}},\ }\bibfield  {title} {\bibinfo {title}
  {Dynamical quantum noise in trapped bose-einstein condensates},\ }\href
  {https://doi.org/10.1103/PhysRevA.58.4824} {\bibfield  {journal} {\bibinfo
  {journal} {Phys. Rev. A}\ }\textbf {\bibinfo {volume} {58}},\ \bibinfo
  {pages} {4824} (\bibinfo {year} {1998})}\BibitemShut {NoStop}%
\bibitem [{\citenamefont {Drummond}(1986)}]{Drummond:PRA1986}%
  \BibitemOpen
  \bibfield  {author} {\bibinfo {author} {\bibfnamefont {P.~D.}\ \bibnamefont
  {Drummond}},\ }\bibfield  {title} {\bibinfo {title} {Quantum optical
  tunneling: A representation-free theory valid near the state-equation turning
  points},\ }\href {https://doi.org/10.1103/PhysRevA.33.4462} {\bibfield
  {journal} {\bibinfo  {journal} {Phys. Rev. A}\ }\textbf {\bibinfo {volume}
  {33}},\ \bibinfo {pages} {4462} (\bibinfo {year} {1986})}\BibitemShut
  {NoStop}%
\bibitem [{\citenamefont {Vogel}\ and\ \citenamefont
  {Risken}(1988)}]{Vogel:PRA1988}%
  \BibitemOpen
  \bibfield  {author} {\bibinfo {author} {\bibfnamefont {K.}~\bibnamefont
  {Vogel}}\ and\ \bibinfo {author} {\bibfnamefont {H.}~\bibnamefont {Risken}},\
  }\bibfield  {title} {\bibinfo {title} {Quantum-tunneling rates and stationary
  solutions in dispersive optical bistability},\ }\href
  {https://doi.org/10.1103/PhysRevA.38.2409} {\bibfield  {journal} {\bibinfo
  {journal} {Phys. Rev. A}\ }\textbf {\bibinfo {volume} {38}},\ \bibinfo
  {pages} {2409} (\bibinfo {year} {1988})}\BibitemShut {NoStop}%
\bibitem [{\citenamefont {Van~Regemortel}\ \emph {et~al.}(2017)\citenamefont
  {Van~Regemortel}, \citenamefont {Casteels}, \citenamefont {Carusotto},\ and\
  \citenamefont {Wouters}}]{VanRegemortel:PRA2017}%
  \BibitemOpen
  \bibfield  {author} {\bibinfo {author} {\bibfnamefont {M.}~\bibnamefont
  {Van~Regemortel}}, \bibinfo {author} {\bibfnamefont {W.}~\bibnamefont
  {Casteels}}, \bibinfo {author} {\bibfnamefont {I.}~\bibnamefont
  {Carusotto}},\ and\ \bibinfo {author} {\bibfnamefont {M.}~\bibnamefont
  {Wouters}},\ }\bibfield  {title} {\bibinfo {title} {Spontaneous
  beliaev-landau scattering out of equilibrium},\ }\href@noop {} {\bibfield
  {journal} {\bibinfo  {journal} {Physical Review A}\ }\textbf {\bibinfo
  {volume} {96}},\ \bibinfo {pages} {053854} (\bibinfo {year}
  {2017})}\BibitemShut {NoStop}%
\bibitem [{\citenamefont {Polkovnikov}(2003)}]{Polkovnikov:PRA2003}%
  \BibitemOpen
  \bibfield  {author} {\bibinfo {author} {\bibfnamefont {A.}~\bibnamefont
  {Polkovnikov}},\ }\bibfield  {title} {\bibinfo {title} {Quantum corrections
  to the dynamics of interacting bosons: Beyond the truncated wigner
  approximation},\ }\href@noop {} {\bibfield  {journal} {\bibinfo  {journal}
  {Physical Review A}\ }\textbf {\bibinfo {volume} {68}},\ \bibinfo {pages}
  {053604} (\bibinfo {year} {2003})}\BibitemShut {NoStop}%
\bibitem [{\citenamefont {Busch}\ \emph {et~al.}(2014)\citenamefont {Busch},
  \citenamefont {Carusotto},\ and\ \citenamefont {Parentani}}]{Busch:PRA2014}%
  \BibitemOpen
  \bibfield  {author} {\bibinfo {author} {\bibfnamefont {X.}~\bibnamefont
  {Busch}}, \bibinfo {author} {\bibfnamefont {I.}~\bibnamefont {Carusotto}},\
  and\ \bibinfo {author} {\bibfnamefont {R.}~\bibnamefont {Parentani}},\
  }\bibfield  {title} {\bibinfo {title} {Spectrum and entanglement of phonons
  in quantum fluids of light},\ }\href@noop {} {\bibfield  {journal} {\bibinfo
  {journal} {Physical Review A}\ }\textbf {\bibinfo {volume} {89}},\ \bibinfo
  {pages} {043819} (\bibinfo {year} {2014})}\BibitemShut {NoStop}%
\bibitem [{\citenamefont {Carusotto}\ and\ \citenamefont
  {Ciuti}(2004)}]{Carusotto:PRL2004}%
  \BibitemOpen
  \bibfield  {author} {\bibinfo {author} {\bibfnamefont {I.}~\bibnamefont
  {Carusotto}}\ and\ \bibinfo {author} {\bibfnamefont {C.}~\bibnamefont
  {Ciuti}},\ }\bibfield  {title} {\bibinfo {title} {Probing microcavity
  polariton superfluidity through resonant rayleigh scattering},\ }\href
  {https://doi.org/10.1103/PhysRevLett.93.166401} {\bibfield  {journal}
  {\bibinfo  {journal} {Phys. Rev. Lett.}\ }\textbf {\bibinfo {volume} {93}},\
  \bibinfo {pages} {166401} (\bibinfo {year} {2004})}\BibitemShut {NoStop}%
\bibitem [{\citenamefont {Amo}\ \emph {et~al.}(2009)\citenamefont {Amo},
  \citenamefont {Lefrere}, \citenamefont {Pigeon}, \citenamefont {Adrados},
  \citenamefont {Ciuti}, \citenamefont {Carusotto}, \citenamefont {Houdre},
  \citenamefont {Giacobino},\ and\ \citenamefont {Bramati}}]{Amo:NPhys2009}%
  \BibitemOpen
  \bibfield  {author} {\bibinfo {author} {\bibfnamefont {A.}~\bibnamefont
  {Amo}}, \bibinfo {author} {\bibfnamefont {J.}~\bibnamefont {Lefrere}},
  \bibinfo {author} {\bibfnamefont {S.}~\bibnamefont {Pigeon}}, \bibinfo
  {author} {\bibfnamefont {C.}~\bibnamefont {Adrados}}, \bibinfo {author}
  {\bibfnamefont {C.}~\bibnamefont {Ciuti}}, \bibinfo {author} {\bibfnamefont
  {I.}~\bibnamefont {Carusotto}}, \bibinfo {author} {\bibfnamefont
  {R.}~\bibnamefont {Houdre}}, \bibinfo {author} {\bibfnamefont
  {E.}~\bibnamefont {Giacobino}},\ and\ \bibinfo {author} {\bibfnamefont
  {A.}~\bibnamefont {Bramati}},\ }\bibfield  {title} {\bibinfo {title}
  {Superfluidity of polaritons in semiconductor microcavities},\ }\href
  {https://doi.org/10.1038/NPHYS1364} {\bibfield  {journal} {\bibinfo
  {journal} {Nature Phys.}\ }\textbf {\bibinfo {volume} {5}},\ \bibinfo {pages}
  {805} (\bibinfo {year} {2009})}\BibitemShut {NoStop}%
\bibitem [{\citenamefont {Claude}\ \emph {et~al.}(2022)\citenamefont {Claude},
  \citenamefont {Jacquet}, \citenamefont {Usciati}, \citenamefont {Carusotto},
  \citenamefont {Giacobino}, \citenamefont {Bramati},\ and\ \citenamefont
  {Glorieux}}]{Claude:PRL2022}%
  \BibitemOpen
  \bibfield  {author} {\bibinfo {author} {\bibfnamefont {F.}~\bibnamefont
  {Claude}}, \bibinfo {author} {\bibfnamefont {M.~J.}\ \bibnamefont {Jacquet}},
  \bibinfo {author} {\bibfnamefont {R.}~\bibnamefont {Usciati}}, \bibinfo
  {author} {\bibfnamefont {I.}~\bibnamefont {Carusotto}}, \bibinfo {author}
  {\bibfnamefont {E.}~\bibnamefont {Giacobino}}, \bibinfo {author}
  {\bibfnamefont {A.}~\bibnamefont {Bramati}},\ and\ \bibinfo {author}
  {\bibfnamefont {Q.}~\bibnamefont {Glorieux}},\ }\bibfield  {title} {\bibinfo
  {title} {High-resolution coherent probe spectroscopy of a polariton quantum
  fluid},\ }\href@noop {} {\bibfield  {journal} {\bibinfo  {journal} {Physical
  Review Letters}\ }\textbf {\bibinfo {volume} {129}},\ \bibinfo {pages}
  {103601} (\bibinfo {year} {2022})}\BibitemShut {NoStop}%
\bibitem [{\citenamefont {Boulier}\ \emph {et~al.}(2016)\citenamefont
  {Boulier}, \citenamefont {Cancellieri}, \citenamefont {Sangouard},
  \citenamefont {Hivet}, \citenamefont {Glorieux}, \citenamefont {Giacobino},\
  and\ \citenamefont {Bramati}}]{Boulier:CRAS2016}%
  \BibitemOpen
  \bibfield  {author} {\bibinfo {author} {\bibfnamefont {T.}~\bibnamefont
  {Boulier}}, \bibinfo {author} {\bibfnamefont {E.}~\bibnamefont
  {Cancellieri}}, \bibinfo {author} {\bibfnamefont {N.~D.}\ \bibnamefont
  {Sangouard}}, \bibinfo {author} {\bibfnamefont {R.}~\bibnamefont {Hivet}},
  \bibinfo {author} {\bibfnamefont {Q.}~\bibnamefont {Glorieux}}, \bibinfo
  {author} {\bibfnamefont {{\'E}.}~\bibnamefont {Giacobino}},\ and\ \bibinfo
  {author} {\bibfnamefont {A.}~\bibnamefont {Bramati}},\ }\bibfield  {title}
  {\bibinfo {title} {Lattices of quantized vortices in polariton superfluids},\
  }\href@noop {} {\bibfield  {journal} {\bibinfo  {journal} {Comptes Rendus.
  Physique}\ }\textbf {\bibinfo {volume} {17}},\ \bibinfo {pages} {893}
  (\bibinfo {year} {2016})}\BibitemShut {NoStop}%
\bibitem [{\citenamefont {Barcelo}\ \emph {et~al.}(2011)\citenamefont
  {Barcelo}, \citenamefont {Liberati},\ and\ \citenamefont
  {Visser}}]{barcelo2011analogue}%
  \BibitemOpen
  \bibfield  {author} {\bibinfo {author} {\bibfnamefont {C.}~\bibnamefont
  {Barcelo}}, \bibinfo {author} {\bibfnamefont {S.}~\bibnamefont {Liberati}},\
  and\ \bibinfo {author} {\bibfnamefont {M.}~\bibnamefont {Visser}},\
  }\bibfield  {title} {\bibinfo {title} {Analogue gravity},\ }\href@noop {}
  {\bibfield  {journal} {\bibinfo  {journal} {Living reviews in relativity}\
  }\textbf {\bibinfo {volume} {14}},\ \bibinfo {pages} {1} (\bibinfo {year}
  {2011})}\BibitemShut {NoStop}%
\bibitem [{\citenamefont {Gerace}\ and\ \citenamefont
  {Carusotto}(2012)}]{Gerace:PRB2012}%
  \BibitemOpen
  \bibfield  {author} {\bibinfo {author} {\bibfnamefont {D.}~\bibnamefont
  {Gerace}}\ and\ \bibinfo {author} {\bibfnamefont {I.}~\bibnamefont
  {Carusotto}},\ }\bibfield  {title} {\bibinfo {title} {Analog hawking
  radiation from an acoustic black hole in a flowing polariton superfluid},\
  }\href@noop {} {\bibfield  {journal} {\bibinfo  {journal} {Physical Review
  B—Condensed Matter and Materials Physics}\ }\textbf {\bibinfo {volume}
  {86}},\ \bibinfo {pages} {144505} (\bibinfo {year} {2012})}\BibitemShut
  {NoStop}%
\bibitem [{\citenamefont {Nguyen}\ \emph {et~al.}(2015)\citenamefont {Nguyen},
  \citenamefont {Gerace}, \citenamefont {Carusotto}, \citenamefont {Sanvitto},
  \citenamefont {Galopin}, \citenamefont {Lema{\^\i}tre}, \citenamefont
  {Sagnes}, \citenamefont {Bloch},\ and\ \citenamefont {Amo}}]{Nguyen:PRL2015}%
  \BibitemOpen
  \bibfield  {author} {\bibinfo {author} {\bibfnamefont {H.~S.}\ \bibnamefont
  {Nguyen}}, \bibinfo {author} {\bibfnamefont {D.}~\bibnamefont {Gerace}},
  \bibinfo {author} {\bibfnamefont {I.}~\bibnamefont {Carusotto}}, \bibinfo
  {author} {\bibfnamefont {D.}~\bibnamefont {Sanvitto}}, \bibinfo {author}
  {\bibfnamefont {E.}~\bibnamefont {Galopin}}, \bibinfo {author} {\bibfnamefont
  {A.}~\bibnamefont {Lema{\^\i}tre}}, \bibinfo {author} {\bibfnamefont
  {I.}~\bibnamefont {Sagnes}}, \bibinfo {author} {\bibfnamefont
  {J.}~\bibnamefont {Bloch}},\ and\ \bibinfo {author} {\bibfnamefont
  {A.}~\bibnamefont {Amo}},\ }\bibfield  {title} {\bibinfo {title} {Acoustic
  black hole in a stationary hydrodynamic flow of microcavity polaritons},\
  }\href@noop {} {\bibfield  {journal} {\bibinfo  {journal} {Physical review
  letters}\ }\textbf {\bibinfo {volume} {114}},\ \bibinfo {pages} {036402}
  (\bibinfo {year} {2015})}\BibitemShut {NoStop}%
\bibitem [{\citenamefont {Falque}\ \emph {et~al.}(2023)\citenamefont {Falque},
  \citenamefont {Delhom}, \citenamefont {Glorieux}, \citenamefont {Giacobino},
  \citenamefont {Bramati},\ and\ \citenamefont {Jacquet}}]{Falque:arXiv2023}%
  \BibitemOpen
  \bibfield  {author} {\bibinfo {author} {\bibfnamefont {K.}~\bibnamefont
  {Falque}}, \bibinfo {author} {\bibfnamefont {A.}~\bibnamefont {Delhom}},
  \bibinfo {author} {\bibfnamefont {Q.}~\bibnamefont {Glorieux}}, \bibinfo
  {author} {\bibfnamefont {E.}~\bibnamefont {Giacobino}}, \bibinfo {author}
  {\bibfnamefont {A.}~\bibnamefont {Bramati}},\ and\ \bibinfo {author}
  {\bibfnamefont {M.~J.}\ \bibnamefont {Jacquet}},\ }\bibfield  {title}
  {\bibinfo {title} {Polariton fluids as quantum field theory simulators on
  tailored curved spacetimes},\ }\href@noop {} {\bibfield  {journal} {\bibinfo
  {journal} {arXiv preprint arXiv:2311.01392}\ } (\bibinfo {year}
  {2023})}\BibitemShut {NoStop}%
\bibitem [{\citenamefont {Mu{\~n}oz~de Nova}\ \emph {et~al.}(2019)\citenamefont
  {Mu{\~n}oz~de Nova}, \citenamefont {Golubkov}, \citenamefont {Kolobov},\ and\
  \citenamefont {Steinhauer}}]{Munoz:Nature2019}%
  \BibitemOpen
  \bibfield  {author} {\bibinfo {author} {\bibfnamefont {J.~R.}\ \bibnamefont
  {Mu{\~n}oz~de Nova}}, \bibinfo {author} {\bibfnamefont {K.}~\bibnamefont
  {Golubkov}}, \bibinfo {author} {\bibfnamefont {V.~I.}\ \bibnamefont
  {Kolobov}},\ and\ \bibinfo {author} {\bibfnamefont {J.}~\bibnamefont
  {Steinhauer}},\ }\bibfield  {title} {\bibinfo {title} {Observation of thermal
  hawking radiation and its temperature in an analogue black hole},\
  }\href@noop {} {\bibfield  {journal} {\bibinfo  {journal} {Nature}\ }\textbf
  {\bibinfo {volume} {569}},\ \bibinfo {pages} {688} (\bibinfo {year}
  {2019})}\BibitemShut {NoStop}%
\bibitem [{\citenamefont {{Tong}}(2016)}]{Tong:QHbook}%
  \BibitemOpen
  \bibfield  {author} {\bibinfo {author} {\bibfnamefont {D.}~\bibnamefont
  {{Tong}}},\ }\bibfield  {title} {\bibinfo {title} {Lectures on the quantum
  hall effect},\ }\href@noop {} {\bibfield  {journal} {\bibinfo  {journal}
  {ArXiv e-prints}\ } (\bibinfo {year} {2016})},\ \Eprint
  {https://arxiv.org/abs/1606.06687} {arXiv:1606.06687 [hep-th]} \BibitemShut
  {NoStop}%
\bibitem [{\citenamefont {Ozawa}\ and\ \citenamefont
  {Carusotto}(2014)}]{Ozawa:PRL2014}%
  \BibitemOpen
  \bibfield  {author} {\bibinfo {author} {\bibfnamefont {T.}~\bibnamefont
  {Ozawa}}\ and\ \bibinfo {author} {\bibfnamefont {I.}~\bibnamefont
  {Carusotto}},\ }\bibfield  {title} {\bibinfo {title} {Anomalous and quantum
  hall effects in lossy photonic lattices},\ }\href@noop {} {\bibfield
  {journal} {\bibinfo  {journal} {Physical review letters}\ }\textbf {\bibinfo
  {volume} {112}},\ \bibinfo {pages} {133902} (\bibinfo {year}
  {2014})}\BibitemShut {NoStop}%
\bibitem [{\citenamefont {Ozawa}\ \emph {et~al.}(2019)\citenamefont {Ozawa},
  \citenamefont {Price}, \citenamefont {Amo}, \citenamefont {Goldman},
  \citenamefont {Hafezi}, \citenamefont {Lu}, \citenamefont {Rechtsman},
  \citenamefont {Schuster}, \citenamefont {Simon}, \citenamefont {Zilberberg}
  \emph {et~al.}}]{ozawaRMP2019topological}%
  \BibitemOpen
  \bibfield  {author} {\bibinfo {author} {\bibfnamefont {T.}~\bibnamefont
  {Ozawa}}, \bibinfo {author} {\bibfnamefont {H.~M.}\ \bibnamefont {Price}},
  \bibinfo {author} {\bibfnamefont {A.}~\bibnamefont {Amo}}, \bibinfo {author}
  {\bibfnamefont {N.}~\bibnamefont {Goldman}}, \bibinfo {author} {\bibfnamefont
  {M.}~\bibnamefont {Hafezi}}, \bibinfo {author} {\bibfnamefont
  {L.}~\bibnamefont {Lu}}, \bibinfo {author} {\bibfnamefont {M.~C.}\
  \bibnamefont {Rechtsman}}, \bibinfo {author} {\bibfnamefont {D.}~\bibnamefont
  {Schuster}}, \bibinfo {author} {\bibfnamefont {J.}~\bibnamefont {Simon}},
  \bibinfo {author} {\bibfnamefont {O.}~\bibnamefont {Zilberberg}}, \emph
  {et~al.},\ }\bibfield  {title} {\bibinfo {title} {Topological photonics},\
  }\href@noop {} {\bibfield  {journal} {\bibinfo  {journal} {Reviews of Modern
  Physics}\ }\textbf {\bibinfo {volume} {91}},\ \bibinfo {pages} {015006}
  (\bibinfo {year} {2019})}\BibitemShut {NoStop}%
\bibitem [{\citenamefont {Ch{\'e}nier}\ \emph {et~al.}(2024)\citenamefont
  {Ch{\'e}nier}, \citenamefont {d'Aligny}, \citenamefont {Pellerin},
  \citenamefont {Blanchard}, \citenamefont {Ozawa}, \citenamefont {Carusotto},\
  and\ \citenamefont {St-Jean}}]{chenier2024quantized}%
  \BibitemOpen
  \bibfield  {author} {\bibinfo {author} {\bibfnamefont {A.}~\bibnamefont
  {Ch{\'e}nier}}, \bibinfo {author} {\bibfnamefont {B.}~\bibnamefont
  {d'Aligny}}, \bibinfo {author} {\bibfnamefont {F.}~\bibnamefont {Pellerin}},
  \bibinfo {author} {\bibfnamefont {P.-{\'E}.}\ \bibnamefont {Blanchard}},
  \bibinfo {author} {\bibfnamefont {T.}~\bibnamefont {Ozawa}}, \bibinfo
  {author} {\bibfnamefont {I.}~\bibnamefont {Carusotto}},\ and\ \bibinfo
  {author} {\bibfnamefont {P.}~\bibnamefont {St-Jean}},\ }\bibfield  {title}
  {\bibinfo {title} {Quantized hall drift in a frequency-encoded photonic chern
  insulator},\ }\href@noop {} {\bibfield  {journal} {\bibinfo  {journal} {arXiv
  preprint arXiv:2412.04347}\ } (\bibinfo {year} {2024})}\BibitemShut {NoStop}%
\bibitem [{\citenamefont {Claude}\ \emph {et~al.}(2023)\citenamefont {Claude},
  \citenamefont {Jacquet}, \citenamefont {Wouters}, \citenamefont {Giacobino},
  \citenamefont {Glorieux}, \citenamefont {Carusotto},\ and\ \citenamefont
  {Bramati}}]{claude2023observation}%
  \BibitemOpen
  \bibfield  {author} {\bibinfo {author} {\bibfnamefont {F.}~\bibnamefont
  {Claude}}, \bibinfo {author} {\bibfnamefont {M.~J.}\ \bibnamefont {Jacquet}},
  \bibinfo {author} {\bibfnamefont {M.}~\bibnamefont {Wouters}}, \bibinfo
  {author} {\bibfnamefont {E.}~\bibnamefont {Giacobino}}, \bibinfo {author}
  {\bibfnamefont {Q.}~\bibnamefont {Glorieux}}, \bibinfo {author}
  {\bibfnamefont {I.}~\bibnamefont {Carusotto}},\ and\ \bibinfo {author}
  {\bibfnamefont {A.}~\bibnamefont {Bramati}},\ }\bibfield  {title} {\bibinfo
  {title} {Observation of the diffusive nambu-goldstone mode of a
  non-equilibrium phase transition},\ }\href@noop {} {\bibfield  {journal}
  {\bibinfo  {journal} {arXiv preprint arXiv:2310.11903}\ } (\bibinfo {year}
  {2023})}\BibitemShut {NoStop}%
\bibitem [{\citenamefont {Huang}(1987)}]{Huang}%
  \BibitemOpen
  \bibfield  {author} {\bibinfo {author} {\bibfnamefont {K.}~\bibnamefont
  {Huang}},\ }\href@noop {} {\emph {\bibinfo {title} {Statistical Mechanics}}}\
  (\bibinfo  {publisher} {Wiley},\ \bibinfo {year} {1987})\BibitemShut
  {NoStop}%
\bibitem [{\citenamefont {Richard}\ \emph {et~al.}(2005)\citenamefont
  {Richard}, \citenamefont {Kasprzak}, \citenamefont {Romestain}, \citenamefont
  {Andre},\ and\ \citenamefont {Dang}}]{Richard:PRL2005}%
  \BibitemOpen
  \bibfield  {author} {\bibinfo {author} {\bibfnamefont {M.}~\bibnamefont
  {Richard}}, \bibinfo {author} {\bibfnamefont {J.}~\bibnamefont {Kasprzak}},
  \bibinfo {author} {\bibfnamefont {R.}~\bibnamefont {Romestain}}, \bibinfo
  {author} {\bibfnamefont {R.}~\bibnamefont {Andre}},\ and\ \bibinfo {author}
  {\bibfnamefont {L.}~\bibnamefont {Dang}},\ }\bibfield  {title} {\bibinfo
  {title} {Spontaneous coherent phase transition of polaritons in cdte
  microcavities},\ }\href {https://doi.org/10.1103/PhysRevLett.94.187401}
  {\bibfield  {journal} {\bibinfo  {journal} {Phys. Rev. Lett.}\ }\textbf
  {\bibinfo {volume} {94}},\ \bibinfo {pages} {187401} (\bibinfo {year}
  {2005})}\BibitemShut {NoStop}%
\bibitem [{\citenamefont {Wouters}\ \emph {et~al.}(2008)\citenamefont
  {Wouters}, \citenamefont {Carusotto},\ and\ \citenamefont
  {Ciuti}}]{Wouters:PRB2008}%
  \BibitemOpen
  \bibfield  {author} {\bibinfo {author} {\bibfnamefont {M.}~\bibnamefont
  {Wouters}}, \bibinfo {author} {\bibfnamefont {I.}~\bibnamefont {Carusotto}},\
  and\ \bibinfo {author} {\bibfnamefont {C.}~\bibnamefont {Ciuti}},\ }\bibfield
   {title} {\bibinfo {title} {Spatial and spectral shape of inhomogeneous
  nonequilibrium exciton-polariton condensates},\ }\href
  {https://doi.org/10.1103/PhysRevB.77.115340} {\bibfield  {journal} {\bibinfo
  {journal} {Phys. Rev. B}\ }\textbf {\bibinfo {volume} {77}},\ \bibinfo
  {pages} {115340} (\bibinfo {year} {2008})}\BibitemShut {NoStop}%
\bibitem [{\citenamefont {Wertz}\ \emph {et~al.}(2010)\citenamefont {Wertz},
  \citenamefont {Ferrier}, \citenamefont {Solnyshkov}, \citenamefont {Johne},
  \citenamefont {Sanvitto}, \citenamefont {Lemaitre}, \citenamefont {Sagnes},
  \citenamefont {Grousson}, \citenamefont {Kavokin}, \citenamefont {Senellart},
  \citenamefont {Malpuech},\ and\ \citenamefont {Bloch}}]{Wertz:NatPhys2010}%
  \BibitemOpen
  \bibfield  {author} {\bibinfo {author} {\bibfnamefont {E.}~\bibnamefont
  {Wertz}}, \bibinfo {author} {\bibfnamefont {L.}~\bibnamefont {Ferrier}},
  \bibinfo {author} {\bibfnamefont {D.~D.}\ \bibnamefont {Solnyshkov}},
  \bibinfo {author} {\bibfnamefont {R.}~\bibnamefont {Johne}}, \bibinfo
  {author} {\bibfnamefont {D.}~\bibnamefont {Sanvitto}}, \bibinfo {author}
  {\bibfnamefont {A.}~\bibnamefont {Lemaitre}}, \bibinfo {author}
  {\bibfnamefont {I.}~\bibnamefont {Sagnes}}, \bibinfo {author} {\bibfnamefont
  {R.}~\bibnamefont {Grousson}}, \bibinfo {author} {\bibfnamefont {A.~V.}\
  \bibnamefont {Kavokin}}, \bibinfo {author} {\bibfnamefont {P.}~\bibnamefont
  {Senellart}}, \bibinfo {author} {\bibfnamefont {G.}~\bibnamefont
  {Malpuech}},\ and\ \bibinfo {author} {\bibfnamefont {J.}~\bibnamefont
  {Bloch}},\ }\bibfield  {title} {\bibinfo {title} {Spontaneous formation and
  optical manipulation of extended polariton condensates},\ }\href
  {https://doi.org/10.1038/NPHYS1750} {\bibfield  {journal} {\bibinfo
  {journal} {Nat. Phys.}\ }\textbf {\bibinfo {volume} {6}},\ \bibinfo {pages}
  {860} (\bibinfo {year} {2010})}\BibitemShut {NoStop}%
\bibitem [{\citenamefont {Tanese}\ \emph {et~al.}(2013)\citenamefont {Tanese},
  \citenamefont {Flayac}, \citenamefont {Solnyshkov}, \citenamefont {Amo},
  \citenamefont {Lema{\^\i}tre}, \citenamefont {Galopin}, \citenamefont
  {Braive}, \citenamefont {Senellart}, \citenamefont {Sagnes}, \citenamefont
  {Malpuech} \emph {et~al.}}]{Tanese:NatComm2013}%
  \BibitemOpen
  \bibfield  {author} {\bibinfo {author} {\bibfnamefont {D.}~\bibnamefont
  {Tanese}}, \bibinfo {author} {\bibfnamefont {H.}~\bibnamefont {Flayac}},
  \bibinfo {author} {\bibfnamefont {D.}~\bibnamefont {Solnyshkov}}, \bibinfo
  {author} {\bibfnamefont {A.}~\bibnamefont {Amo}}, \bibinfo {author}
  {\bibfnamefont {A.}~\bibnamefont {Lema{\^\i}tre}}, \bibinfo {author}
  {\bibfnamefont {E.}~\bibnamefont {Galopin}}, \bibinfo {author} {\bibfnamefont
  {R.}~\bibnamefont {Braive}}, \bibinfo {author} {\bibfnamefont
  {P.}~\bibnamefont {Senellart}}, \bibinfo {author} {\bibfnamefont
  {I.}~\bibnamefont {Sagnes}}, \bibinfo {author} {\bibfnamefont
  {G.}~\bibnamefont {Malpuech}}, \emph {et~al.},\ }\bibfield  {title} {\bibinfo
  {title} {Polariton condensation in solitonic gap states in a one-dimensional
  periodic potential},\ }\href@noop {} {\bibfield  {journal} {\bibinfo
  {journal} {Nature communications}\ }\textbf {\bibinfo {volume} {4}},\
  \bibinfo {pages} {1749} (\bibinfo {year} {2013})}\BibitemShut {NoStop}%
\bibitem [{\citenamefont {Baboux}\ \emph {et~al.}(2018)\citenamefont {Baboux},
  \citenamefont {De~Bernardis}, \citenamefont {Goblot}, \citenamefont
  {Gladilin}, \citenamefont {Gomez}, \citenamefont {Galopin}, \citenamefont
  {Le~Gratiet}, \citenamefont {Lema{\^\i}tre}, \citenamefont {Sagnes},
  \citenamefont {Carusotto} \emph {et~al.}}]{Baboux:Optica2018}%
  \BibitemOpen
  \bibfield  {author} {\bibinfo {author} {\bibfnamefont {F.}~\bibnamefont
  {Baboux}}, \bibinfo {author} {\bibfnamefont {D.}~\bibnamefont
  {De~Bernardis}}, \bibinfo {author} {\bibfnamefont {V.}~\bibnamefont
  {Goblot}}, \bibinfo {author} {\bibfnamefont {V.}~\bibnamefont {Gladilin}},
  \bibinfo {author} {\bibfnamefont {C.}~\bibnamefont {Gomez}}, \bibinfo
  {author} {\bibfnamefont {E.}~\bibnamefont {Galopin}}, \bibinfo {author}
  {\bibfnamefont {L.}~\bibnamefont {Le~Gratiet}}, \bibinfo {author}
  {\bibfnamefont {A.}~\bibnamefont {Lema{\^\i}tre}}, \bibinfo {author}
  {\bibfnamefont {I.}~\bibnamefont {Sagnes}}, \bibinfo {author} {\bibfnamefont
  {I.}~\bibnamefont {Carusotto}}, \emph {et~al.},\ }\bibfield  {title}
  {\bibinfo {title} {Unstable and stable regimes of polariton condensation},\
  }\href@noop {} {\bibfield  {journal} {\bibinfo  {journal} {Optica}\ }\textbf
  {\bibinfo {volume} {5}},\ \bibinfo {pages} {1163} (\bibinfo {year}
  {2018})}\BibitemShut {NoStop}%
\bibitem [{\citenamefont {Nigro}\ \emph {et~al.}(2025)\citenamefont {Nigro},
  \citenamefont {Trypogeorgos}, \citenamefont {Gianfrate}, \citenamefont
  {Sanvitto}, \citenamefont {Carusotto},\ and\ \citenamefont
  {Gerace}}]{Nigro:PRL2025}%
  \BibitemOpen
  \bibfield  {author} {\bibinfo {author} {\bibfnamefont {D.}~\bibnamefont
  {Nigro}}, \bibinfo {author} {\bibfnamefont {D.}~\bibnamefont {Trypogeorgos}},
  \bibinfo {author} {\bibfnamefont {A.}~\bibnamefont {Gianfrate}}, \bibinfo
  {author} {\bibfnamefont {D.}~\bibnamefont {Sanvitto}}, \bibinfo {author}
  {\bibfnamefont {I.}~\bibnamefont {Carusotto}},\ and\ \bibinfo {author}
  {\bibfnamefont {D.}~\bibnamefont {Gerace}},\ }\bibfield  {title} {\bibinfo
  {title} {Supersolidity of polariton condensates in photonic crystal
  waveguides},\ }\href@noop {} {\bibfield  {journal} {\bibinfo  {journal}
  {Physical Review Letters}\ }\textbf {\bibinfo {volume} {134}},\ \bibinfo
  {pages} {056002} (\bibinfo {year} {2025})}\BibitemShut {NoStop}%
\bibitem [{\citenamefont {Trypogeorgos}\ \emph {et~al.}(2025)\citenamefont
  {Trypogeorgos}, \citenamefont {Gianfrate}, \citenamefont {Landini},
  \citenamefont {Nigro}, \citenamefont {Gerace}, \citenamefont {Carusotto},
  \citenamefont {Riminucci}, \citenamefont {Baldwin}, \citenamefont {Pfeiffer},
  \citenamefont {Martone} \emph {et~al.}}]{trypogeorgos2025emerging}%
  \BibitemOpen
  \bibfield  {author} {\bibinfo {author} {\bibfnamefont {D.}~\bibnamefont
  {Trypogeorgos}}, \bibinfo {author} {\bibfnamefont {A.}~\bibnamefont
  {Gianfrate}}, \bibinfo {author} {\bibfnamefont {M.}~\bibnamefont {Landini}},
  \bibinfo {author} {\bibfnamefont {D.}~\bibnamefont {Nigro}}, \bibinfo
  {author} {\bibfnamefont {D.}~\bibnamefont {Gerace}}, \bibinfo {author}
  {\bibfnamefont {I.}~\bibnamefont {Carusotto}}, \bibinfo {author}
  {\bibfnamefont {F.}~\bibnamefont {Riminucci}}, \bibinfo {author}
  {\bibfnamefont {K.~W.}\ \bibnamefont {Baldwin}}, \bibinfo {author}
  {\bibfnamefont {L.~N.}\ \bibnamefont {Pfeiffer}}, \bibinfo {author}
  {\bibfnamefont {G.~I.}\ \bibnamefont {Martone}}, \emph {et~al.},\ }\bibfield
  {title} {\bibinfo {title} {Emerging supersolidity in photonic-crystal
  polariton condensates},\ }\href@noop {} {\bibfield  {journal} {\bibinfo
  {journal} {Nature}\ ,\ \bibinfo {pages} {1}} (\bibinfo {year}
  {2025})}\BibitemShut {NoStop}%
\bibitem [{\citenamefont {Recati}\ and\ \citenamefont
  {Stringari}(2023)}]{recati2023supersolidity}%
  \BibitemOpen
  \bibfield  {author} {\bibinfo {author} {\bibfnamefont {A.}~\bibnamefont
  {Recati}}\ and\ \bibinfo {author} {\bibfnamefont {S.}~\bibnamefont
  {Stringari}},\ }\bibfield  {title} {\bibinfo {title} {Supersolidity in
  ultracold dipolar gases},\ }\href@noop {} {\bibfield  {journal} {\bibinfo
  {journal} {Nature Reviews Physics}\ }\textbf {\bibinfo {volume} {5}},\
  \bibinfo {pages} {735} (\bibinfo {year} {2023})}\BibitemShut {NoStop}%
\bibitem [{\citenamefont {Diddams}\ \emph {et~al.}(2020)\citenamefont
  {Diddams}, \citenamefont {Vahala},\ and\ \citenamefont
  {Udem}}]{diddams2020optical}%
  \BibitemOpen
  \bibfield  {author} {\bibinfo {author} {\bibfnamefont {S.~A.}\ \bibnamefont
  {Diddams}}, \bibinfo {author} {\bibfnamefont {K.}~\bibnamefont {Vahala}},\
  and\ \bibinfo {author} {\bibfnamefont {T.}~\bibnamefont {Udem}},\ }\bibfield
  {title} {\bibinfo {title} {Optical frequency combs: Coherently uniting the
  electromagnetic spectrum},\ }\href@noop {} {\bibfield  {journal} {\bibinfo
  {journal} {Science}\ }\textbf {\bibinfo {volume} {369}},\ \bibinfo {pages}
  {eaay3676} (\bibinfo {year} {2020})}\BibitemShut {NoStop}%
\bibitem [{\citenamefont {Chang}\ \emph {et~al.}(2022)\citenamefont {Chang},
  \citenamefont {Liu},\ and\ \citenamefont {Bowers}}]{chang2022integrated}%
  \BibitemOpen
  \bibfield  {author} {\bibinfo {author} {\bibfnamefont {L.}~\bibnamefont
  {Chang}}, \bibinfo {author} {\bibfnamefont {S.}~\bibnamefont {Liu}},\ and\
  \bibinfo {author} {\bibfnamefont {J.~E.}\ \bibnamefont {Bowers}},\ }\bibfield
   {title} {\bibinfo {title} {Integrated optical frequency comb technologies},\
  }\href@noop {} {\bibfield  {journal} {\bibinfo  {journal} {Nature Photonics}\
  }\textbf {\bibinfo {volume} {16}},\ \bibinfo {pages} {95} (\bibinfo {year}
  {2022})}\BibitemShut {NoStop}%
\bibitem [{\citenamefont {Cohen-Tannoudji}(01 2)}]{CCT:CdF}%
  \BibitemOpen
  \bibfield  {author} {\bibinfo {author} {\bibfnamefont {C.}~\bibnamefont
  {Cohen-Tannoudji}},\ }\href {http://www.phys.ens.fr/cours/college-de-france/}
  {\bibfield  {journal} {\bibinfo  {journal} {Lectures at Coll\`ege de France}\
  } (\bibinfo {year} {2001-2})}\BibitemShut {NoStop}%
\bibitem [{Note2()}]{Note2}%
  \BibitemOpen
  \bibinfo {note} {It is interesting to note that a three-dimensional
  easy-plane (in the $xy$ plane) model of ferromagnetism subject to an
  orthogonal magnetic field (along $z$) is expected to displays a steady
  rotation of the magnetization rotation along $z$. As a result, the $x$ and
  $y$ components of the magnetization display monochromatic oscillations as it
  happens to the field $\psi $ of the optical model.}\BibitemShut {Stop}%
\bibitem [{\citenamefont {Wouters}\ and\ \citenamefont
  {Carusotto}(2007)}]{Wouters:PRA2007}%
  \BibitemOpen
  \bibfield  {author} {\bibinfo {author} {\bibfnamefont {M.}~\bibnamefont
  {Wouters}}\ and\ \bibinfo {author} {\bibfnamefont {I.}~\bibnamefont
  {Carusotto}},\ }\bibfield  {title} {\bibinfo {title} {Goldstone mode of
  optical parametric oscillators in planar semiconductor microcavities in the
  strong-coupling regime},\ }\href {https://doi.org/10.1103/PhysRevA.76.043807}
  {\bibfield  {journal} {\bibinfo  {journal} {Phys. Rev. A}\ }\textbf {\bibinfo
  {volume} {76}},\ \bibinfo {pages} {043807} (\bibinfo {year}
  {2007})}\BibitemShut {NoStop}%
\bibitem [{\citenamefont {Graham}\ and\ \citenamefont
  {Haken}(1970)}]{Graham:ZPhys1970}%
  \BibitemOpen
  \bibfield  {author} {\bibinfo {author} {\bibfnamefont {R.}~\bibnamefont
  {Graham}}\ and\ \bibinfo {author} {\bibfnamefont {H.}~\bibnamefont {Haken}},\
  }\bibfield  {title} {\bibinfo {title} {Laserlight--first example of a
  second-order phase transition far away from thermal equilibrium},\ }\href
  {http://dx.doi.org/10.1007/BF01400474} {\bibfield  {journal} {\bibinfo
  {journal} {Zeitschrift f\"ur Physik A Hadrons and Nuclei}\ }\textbf {\bibinfo
  {volume} {237}},\ \bibinfo {pages} {31} (\bibinfo {year} {1970})},\ \bibinfo
  {note} {10.1007/BF01400474}\BibitemShut {NoStop}%
\bibitem [{\citenamefont {Wouters}\ and\ \citenamefont
  {Carusotto}(2006)}]{Wouters:PRB2006}%
  \BibitemOpen
  \bibfield  {author} {\bibinfo {author} {\bibfnamefont {M.}~\bibnamefont
  {Wouters}}\ and\ \bibinfo {author} {\bibfnamefont {I.}~\bibnamefont
  {Carusotto}},\ }\bibfield  {title} {\bibinfo {title} {Absence of long-range
  coherence in the parametric emission of photonic wires},\ }\href
  {https://doi.org/10.1103/PhysRevB.74.245316} {\bibfield  {journal} {\bibinfo
  {journal} {Phys. Rev. B}\ }\textbf {\bibinfo {volume} {74}},\ \bibinfo
  {pages} {245316} (\bibinfo {year} {2006})}\BibitemShut {NoStop}%
\bibitem [{\citenamefont {Altman}\ \emph {et~al.}(2015)\citenamefont {Altman},
  \citenamefont {Sieberer}, \citenamefont {Chen}, \citenamefont {Diehl},\ and\
  \citenamefont {Toner}}]{Altman:PRX2015}%
  \BibitemOpen
  \bibfield  {author} {\bibinfo {author} {\bibfnamefont {E.}~\bibnamefont
  {Altman}}, \bibinfo {author} {\bibfnamefont {L.~M.}\ \bibnamefont
  {Sieberer}}, \bibinfo {author} {\bibfnamefont {L.}~\bibnamefont {Chen}},
  \bibinfo {author} {\bibfnamefont {S.}~\bibnamefont {Diehl}},\ and\ \bibinfo
  {author} {\bibfnamefont {J.}~\bibnamefont {Toner}},\ }\bibfield  {title}
  {\bibinfo {title} {Two-dimensional superfluidity of exciton polaritons
  requires strong anisotropy},\ }\href@noop {} {\bibfield  {journal} {\bibinfo
  {journal} {Physical Review X}\ }\textbf {\bibinfo {volume} {5}},\ \bibinfo
  {pages} {011017} (\bibinfo {year} {2015})}\BibitemShut {NoStop}%
\bibitem [{\citenamefont {Sieberer}\ \emph
  {et~al.}(2016{\natexlab{b}})\citenamefont {Sieberer}, \citenamefont
  {Buchhold},\ and\ \citenamefont {Diehl}}]{Sieberer:RPP2016}%
  \BibitemOpen
  \bibfield  {author} {\bibinfo {author} {\bibfnamefont {L.~M.}\ \bibnamefont
  {Sieberer}}, \bibinfo {author} {\bibfnamefont {M.}~\bibnamefont {Buchhold}},\
  and\ \bibinfo {author} {\bibfnamefont {S.}~\bibnamefont {Diehl}},\ }\bibfield
   {title} {\bibinfo {title} {Keldysh field theory for driven open quantum
  systems},\ }\href@noop {} {\bibfield  {journal} {\bibinfo  {journal} {Reports
  on Progress in Physics}\ }\textbf {\bibinfo {volume} {79}},\ \bibinfo {pages}
  {096001} (\bibinfo {year} {2016}{\natexlab{b}})}\BibitemShut {NoStop}%
\bibitem [{\citenamefont {Zamora}\ \emph {et~al.}(2017)\citenamefont {Zamora},
  \citenamefont {Sieberer}, \citenamefont {Dunnett}, \citenamefont {Diehl},\
  and\ \citenamefont {Szyma{\'n}ska}}]{Zamora:PRX2017}%
  \BibitemOpen
  \bibfield  {author} {\bibinfo {author} {\bibfnamefont {A.}~\bibnamefont
  {Zamora}}, \bibinfo {author} {\bibfnamefont {L.}~\bibnamefont {Sieberer}},
  \bibinfo {author} {\bibfnamefont {K.}~\bibnamefont {Dunnett}}, \bibinfo
  {author} {\bibfnamefont {S.}~\bibnamefont {Diehl}},\ and\ \bibinfo {author}
  {\bibfnamefont {M.}~\bibnamefont {Szyma{\'n}ska}},\ }\bibfield  {title}
  {\bibinfo {title} {Tuning across universalities with a driven open
  condensate},\ }\href@noop {} {\bibfield  {journal} {\bibinfo  {journal}
  {Physical Review X}\ }\textbf {\bibinfo {volume} {7}},\ \bibinfo {pages}
  {041006} (\bibinfo {year} {2017})}\BibitemShut {NoStop}%
\bibitem [{\citenamefont {Fontaine}\ \emph {et~al.}(2022)\citenamefont
  {Fontaine}, \citenamefont {Squizzato}, \citenamefont {Baboux}, \citenamefont
  {Amelio}, \citenamefont {Lema{\^\i}tre}, \citenamefont {Morassi},
  \citenamefont {Sagnes}, \citenamefont {Le~Gratiet}, \citenamefont {Harouri},
  \citenamefont {Wouters} \emph {et~al.}}]{fontaine2022kardar}%
  \BibitemOpen
  \bibfield  {author} {\bibinfo {author} {\bibfnamefont {Q.}~\bibnamefont
  {Fontaine}}, \bibinfo {author} {\bibfnamefont {D.}~\bibnamefont {Squizzato}},
  \bibinfo {author} {\bibfnamefont {F.}~\bibnamefont {Baboux}}, \bibinfo
  {author} {\bibfnamefont {I.}~\bibnamefont {Amelio}}, \bibinfo {author}
  {\bibfnamefont {A.}~\bibnamefont {Lema{\^\i}tre}}, \bibinfo {author}
  {\bibfnamefont {M.}~\bibnamefont {Morassi}}, \bibinfo {author} {\bibfnamefont
  {I.}~\bibnamefont {Sagnes}}, \bibinfo {author} {\bibfnamefont
  {L.}~\bibnamefont {Le~Gratiet}}, \bibinfo {author} {\bibfnamefont
  {A.}~\bibnamefont {Harouri}}, \bibinfo {author} {\bibfnamefont
  {M.}~\bibnamefont {Wouters}}, \emph {et~al.},\ }\bibfield  {title} {\bibinfo
  {title} {Kardar--parisi--zhang universality in a one-dimensional polariton
  condensate},\ }\href@noop {} {\bibfield  {journal} {\bibinfo  {journal}
  {Nature}\ }\textbf {\bibinfo {volume} {608}},\ \bibinfo {pages} {687}
  (\bibinfo {year} {2022})}\BibitemShut {NoStop}%
\bibitem [{\citenamefont {Ozawa}\ \emph {et~al.}(2016)\citenamefont {Ozawa},
  \citenamefont {Price}, \citenamefont {Goldman}, \citenamefont {Zilberberg},\
  and\ \citenamefont {Carusotto}}]{Ozawa:PRA2016}%
  \BibitemOpen
  \bibfield  {author} {\bibinfo {author} {\bibfnamefont {T.}~\bibnamefont
  {Ozawa}}, \bibinfo {author} {\bibfnamefont {H.~M.}\ \bibnamefont {Price}},
  \bibinfo {author} {\bibfnamefont {N.}~\bibnamefont {Goldman}}, \bibinfo
  {author} {\bibfnamefont {O.}~\bibnamefont {Zilberberg}},\ and\ \bibinfo
  {author} {\bibfnamefont {I.}~\bibnamefont {Carusotto}},\ }\bibfield  {title}
  {\bibinfo {title} {Synthetic dimensions in integrated photonics: From optical
  isolation to four-dimensional quantum hall physics},\ }\href@noop {}
  {\bibfield  {journal} {\bibinfo  {journal} {Physical Review A}\ }\textbf
  {\bibinfo {volume} {93}},\ \bibinfo {pages} {043827} (\bibinfo {year}
  {2016})}\BibitemShut {NoStop}%
\bibitem [{\citenamefont {Ehrhardt}\ \emph {et~al.}(2023)\citenamefont
  {Ehrhardt}, \citenamefont {Weidemann}, \citenamefont {Maczewsky},
  \citenamefont {Heinrich},\ and\ \citenamefont {Szameit}}]{Ehrhardt:LPR2023}%
  \BibitemOpen
  \bibfield  {author} {\bibinfo {author} {\bibfnamefont {M.}~\bibnamefont
  {Ehrhardt}}, \bibinfo {author} {\bibfnamefont {S.}~\bibnamefont {Weidemann}},
  \bibinfo {author} {\bibfnamefont {L.~J.}\ \bibnamefont {Maczewsky}}, \bibinfo
  {author} {\bibfnamefont {M.}~\bibnamefont {Heinrich}},\ and\ \bibinfo
  {author} {\bibfnamefont {A.}~\bibnamefont {Szameit}},\ }\bibfield  {title}
  {\bibinfo {title} {A perspective on synthetic dimensions in photonics},\
  }\href@noop {} {\bibfield  {journal} {\bibinfo  {journal} {Laser \& Photonics
  Reviews}\ }\textbf {\bibinfo {volume} {17}},\ \bibinfo {pages} {2200518}
  (\bibinfo {year} {2023})}\BibitemShut {NoStop}%
\bibitem [{\citenamefont {Yuan}\ \emph {et~al.}(2018)\citenamefont {Yuan},
  \citenamefont {Lin}, \citenamefont {Xiao},\ and\ \citenamefont
  {Fan}}]{Yuan:Optica2018}%
  \BibitemOpen
  \bibfield  {author} {\bibinfo {author} {\bibfnamefont {L.}~\bibnamefont
  {Yuan}}, \bibinfo {author} {\bibfnamefont {Q.}~\bibnamefont {Lin}}, \bibinfo
  {author} {\bibfnamefont {M.}~\bibnamefont {Xiao}},\ and\ \bibinfo {author}
  {\bibfnamefont {S.}~\bibnamefont {Fan}},\ }\bibfield  {title} {\bibinfo
  {title} {Synthetic dimension in photonics},\ }\href@noop {} {\bibfield
  {journal} {\bibinfo  {journal} {optica}\ }\textbf {\bibinfo {volume} {5}},\
  \bibinfo {pages} {1396} (\bibinfo {year} {2018})}\BibitemShut {NoStop}%
\bibitem [{\citenamefont {Takeuchi}(2018)}]{TAKEUCHI201877}%
  \BibitemOpen
  \bibfield  {author} {\bibinfo {author} {\bibfnamefont {K.~A.}\ \bibnamefont
  {Takeuchi}},\ }\bibfield  {title} {\bibinfo {title} {An appetizer to modern
  developments on the kardar–parisi–zhang universality class},\ }\href
  {https://doi.org/https://doi.org/10.1016/j.physa.2018.03.009} {\bibfield
  {journal} {\bibinfo  {journal} {Physica A: Statistical Mechanics and its
  Applications}\ }\textbf {\bibinfo {volume} {504}},\ \bibinfo {pages} {77}
  (\bibinfo {year} {2018})},\ \bibinfo {note} {lecture Notes of the 14th
  International Summer School on Fundamental Problems in Statistical
  Physics}\BibitemShut {NoStop}%
\bibitem [{\citenamefont {Wiese}(1998)}]{wiese1998perturbation}%
  \BibitemOpen
  \bibfield  {author} {\bibinfo {author} {\bibfnamefont {K.~J.}\ \bibnamefont
  {Wiese}},\ }\bibfield  {title} {\bibinfo {title} {On the perturbation
  expansion of the kpz equation},\ }\href@noop {} {\bibfield  {journal}
  {\bibinfo  {journal} {Journal of statistical physics}\ }\textbf {\bibinfo
  {volume} {93}},\ \bibinfo {pages} {143} (\bibinfo {year} {1998})}\BibitemShut
  {NoStop}%
\bibitem [{\citenamefont {Marinari}\ \emph {et~al.}(2000)\citenamefont
  {Marinari}, \citenamefont {Pagnani},\ and\ \citenamefont
  {Parisi}}]{marinari2000critical}%
  \BibitemOpen
  \bibfield  {author} {\bibinfo {author} {\bibfnamefont {E.}~\bibnamefont
  {Marinari}}, \bibinfo {author} {\bibfnamefont {A.}~\bibnamefont {Pagnani}},\
  and\ \bibinfo {author} {\bibfnamefont {G.}~\bibnamefont {Parisi}},\
  }\bibfield  {title} {\bibinfo {title} {Critical exponents of the kpz equation
  via multi-surfacecoding numerical simulations},\ }\href@noop {} {\bibfield
  {journal} {\bibinfo  {journal} {Journal of Physics A: Mathematical and
  General}\ }\textbf {\bibinfo {volume} {33}},\ \bibinfo {pages} {8181}
  (\bibinfo {year} {2000})}\BibitemShut {NoStop}%
\bibitem [{\citenamefont {Canet}\ \emph {et~al.}(2010)\citenamefont {Canet},
  \citenamefont {Chat\'e}, \citenamefont {Delamotte},\ and\ \citenamefont
  {Wschebor}}]{Canet:PRL2010}%
  \BibitemOpen
  \bibfield  {author} {\bibinfo {author} {\bibfnamefont {L.}~\bibnamefont
  {Canet}}, \bibinfo {author} {\bibfnamefont {H.}~\bibnamefont {Chat\'e}},
  \bibinfo {author} {\bibfnamefont {B.}~\bibnamefont {Delamotte}},\ and\
  \bibinfo {author} {\bibfnamefont {N.}~\bibnamefont {Wschebor}},\ }\bibfield
  {title} {\bibinfo {title} {Nonperturbative renormalization group for the
  kardar-parisi-zhang equation},\ }\href
  {https://doi.org/10.1103/PhysRevLett.104.150601} {\bibfield  {journal}
  {\bibinfo  {journal} {Phys. Rev. Lett.}\ }\textbf {\bibinfo {volume} {104}},\
  \bibinfo {pages} {150601} (\bibinfo {year} {2010})}\BibitemShut {NoStop}%
\bibitem [{\citenamefont {Imamoglu}\ \emph {et~al.}(1997)\citenamefont
  {Imamoglu}, \citenamefont {Schmidt}, \citenamefont {Woods},\ and\
  \citenamefont {Deutsch}}]{Imamoglu:PRL97}%
  \BibitemOpen
  \bibfield  {author} {\bibinfo {author} {\bibfnamefont {A.}~\bibnamefont
  {Imamoglu}}, \bibinfo {author} {\bibfnamefont {H.}~\bibnamefont {Schmidt}},
  \bibinfo {author} {\bibfnamefont {G.}~\bibnamefont {Woods}},\ and\ \bibinfo
  {author} {\bibfnamefont {M.}~\bibnamefont {Deutsch}},\ }\bibfield  {title}
  {\bibinfo {title} {Strongly interacting photons in a nonlinear cavity},\
  }\href@noop {} {\bibfield  {journal} {\bibinfo  {journal} {Phys. Rev. Lett.}\
  }\textbf {\bibinfo {volume} {79}},\ \bibinfo {pages} {1467} (\bibinfo {year}
  {1997})}\BibitemShut {NoStop}%
\bibitem [{\citenamefont {Carusotto}\ \emph {et~al.}(2009)\citenamefont
  {Carusotto}, \citenamefont {Gerace}, \citenamefont {Tureci}, \citenamefont
  {De~Liberato}, \citenamefont {Ciuti},\ and\ \citenamefont
  {Imamo\ifmmode~\check{g}\else \v{g}\fi{}lu}}]{Carusotto:PRL2009}%
  \BibitemOpen
  \bibfield  {author} {\bibinfo {author} {\bibfnamefont {I.}~\bibnamefont
  {Carusotto}}, \bibinfo {author} {\bibfnamefont {D.}~\bibnamefont {Gerace}},
  \bibinfo {author} {\bibfnamefont {H.~E.}\ \bibnamefont {Tureci}}, \bibinfo
  {author} {\bibfnamefont {S.}~\bibnamefont {De~Liberato}}, \bibinfo {author}
  {\bibfnamefont {C.}~\bibnamefont {Ciuti}},\ and\ \bibinfo {author}
  {\bibfnamefont {A.}~\bibnamefont {Imamo\ifmmode~\check{g}\else
  \v{g}\fi{}lu}},\ }\bibfield  {title} {\bibinfo {title} {Fermionized photons
  in an array of driven dissipative nonlinear cavities},\ }\href
  {https://doi.org/10.1103/PhysRevLett.103.033601} {\bibfield  {journal}
  {\bibinfo  {journal} {Phys. Rev. Lett.}\ }\textbf {\bibinfo {volume} {103}},\
  \bibinfo {pages} {033601} (\bibinfo {year} {2009})}\BibitemShut {NoStop}%
\bibitem [{\citenamefont {Johansson}\ \emph {et~al.}(2012)\citenamefont
  {Johansson}, \citenamefont {Nation},\ and\ \citenamefont {Nori}}]{qutip}%
  \BibitemOpen
  \bibfield  {author} {\bibinfo {author} {\bibfnamefont {J.~R.}\ \bibnamefont
  {Johansson}}, \bibinfo {author} {\bibfnamefont {P.~D.}\ \bibnamefont
  {Nation}},\ and\ \bibinfo {author} {\bibfnamefont {F.}~\bibnamefont {Nori}},\
  }\bibfield  {title} {\bibinfo {title} {Qutip: An open-source python framework
  for the dynamics of open quantum systems},\ }\href@noop {} {\bibfield
  {journal} {\bibinfo  {journal} {Computer physics communications}\ }\textbf
  {\bibinfo {volume} {183}},\ \bibinfo {pages} {1760} (\bibinfo {year}
  {2012})}\BibitemShut {NoStop}%
\bibitem [{\citenamefont {Castin}(2004)}]{CastinLectures2004}%
  \BibitemOpen
  \bibfield  {author} {\bibinfo {author} {\bibfnamefont {Y.}~\bibnamefont
  {Castin}},\ }\bibfield  {title} {\bibinfo {title} {Simple theoretical tools
  for low dimension bose gases},\ }in\ \href@noop {} {\emph {\bibinfo
  {booktitle} {Journal de Physique IV (Proceedings)}}},\ Vol.\ \bibinfo
  {volume} {116}\ (\bibinfo {organization} {EDP sciences},\ \bibinfo {year}
  {2004})\ pp.\ \bibinfo {pages} {89--132}\BibitemShut {NoStop}%
\bibitem [{\citenamefont {Cohen-Tannoudji}\ \emph {et~al.}(1998)\citenamefont
  {Cohen-Tannoudji}, \citenamefont {Dupont-Roc},\ and\ \citenamefont
  {Grynberg}}]{CCT4}%
  \BibitemOpen
  \bibfield  {author} {\bibinfo {author} {\bibfnamefont {C.}~\bibnamefont
  {Cohen-Tannoudji}}, \bibinfo {author} {\bibfnamefont {J.}~\bibnamefont
  {Dupont-Roc}},\ and\ \bibinfo {author} {\bibfnamefont {G.}~\bibnamefont
  {Grynberg}},\ }\href@noop {} {\emph {\bibinfo {title} {Atom-photon
  interactions: basic processes and applications}}}\ (\bibinfo  {publisher}
  {John Wiley \& Sons},\ \bibinfo {year} {1998})\BibitemShut {NoStop}%
\bibitem [{\citenamefont {Fedorov}\ \emph {et~al.}(2021)\citenamefont
  {Fedorov}, \citenamefont {Remizov}, \citenamefont {Shapiro}, \citenamefont
  {Pogosov}, \citenamefont {Egorova}, \citenamefont {Tsitsilin}, \citenamefont
  {Andronik}, \citenamefont {Dobronosova}, \citenamefont {Rodionov},
  \citenamefont {Astafiev},\ and\ \citenamefont {Ustinov}}]{Fedorov:PRL2021}%
  \BibitemOpen
  \bibfield  {author} {\bibinfo {author} {\bibfnamefont {G.~P.}\ \bibnamefont
  {Fedorov}}, \bibinfo {author} {\bibfnamefont {S.~V.}\ \bibnamefont
  {Remizov}}, \bibinfo {author} {\bibfnamefont {D.~S.}\ \bibnamefont
  {Shapiro}}, \bibinfo {author} {\bibfnamefont {W.~V.}\ \bibnamefont
  {Pogosov}}, \bibinfo {author} {\bibfnamefont {E.}~\bibnamefont {Egorova}},
  \bibinfo {author} {\bibfnamefont {I.}~\bibnamefont {Tsitsilin}}, \bibinfo
  {author} {\bibfnamefont {M.}~\bibnamefont {Andronik}}, \bibinfo {author}
  {\bibfnamefont {A.~A.}\ \bibnamefont {Dobronosova}}, \bibinfo {author}
  {\bibfnamefont {I.~A.}\ \bibnamefont {Rodionov}}, \bibinfo {author}
  {\bibfnamefont {O.~V.}\ \bibnamefont {Astafiev}},\ and\ \bibinfo {author}
  {\bibfnamefont {A.~V.}\ \bibnamefont {Ustinov}},\ }\bibfield  {title}
  {\bibinfo {title} {Photon transport in a bose-hubbard chain of
  superconducting artificial atoms},\ }\href
  {https://doi.org/10.1103/PhysRevLett.126.180503} {\bibfield  {journal}
  {\bibinfo  {journal} {Phys. Rev. Lett.}\ }\textbf {\bibinfo {volume} {126}},\
  \bibinfo {pages} {180503} (\bibinfo {year} {2021})}\BibitemShut {NoStop}%
\bibitem [{\citenamefont {Umucal{\i}lar}\ and\ \citenamefont
  {Carusotto}(2012)}]{Umucalilar:PRL2012}%
  \BibitemOpen
  \bibfield  {author} {\bibinfo {author} {\bibfnamefont {R.~O.}\ \bibnamefont
  {Umucal{\i}lar}}\ and\ \bibinfo {author} {\bibfnamefont {I.}~\bibnamefont
  {Carusotto}},\ }\bibfield  {title} {\bibinfo {title} {Fractional quantum hall
  states of photons in an array of dissipative coupled cavities},\ }\href
  {https://doi.org/10.1103/PhysRevLett.108.206809} {\bibfield  {journal}
  {\bibinfo  {journal} {Phys. Rev. Lett.}\ }\textbf {\bibinfo {volume} {108}},\
  \bibinfo {pages} {206809} (\bibinfo {year} {2012})}\BibitemShut {NoStop}%
\bibitem [{\citenamefont {Umucal{\i}lar}\ and\ \citenamefont
  {Carusotto}(2013)}]{Umucalilar:PLA2013}%
  \BibitemOpen
  \bibfield  {author} {\bibinfo {author} {\bibfnamefont {R.}~\bibnamefont
  {Umucal{\i}lar}}\ and\ \bibinfo {author} {\bibfnamefont {I.}~\bibnamefont
  {Carusotto}},\ }\bibfield  {title} {\bibinfo {title} {Many-body braiding
  phases in a rotating strongly correlated photon gas},\ }\href@noop {}
  {\bibfield  {journal} {\bibinfo  {journal} {Physics Letters A}\ }\textbf
  {\bibinfo {volume} {377}},\ \bibinfo {pages} {2074} (\bibinfo {year}
  {2013})}\BibitemShut {NoStop}%
\bibitem [{\citenamefont {Clark}\ \emph {et~al.}(2020)\citenamefont {Clark},
  \citenamefont {Schine}, \citenamefont {Baum}, \citenamefont {Jia},\ and\
  \citenamefont {Simon}}]{clark2020observation}%
  \BibitemOpen
  \bibfield  {author} {\bibinfo {author} {\bibfnamefont {L.~W.}\ \bibnamefont
  {Clark}}, \bibinfo {author} {\bibfnamefont {N.}~\bibnamefont {Schine}},
  \bibinfo {author} {\bibfnamefont {C.}~\bibnamefont {Baum}}, \bibinfo {author}
  {\bibfnamefont {N.}~\bibnamefont {Jia}},\ and\ \bibinfo {author}
  {\bibfnamefont {J.}~\bibnamefont {Simon}},\ }\bibfield  {title} {\bibinfo
  {title} {Observation of laughlin states made of light},\ }\href@noop {}
  {\bibfield  {journal} {\bibinfo  {journal} {Nature}\ }\textbf {\bibinfo
  {volume} {582}},\ \bibinfo {pages} {41} (\bibinfo {year} {2020})}\BibitemShut
  {NoStop}%
\bibitem [{\citenamefont {Gorshkov}\ \emph {et~al.}(2011)\citenamefont
  {Gorshkov}, \citenamefont {Otterbach}, \citenamefont {Fleischhauer},
  \citenamefont {Pohl},\ and\ \citenamefont {Lukin}}]{Gorshkov:PRL2011}%
  \BibitemOpen
  \bibfield  {author} {\bibinfo {author} {\bibfnamefont {A.~V.}\ \bibnamefont
  {Gorshkov}}, \bibinfo {author} {\bibfnamefont {J.}~\bibnamefont {Otterbach}},
  \bibinfo {author} {\bibfnamefont {M.}~\bibnamefont {Fleischhauer}}, \bibinfo
  {author} {\bibfnamefont {T.}~\bibnamefont {Pohl}},\ and\ \bibinfo {author}
  {\bibfnamefont {M.~D.}\ \bibnamefont {Lukin}},\ }\bibfield  {title} {\bibinfo
  {title} {Photon-photon interactions via rydberg blockade},\ }\href
  {https://doi.org/10.1103/PhysRevLett.107.133602} {\bibfield  {journal}
  {\bibinfo  {journal} {Phys. Rev. Lett.}\ }\textbf {\bibinfo {volume} {107}},\
  \bibinfo {pages} {133602} (\bibinfo {year} {2011})}\BibitemShut {NoStop}%
\bibitem [{\citenamefont {Schine}\ \emph {et~al.}(2016)\citenamefont {Schine},
  \citenamefont {Ryou}, \citenamefont {Gromov}, \citenamefont {Sommer},\ and\
  \citenamefont {Simon}}]{schine2016synthetic}%
  \BibitemOpen
  \bibfield  {author} {\bibinfo {author} {\bibfnamefont {N.}~\bibnamefont
  {Schine}}, \bibinfo {author} {\bibfnamefont {A.}~\bibnamefont {Ryou}},
  \bibinfo {author} {\bibfnamefont {A.}~\bibnamefont {Gromov}}, \bibinfo
  {author} {\bibfnamefont {A.}~\bibnamefont {Sommer}},\ and\ \bibinfo {author}
  {\bibfnamefont {J.}~\bibnamefont {Simon}},\ }\bibfield  {title} {\bibinfo
  {title} {Synthetic landau levels for photons},\ }\href@noop {} {\bibfield
  {journal} {\bibinfo  {journal} {Nature}\ }\textbf {\bibinfo {volume} {534}},\
  \bibinfo {pages} {671} (\bibinfo {year} {2016})}\BibitemShut {NoStop}%
\bibitem [{\citenamefont {Stern}(2008)}]{stern2008anyons}%
  \BibitemOpen
  \bibfield  {author} {\bibinfo {author} {\bibfnamefont {A.}~\bibnamefont
  {Stern}},\ }\bibfield  {title} {\bibinfo {title} {Anyons and the quantum hall
  effect—a pedagogical review},\ }\href@noop {} {\bibfield  {journal}
  {\bibinfo  {journal} {Annals of Physics}\ }\textbf {\bibinfo {volume}
  {323}},\ \bibinfo {pages} {204} (\bibinfo {year} {2008})}\BibitemShut
  {NoStop}%
\bibitem [{\citenamefont {Paredes}\ \emph {et~al.}(2001)\citenamefont
  {Paredes}, \citenamefont {Fedichev}, \citenamefont {Cirac},\ and\
  \citenamefont {Zoller}}]{Paredes:PRL2001}%
  \BibitemOpen
  \bibfield  {author} {\bibinfo {author} {\bibfnamefont {B.}~\bibnamefont
  {Paredes}}, \bibinfo {author} {\bibfnamefont {P.}~\bibnamefont {Fedichev}},
  \bibinfo {author} {\bibfnamefont {J.}~\bibnamefont {Cirac}},\ and\ \bibinfo
  {author} {\bibfnamefont {P.}~\bibnamefont {Zoller}},\ }\bibfield  {title}
  {\bibinfo {title} {1 2-anyons in small atomic bose-einstein condensates},\
  }\href@noop {} {\bibfield  {journal} {\bibinfo  {journal} {Physical review
  letters}\ }\textbf {\bibinfo {volume} {87}},\ \bibinfo {pages} {010402}
  (\bibinfo {year} {2001})}\BibitemShut {NoStop}%
\bibitem [{\citenamefont {Grusdt}\ \emph {et~al.}(2016)\citenamefont {Grusdt},
  \citenamefont {Yao}, \citenamefont {Abanin}, \citenamefont {Fleischhauer},\
  and\ \citenamefont {Demler}}]{Grusdt:NatComm2016}%
  \BibitemOpen
  \bibfield  {author} {\bibinfo {author} {\bibfnamefont {F.}~\bibnamefont
  {Grusdt}}, \bibinfo {author} {\bibfnamefont {N.~Y.}\ \bibnamefont {Yao}},
  \bibinfo {author} {\bibfnamefont {D.}~\bibnamefont {Abanin}}, \bibinfo
  {author} {\bibfnamefont {M.}~\bibnamefont {Fleischhauer}},\ and\ \bibinfo
  {author} {\bibfnamefont {E.}~\bibnamefont {Demler}},\ }\bibfield  {title}
  {\bibinfo {title} {Interferometric measurements of many-body topological
  invariants using mobile impurities},\ }\href@noop {} {\bibfield  {journal}
  {\bibinfo  {journal} {Nature communications}\ }\textbf {\bibinfo {volume}
  {7}},\ \bibinfo {pages} {11994} (\bibinfo {year} {2016})}\BibitemShut
  {NoStop}%
\bibitem [{\citenamefont {de~las Heras}\ \emph {et~al.}(2020)\citenamefont
  {de~las Heras}, \citenamefont {Macaluso},\ and\ \citenamefont
  {Carusotto}}]{de2020anyonic}%
  \BibitemOpen
  \bibfield  {author} {\bibinfo {author} {\bibfnamefont {A.~M.}\ \bibnamefont
  {de~las Heras}}, \bibinfo {author} {\bibfnamefont {E.}~\bibnamefont
  {Macaluso}},\ and\ \bibinfo {author} {\bibfnamefont {I.}~\bibnamefont
  {Carusotto}},\ }\bibfield  {title} {\bibinfo {title} {Anyonic molecules in
  atomic fractional quantum hall liquids: a quantitative probe of fractional
  charge and anyonic statistics},\ }\href@noop {} {\bibfield  {journal}
  {\bibinfo  {journal} {Physical Review X}\ }\textbf {\bibinfo {volume} {10}},\
  \bibinfo {pages} {041058} (\bibinfo {year} {2020})}\BibitemShut {NoStop}%
\bibitem [{Note3()}]{Note3}%
  \BibitemOpen
  \bibinfo {note} {This is easy to see on the analytical form of the Laughlin
  wavefunctions of FQH states~\cite {Tong:QHbook}, $$\psi (z_1,\protect \ldots
  ,z_N)=P(z_1,\protect \ldots ,z_N)\protect \,\Pi _{i>j}(z_i-z_j)^p\protect
  \,e^{-\DOTSB \sum@ \slimits@ _i |z_i|^2}\protect \,, $$ where $P$ is an
  arbitrary symmetric polynomial in the normalized complex $z_i=(x_i+i\protect
  \,y_i)/\ell _B$ coordinates of the $i=1,\protect \ldots ,N$ particles and $p$
  is an integer (even for bosonic particles like photons and odd for fermionic
  particles like electrons) describing the specific FQH state under
  consideration. The non-interacting nature of this state is enforced by the
  presence of factors $(z_i-z_j)$ for any pair $i,j$ of particles, which give
  zeros in the wavefunction for overlapping particles. This impenetrability
  condition is preserved when a particle is removed from the
  system.}\BibitemShut {Stop}%
\bibitem [{\citenamefont {Macaluso}\ and\ \citenamefont
  {Carusotto}(2017)}]{Macaluso:PRA2017}%
  \BibitemOpen
  \bibfield  {author} {\bibinfo {author} {\bibfnamefont {E.}~\bibnamefont
  {Macaluso}}\ and\ \bibinfo {author} {\bibfnamefont {I.}~\bibnamefont
  {Carusotto}},\ }\bibfield  {title} {\bibinfo {title} {Hard-wall confinement
  of a fractional quantum hall liquid},\ }\href@noop {} {\bibfield  {journal}
  {\bibinfo  {journal} {Physical Review A}\ }\textbf {\bibinfo {volume} {96}},\
  \bibinfo {pages} {043607} (\bibinfo {year} {2017})}\BibitemShut {NoStop}%
\bibitem [{\citenamefont {Nardin}\ \emph {et~al.}(2024)\citenamefont {Nardin},
  \citenamefont {De~Bernardis}, \citenamefont {Umucal{\i}lar}, \citenamefont
  {Mazza}, \citenamefont {Rizzi},\ and\ \citenamefont
  {Carusotto}}]{Nardin:PRL2024}%
  \BibitemOpen
  \bibfield  {author} {\bibinfo {author} {\bibfnamefont {A.}~\bibnamefont
  {Nardin}}, \bibinfo {author} {\bibfnamefont {D.}~\bibnamefont
  {De~Bernardis}}, \bibinfo {author} {\bibfnamefont {R.~O.}\ \bibnamefont
  {Umucal{\i}lar}}, \bibinfo {author} {\bibfnamefont {L.}~\bibnamefont
  {Mazza}}, \bibinfo {author} {\bibfnamefont {M.}~\bibnamefont {Rizzi}},\ and\
  \bibinfo {author} {\bibfnamefont {I.}~\bibnamefont {Carusotto}},\ }\bibfield
  {title} {\bibinfo {title} {Quantum nonlinear optics on the edge of a
  few-particle fractional quantum hall fluid in a small lattice},\ }\href@noop
  {} {\bibfield  {journal} {\bibinfo  {journal} {Physical Review Letters}\
  }\textbf {\bibinfo {volume} {133}},\ \bibinfo {pages} {183401} (\bibinfo
  {year} {2024})}\BibitemShut {NoStop}%
\bibitem [{\citenamefont {Binanti}\ \emph {et~al.}(2024)\citenamefont
  {Binanti}, \citenamefont {Goldman},\ and\ \citenamefont
  {Repellin}}]{Binanti:PRR2024}%
  \BibitemOpen
  \bibfield  {author} {\bibinfo {author} {\bibfnamefont {F.}~\bibnamefont
  {Binanti}}, \bibinfo {author} {\bibfnamefont {N.}~\bibnamefont {Goldman}},\
  and\ \bibinfo {author} {\bibfnamefont {C.}~\bibnamefont {Repellin}},\
  }\bibfield  {title} {\bibinfo {title} {Spectroscopy of edge and bulk
  collective modes in fractional chern insulators},\ }\href@noop {} {\bibfield
  {journal} {\bibinfo  {journal} {Physical Review Research}\ }\textbf {\bibinfo
  {volume} {6}},\ \bibinfo {pages} {L012054} (\bibinfo {year}
  {2024})}\BibitemShut {NoStop}%
\bibitem [{\citenamefont {Biella}\ \emph {et~al.}(2017)\citenamefont {Biella},
  \citenamefont {Storme}, \citenamefont {Lebreuilly}, \citenamefont {Rossini},
  \citenamefont {Fazio}, \citenamefont {Carusotto},\ and\ \citenamefont
  {Ciuti}}]{Biella:PRA2017}%
  \BibitemOpen
  \bibfield  {author} {\bibinfo {author} {\bibfnamefont {A.}~\bibnamefont
  {Biella}}, \bibinfo {author} {\bibfnamefont {F.}~\bibnamefont {Storme}},
  \bibinfo {author} {\bibfnamefont {J.}~\bibnamefont {Lebreuilly}}, \bibinfo
  {author} {\bibfnamefont {D.}~\bibnamefont {Rossini}}, \bibinfo {author}
  {\bibfnamefont {R.}~\bibnamefont {Fazio}}, \bibinfo {author} {\bibfnamefont
  {I.}~\bibnamefont {Carusotto}},\ and\ \bibinfo {author} {\bibfnamefont
  {C.}~\bibnamefont {Ciuti}},\ }\bibfield  {title} {\bibinfo {title} {Phase
  diagram of incoherently driven strongly correlated photonic lattices},\
  }\href@noop {} {\bibfield  {journal} {\bibinfo  {journal} {Physical Review
  A}\ }\textbf {\bibinfo {volume} {96}},\ \bibinfo {pages} {023839} (\bibinfo
  {year} {2017})}\BibitemShut {NoStop}%
\bibitem [{\citenamefont {Umucal{\i}lar}\ and\ \citenamefont
  {Carusotto}(2017)}]{Umucalilar:PRA2017}%
  \BibitemOpen
  \bibfield  {author} {\bibinfo {author} {\bibfnamefont {R.}~\bibnamefont
  {Umucal{\i}lar}}\ and\ \bibinfo {author} {\bibfnamefont {I.}~\bibnamefont
  {Carusotto}},\ }\bibfield  {title} {\bibinfo {title} {Generation and
  spectroscopic signatures of a fractional quantum hall liquid of photons in an
  incoherently pumped optical cavity},\ }\href@noop {} {\bibfield  {journal}
  {\bibinfo  {journal} {Physical Review A}\ }\textbf {\bibinfo {volume} {96}},\
  \bibinfo {pages} {053808} (\bibinfo {year} {2017})}\BibitemShut {NoStop}%
\bibitem [{\citenamefont {Umucal{\i}lar}\ \emph {et~al.}(2021)\citenamefont
  {Umucal{\i}lar}, \citenamefont {Simon},\ and\ \citenamefont
  {Carusotto}}]{umucalilar2021autonomous}%
  \BibitemOpen
  \bibfield  {author} {\bibinfo {author} {\bibfnamefont {R.}~\bibnamefont
  {Umucal{\i}lar}}, \bibinfo {author} {\bibfnamefont {J.}~\bibnamefont
  {Simon}},\ and\ \bibinfo {author} {\bibfnamefont {I.}~\bibnamefont
  {Carusotto}},\ }\bibfield  {title} {\bibinfo {title} {Autonomous
  stabilization of photonic laughlin states through angular momentum
  potentials},\ }\href@noop {} {\bibfield  {journal} {\bibinfo  {journal}
  {Physical Review A}\ }\textbf {\bibinfo {volume} {104}},\ \bibinfo {pages}
  {023704} (\bibinfo {year} {2021})}\BibitemShut {NoStop}%
\bibitem [{\citenamefont {Kurilovich}\ \emph {et~al.}(2022)\citenamefont
  {Kurilovich}, \citenamefont {Kurilovich}, \citenamefont {Lebreuilly},\ and\
  \citenamefont {Girvin}}]{kurilovich2022stabilizing}%
  \BibitemOpen
  \bibfield  {author} {\bibinfo {author} {\bibfnamefont {P.}~\bibnamefont
  {Kurilovich}}, \bibinfo {author} {\bibfnamefont {V.~D.}\ \bibnamefont
  {Kurilovich}}, \bibinfo {author} {\bibfnamefont {J.}~\bibnamefont
  {Lebreuilly}},\ and\ \bibinfo {author} {\bibfnamefont {S.~M.}\ \bibnamefont
  {Girvin}},\ }\bibfield  {title} {\bibinfo {title} {Stabilizing the laughlin
  state of light: Dynamics of hole fractionalization},\ }\href@noop {}
  {\bibfield  {journal} {\bibinfo  {journal} {SciPost Physics}\ }\textbf
  {\bibinfo {volume} {13}},\ \bibinfo {pages} {107} (\bibinfo {year}
  {2022})}\BibitemShut {NoStop}%
\bibitem [{\citenamefont {Hafezi}\ \emph {et~al.}(2015)\citenamefont {Hafezi},
  \citenamefont {Adhikari},\ and\ \citenamefont {Taylor}}]{Hafezi:PRB2015}%
  \BibitemOpen
  \bibfield  {author} {\bibinfo {author} {\bibfnamefont {M.}~\bibnamefont
  {Hafezi}}, \bibinfo {author} {\bibfnamefont {P.}~\bibnamefont {Adhikari}},\
  and\ \bibinfo {author} {\bibfnamefont {J.~M.}\ \bibnamefont {Taylor}},\
  }\bibfield  {title} {\bibinfo {title} {Chemical potential for light by
  parametric coupling},\ }\href {https://doi.org/10.1103/PhysRevB.92.174305}
  {\bibfield  {journal} {\bibinfo  {journal} {Phys. Rev. B}\ }\textbf {\bibinfo
  {volume} {92}},\ \bibinfo {pages} {174305} (\bibinfo {year}
  {2015})}\BibitemShut {NoStop}%
\bibitem [{\citenamefont {Lebreuilly}\ \emph {et~al.}(2017)\citenamefont
  {Lebreuilly}, \citenamefont {Biella}, \citenamefont {Storme}, \citenamefont
  {Rossini}, \citenamefont {Fazio}, \citenamefont {Ciuti},\ and\ \citenamefont
  {Carusotto}}]{Lebreuilly:PRA2017}%
  \BibitemOpen
  \bibfield  {author} {\bibinfo {author} {\bibfnamefont {J.}~\bibnamefont
  {Lebreuilly}}, \bibinfo {author} {\bibfnamefont {A.}~\bibnamefont {Biella}},
  \bibinfo {author} {\bibfnamefont {F.}~\bibnamefont {Storme}}, \bibinfo
  {author} {\bibfnamefont {D.}~\bibnamefont {Rossini}}, \bibinfo {author}
  {\bibfnamefont {R.}~\bibnamefont {Fazio}}, \bibinfo {author} {\bibfnamefont
  {C.}~\bibnamefont {Ciuti}},\ and\ \bibinfo {author} {\bibfnamefont
  {I.}~\bibnamefont {Carusotto}},\ }\bibfield  {title} {\bibinfo {title}
  {Stabilizing strongly correlated photon fluids with non-markovian
  reservoirs},\ }\href@noop {} {\bibfield  {journal} {\bibinfo  {journal}
  {Physical Review A}\ }\textbf {\bibinfo {volume} {96}},\ \bibinfo {pages}
  {033828} (\bibinfo {year} {2017})}\BibitemShut {NoStop}%
\bibitem [{\citenamefont {Jacquet}\ \emph {et~al.}(2023)\citenamefont
  {Jacquet}, \citenamefont {Giacomelli}, \citenamefont {Valnais}, \citenamefont
  {Joly}, \citenamefont {Claude}, \citenamefont {Giacobino}, \citenamefont
  {Glorieux}, \citenamefont {Carusotto},\ and\ \citenamefont
  {Bramati}}]{Jacquet:PRL2023}%
  \BibitemOpen
  \bibfield  {author} {\bibinfo {author} {\bibfnamefont {M.~J.}\ \bibnamefont
  {Jacquet}}, \bibinfo {author} {\bibfnamefont {L.}~\bibnamefont {Giacomelli}},
  \bibinfo {author} {\bibfnamefont {Q.}~\bibnamefont {Valnais}}, \bibinfo
  {author} {\bibfnamefont {M.}~\bibnamefont {Joly}}, \bibinfo {author}
  {\bibfnamefont {F.}~\bibnamefont {Claude}}, \bibinfo {author} {\bibfnamefont
  {E.}~\bibnamefont {Giacobino}}, \bibinfo {author} {\bibfnamefont
  {Q.}~\bibnamefont {Glorieux}}, \bibinfo {author} {\bibfnamefont
  {I.}~\bibnamefont {Carusotto}},\ and\ \bibinfo {author} {\bibfnamefont
  {A.}~\bibnamefont {Bramati}},\ }\bibfield  {title} {\bibinfo {title} {Quantum
  vacuum excitation of a quasinormal mode in an analog model of black hole
  spacetime},\ }\href@noop {} {\bibfield  {journal} {\bibinfo  {journal}
  {Physical Review Letters}\ }\textbf {\bibinfo {volume} {130}},\ \bibinfo
  {pages} {111501} (\bibinfo {year} {2023})}\BibitemShut {NoStop}%
\bibitem [{\citenamefont {Ozawa}\ and\ \citenamefont
  {Price}(2019)}]{ozawaprice2019topological}%
  \BibitemOpen
  \bibfield  {author} {\bibinfo {author} {\bibfnamefont {T.}~\bibnamefont
  {Ozawa}}\ and\ \bibinfo {author} {\bibfnamefont {H.~M.}\ \bibnamefont
  {Price}},\ }\bibfield  {title} {\bibinfo {title} {Topological quantum matter
  in synthetic dimensions},\ }\href@noop {} {\bibfield  {journal} {\bibinfo
  {journal} {Nature Reviews Physics}\ }\textbf {\bibinfo {volume} {1}},\
  \bibinfo {pages} {349} (\bibinfo {year} {2019})}\BibitemShut {NoStop}%
\end{thebibliography}%


\end{document}